\documentclass[a4paper,aps,pra,showkeys,showpacs
]{revtex4}  

\usepackage[pdftex]{graphics}
\usepackage{graphics} 
\usepackage{epsfig}
\usepackage{epstopdf}
\usepackage{subfigure}
\usepackage{graphicx}  

\usepackage{dcolumn}   
\usepackage{bm}        
\usepackage{amssymb}   
\usepackage{amsmath}
\usepackage{amsthm}
\usepackage{amstext}%

\newtheorem{theorem}{\bf Theorem}
\def\Proof{\par{\sc \noindent Proof~}}
\renewcommand{\theequation}{\thesection.\theequation}
\numberwithin{equation}{section}

\begin{document}
\DeclareGraphicsExtensions{.jpg,.pdf,.mps,.png,} 
\title{
Non-Abelian Charge Transport  in Three-Flavor Gauge Semimetal Model   
with  Braiding  Majoranas 
}

\author{Halina~V.~Grushevskaya}
\email{grushevskaja@bsu.by} \affiliation{Physics Department,
Belarusian State University, 4 Nezalezhnasti Ave., 220030 Minsk,
BELARUS}
\author{George~Krylov}
\email{krylov@bsu.by}
\affiliation{Physics Department, Belarusian
State University, 4 Nezalezhnasti Ave., 220030 Minsk, BELARUS}

\begin{abstract}
Known Majorana fermions models
are considered as promising ones for the purposes of quantum
computing robust to decoherence.
One of the most expecting but  unachieved goals is an effective control
for
braiding of Majoranas.
Another one is to describe ${\mathbb{Z}}_2$ topological semimetals,
APRES spectra of which testify on eight-fold
degenerate chiral fermions with $SU(2)$ holonomy of wave functions,
whereas the last can not be reproduced within existing models.
%
Quasi-relativistic theory of non-abelian quantum charge transport
in topological semimetals is developed for a  model with a number of flavors equal three.
Majorana-like quasi-particle excitations in the model are described with
accounting of dynamic mass term arising due to relativistic exchange
interactions. Such exotic features of $\mathbb{Z}_2$ semimetals as
 splitting zero-conductance peaks, longitudinal magnetoresistance,
minimal direct current conductivity,
negative differential conductivity have been calculated in perfect agreement
with experimental data.
We propose a new scheme of braiding for three flavor Majorana-like fermions with new
non-trivial braiding operator.
We demonstrate that in this model,
the presence of chiral Majorana-like bound states is controlled as
emergence of three pairs of resonance-antiresonance in frequency dependence of
dielectric permeability.
\end{abstract}

 \maketitle

\section{Introduction}

Topological semimetals (SMs) are graphene-like materials with touching and  non-overlapping valence
and conduction bands. SMs can
exhibit unique magneto-electrical properties, including
an angle-dependent magnetoresistance
\cite{Luo-McDonald-Rosa-SciRep2016,Q-Li-ChiralNat-Phys2016,Du-Wang-Chen-Sci.-China-Phys2016,
Yan-Zhang-Liu-Liu-Zhang-Xiu-Zhou,H-ZhLuSh-QShen2017},
ultrahigh electroconductivity 
\cite{Cooper2012,Bolotin-Sikes-SolidStateCom2008,Wang-Weng-Wu-PhysRev2013,Liang-Gibson-Ali-NatMat2015,Son-Spivak-PhysRev2013}
, as well as 
ultrahigh radiation resistance 
 \cite{Huang-GrapheneDamage-SciRep2016}. 
Experiments  the charge transport in Dirac materials (graphene,
two-dimensional (2D) semimetals  and three-dimensional (3D) topological insulators
(TI) or, more precisely, 
chiral edge gapless 3D-TI-modes) have revealed a chiral anomaly 
of their transport properties and simultaneously signatures of
zero-energy Majorana 
modes \cite{Huang-Zhao-Long-PhysRevX2015}. 
Berry curvature 
for the monolayer graphene model 
diverges in  touching points 
between valence  and  conduction bands (Dirac points of the Brillouin zone, also called valleys)
\cite{Xiao-Chang-Niu2010Rev-Mod-Phys}. The existence 
of topological currents in graphene superlattices has been proposed in \cite{Science346-2014Gorbachev}.
In graphene, signatures of Majorana excitation have been experimentally observed in \cite{PhysRevX5-2015San-Jose}.
Majorana zero-energy modes are new type of quasiparticles, which have been theoretically predicted 
in high-$T_c$ topological p-wave superconductors (Majorana zero-energy modes and gapped
Fermi arcs in angle-resolved photoemission
spectra (ARPES) \cite{RepProgPhys-71-2008Lee,SciTechnolAdvMater15-2014Wu})
\cite{Physics-Uspekhi44-2001Kitaev,JPhysB40-2007Semenoff,ElJTherPhys3-2006Semenoff,PhysicsEMajorana-2014Wilczek}
and have been experimentally discovered 
in a ferromagnetic atomic chain  (one-dimensional quantum wire)
placed on the surface of a conventional s-superconductor \cite{Science346-2014Nadj-Perge}.
A Majorana fermion is its own antiparticle $\chi =\chi ^\dag$ \cite{E.Majorana1937}.
A Dirac mass fermion $\psi $ can be represented as a particle formed from two Majorana fermions (MFs)
$\chi_1$, $\chi_2$: $\psi =(\chi_1+i\chi_2)/2$, $\psi^* =(\chi_1-i\chi_2)/2$,
and Majorana, accordingly, can be considered as  a particle--hole pair from the Dirac fermions.
Here "$^*$"\ denotes the complex conjugation.

The construction of quantum devices based on Dirac materials is a challenge
 because of the lack of understanding
of  deep connection of their unique electrical properties with
the Majorana pattern (texture) of Dirac-fermions pairs in SM,
and as a consequence, the deficit of  simulation methods.

Let us assume that a Majorana fermion can 
be self-fissionable 
with subsequent 
self-fusion. 
Provided the presence of such a mechanism of existence of the Majorana particles, an external electromagnetic
field would separate  charged Majorana states in space and time, resulting in  charge transport.
Such representation of Majorana particle has been called a 
braiding representation in \cite{Nayak-Simon-RevModPhys2008}.
Braiding operators on a set of Majoranas exist 
and form a unitary representation of the circular Artin braid group \cite{Kauffman-Lomonaco-Jr2016ArXiv}.
The materials with braiding Majorana excitations are known to be
promising materials for quantum computing by quantum tunneling
\cite{Physics-Uspekhi44-2001Kitaev,JPhysB40-2007Semenoff,JETPLetters103-2016Zyuzin,arXiv1612-09276v1-2016}.
Unfortunately,  physical grounds of the 2D-braiding are unknown,
in spite of the fact that
braiding Majorana fermions  by Ising spin 1D-chains have been proposed in 
\cite{Backens-Shnirman-Makhlin2017ArXiv}. 

Particles and holes in 2D and 3D Dirac semimetals  (graphene
\cite{Semenoff1984,Novoselov2005,Neto2009,Cooper2012}, Na$_3$Bi \cite{Science343-2014Liu,Science347-2015Xu}, Cd$_3$As$_2$
\cite{NatCommun5-2014Neupane,PhyRevLett113-2014LBorisenko,NatMater13-2014Jeon}, perovskite SM
\cite{Yan-Zhang-Liu-Liu-Zhang-Xiu-Zhou}) 
are massless. But  Majorana equation of motion  for ultrarelativistic massless   fermions
is not an oscillatory 
one \cite{E.Pessa2006JTheorPhys}, 
and,
 \ accordingly, massless Majorana fermions can not represent
a secondary quantized 
field. One can overcome this obstacle using Majorana-like equations derived in \cite{mySymmetry2016}
for quasi-particle 
massless excitations in SM.

Degeneration of the Dirac 
bands is removed in the 
 Weyl SM \cite{Lu-Shen2017Front-Phys,{Hubener_Sentef_Giovannini_Kemper_Rubio}}
 with the Dirac point, splitting into two
massless Weyl  nodes,
and with gapless Fermi arcs in  ARPES observations for  TaAs family of such materials
\cite{Science-349-2015,PhysRevX-5-2015,NatPhys-11-2015Lv,NatPhys-11-2015Xu,NatPhys11-2015Yang,
AnnualRevCondMattPhys2017Hasan,Chin-Phys-Lett2015NbP}.
Hypothetically, one can assume  
that after pairing of Weyl nodes,  nontrivial gapless surface states from the Weyl band structure
are transformed into  Majorana surface modes inside the pairing gap as Majorana  and Fermi arcs in ARPES
for a magnetic Weyl semimetal in superconducting state
\cite{Li-Haldane2015arXiv,Scientific-Reports6-2016ChangHasan,BRoy2016arXiv}.

Moreover, some  topological SMs 
with/without spin-orbit coupling (SOC) or when spin-orbit
interaction being neglected are  nodal-line/ring semimetals
\cite{PhysRevLett115-2015Kim,PhysRevB92-2015ChenFang,NatCommunic7-2016Bian}.
The 1D nodal-line states, which  possess
mirror reflection (inversion) coexisting with time reversal and an
additional non-symmorphic symmetries (screw axis), are
symmetry-protected   so they are stable against perturbations, including SOC
\cite{Science353-2016Bradlyn,PhysRevB92-2015ChenFang,PhysRevLett116-2016Wieder,PhysRevX6-2016Muechler,
Yang-Yang-Derunova2017ArXiv}.
The nodal-line state enhances the surface Rashba splitting
\cite{NatComunnic8-2017Hirayama}.
The 2D drumhead-like surface states  inside the closed nodal ring are nested between conduction and valence bands
\cite{Weng-Liang-Xu2015PhysRev,Burkov-Hook2011PhysRev,Hasan-Drumhead2016PhysRev,Schnyder-Ca3P2PhyRev2016,
Heikkila-Volovik2011JETP,Marzari-Mostofi2012RevModPhys,
Xu-Yu-Fang-Dai-Weng2017PhysRev}. 
The flat band surface states, density of which  is very high,
are similar to those  in a high-temperature superconductor
\cite{Kopnin-Heikkil2011PhysRev,Heikkil-Kopnin2011JETP}. 
A nodal-line  no-magnetic semimetal PbTaSe$_2$ where the Pb-conducting orbitals (the 
particle-like 6p-Pb bands around $\overline K$, which inevitably cross the hole-like 5d-Ta bands with  similar energy,
leading to formation of the nodal rings) form the topological nodal-line states, is a candidate to a topological superconductor
with nontrivial gapless surface states \cite{NatCommunic7-2016Bian,PhysRevB93-2016Cheng-LongZhang}. The nodal lines in
PbTaSe$_2$ with strong SOC are protected by a reflection symmetry of the space group \cite{NatCommunic7-2016Bian}.
The massless chiral Majorana modes in PbTaSe$_2$ are placed on 
a 1D  contour contrary to the 
 Majorana zero-energy modes localized at 0D points.

Right- and left-hand particles always appear in pairs in virtue of the helicity conservation law
 and can not change their helicity as they are massless.
Hence, there must be a mechanism that allows particles to exist
only at one of the touching bands.
Therefore, the graphene-like materials with chiral symmetry or
3D-TI-modes can be described using a mass term taking zero values
at Dirac points, provided the chirality of the remaining "single
valley"\ particle is preserved.
Similar problem of chiral anomaly 
in high energy physics has been treated in the following way. 
For "single valley"\ massless Dirac fermions, a similar mass term could correspond to the so-called Wilson mass term
which vanishes at momentum $\vec k = 0$ and frequency  $\omega = 0$. But the Wilson mass term explicitly breaks the chiral
symmetry \cite{Ginsparg-Wilson1982PhysRevD}. 
Known receipt of its restoration 
\cite{Moran-Leinweber2011PhysLett,Kaplan-Sun2012PhysRevLett,Vafek-Vishwanath2013ArXiv}
is to find a lattice Dirac Hamiltonian with  a sign alternating
mass term 
on a space-like surface in such a way that the mass would be
zero-valued on  all lattice sites rather than in the origin only.
The last leads to emergence of 
the zero-energy mode. 
A mass term (eigenvalues of the mass operator) entering into the  equation of motion for
the Majorana particle should be alternating one due to positive and negative values
of the mass for Dirac particles and anti-particles respectively.
Therefore, the use of Majorana representation is the way to the chiral theory of SMs.
In a Dirac Hamiltonian describing 3D materials with 
band inversion 
\cite{H.Zhang2009NatPhys,C.-X.Liu2010PhysRev} 
a mass term is a sign alternating one 
as a Majorana mass term and similar to 
the Wilson term it gains zero value in the 
Dirac-like point. 
The  band inversion  is typical for 
3D TIs, for example 
for Bi$_2$Se$_3$ \cite{Fu-Kane2007PhysRev} 
where the inversion takes place in 
$\Gamma$-point of the Brillouin zone. 


The discovery of Dirac materials 
with nodal line surrounding
drumhead-like surface states has shown that
the construction of a 2D Hamiltonian by adding a sign alternating mass term is limited
to the cases when the entire nodal line can be placed on a zero mass surface.

 SOC in the absence of 
inversion and time reversal symmetries breaks a nodal line into separate Weyl nodes as it has been calculated in
\cite{Scientific-Reports6-2016ChangHasan,Fang2016WengarXiv,PhysRevB93-2016Hung-Liu-Vanderbit,PhysRevB94-2016Narayan,ChinesePhysB25-2016Weng-Dai}.
However SOC and odd-parity pairing can  realize   
MFs in the nodal topological superconductor phase \cite{ScieAdv2-2016Kozil}. Moreover, in the
stoichiometric high-T$_c$ superconductor (SC) YBa$_2$Cu$_4$O$_8$ under pressure, the quasiparticle mass decreases as
the critical temperature T$_c$ increases  \cite{ScieAdv2-2016Putzke}.
Vortex cores in topological SCs host braiding MFs \cite{SciTechnolAdvMater15-2014Wu}.

Thus, though  Dirac materials are  emerging topological phases with
emergent Dirac \cite{Novoselov2005}, Weyl \cite{Nielsen-Ninomiya1983PhysLettB},
and Majorana fermions \cite{Physics-Uspekhi44-2001Kitaev,LiangFu-Kane2008PRL},
Majorana representation should be a 
background for the description of their properties and charge transport. 
This explains the fact that the 
first-principles band-structure calculations, which are based on
the usage of one-particle quasi-relativistic Dirac equation,
 demonstrate bands crossing but
 "cannot serve as a proof of the Fermi arcs" \
as it has been mentioned in \cite{NatureMaterials15-2016ShuangJiaS-YXu-MZHasan}.
There is a necessity in new theoretical approaches
to design nodal-line structures and  topological
phases "linked"\ with them
\cite{FrontPhys12-2017YuFangDai}.

Dichroism of s-polarized ARPES-spectra  has been observed
for TIs Bi$_2$Se$_3$, Bi$_2$Te$_3$ in \cite{Cao2013NatPhys,Zhang-Liu-Zhang2013PRL,Zhu-Layer-by-layer2013PRL}
and for NbP in \cite{Chin-Phys-Lett2015NbP}. 
Dichroism puts forward the problem to find a Hamiltonian preserving the chiral symmetry
of the Dirac cone bands, rather than 
a set of  Dirac cone apexes (the Dirac points in the Brillouin
zone). Hence, the search for the SM Hamiltonian should be based on new approaches such as the addition of an sign alternating mass
term, which gets  zero values on the surface of the Brillouin
zone, rather than in distinct
Dirac points.

Space of 
TI states is a space of even dimensionality 
$d$, $d=2$, in which an infinite number of pairs of
oppositely twisted vortices with nontrivial topological charges emerges
in accord with 
the Nielsen–-Ninomiya "no-go"\ --theorem on the existence of vortex lattices in only even  dimensionality of the space
\cite{Nielsen-Ninomiya1NuclPhys1981,Nielsen-Ninomiya2NuclPhys1981,MontvayMunster1997}.
The proliferation of these vortices drives the Berezinskii–-Kosterlitz–-Thouless transition in a standard way
\cite{JPhysC6-1973KosterlitzThouless}.
The only way to preserve the chirality of all the vortex "single valley"\
nodes simultaneously with introducing the sign alternating
mass term, is to use a singular (divergent) mass operator.
The massless Weyl nodes in Weyl SMs  were considered 
as topological defect structures (vortices) \cite{JPhysC5-1972KosterlitzThouless} of the type of 
$U(1)$ gauge fields named as skyrmions, or a $SU(2)\ ($O(4)$)$ gauge fields named as merons (half skyrmions)
in \cite{Supercond-Sci-Technol1988}, or $O(3)$  non-linear sigma models
in \cite{NuclPhys336-1990Shankar}. There is a good coincidence of experimental data with
theoretical predictions on topological 
quantum  phase transitions in 
1D XY universality class, which includes different realizations of 
vortex-chain configurations in the continuum limit for free fermions 
\cite{JPhysC6-1973KosterlitzThouless,
PhysicsLettA93-1983Haldane,PhysRevLett50-1983Haldane,RevModPhys69-1997Sondhi,PhysRevB81-2010Pollmann}.
Multiple vortices creation is observed in graphene and 
topological insulators in an electromagnetic field at lowering the 
symmetry of the structure 
\cite{Science340-2013Hunt,SciAdv2-2016SanfengWu}.

A resonating-valence-bond (RVB) picture
\cite{Anderson1973,Fazekas-Anderson1974,Affleck-Marston1988,MySupercinductivity2010}
and its quantum mechanical 1D-formulation \cite{MySupercinductivity2010} model a Fermi arc in ARPES spectrum
as a  break of the double conjugated chemical bond located on the
left (right) side of a certain lattice site with the subsequent
formation of the same bond between  electrons of this site and
electrons of the right(left) site to the considered one.
The RVB picture has been formulated statistically  in terms of skyrmions or  merons
\cite{Sachdev1992,Nagaosa-Lee1992,Wen-Lee1996}. In these field theories, defect staggered spin flows
\cite{Hsu-Marston-Affleck1991} originated from RVB breaks are 
cores of vortices.
 But, skyrmions do not carry
an electric charge, therefore, to overcome the obstacle,
complex constructions with additional quasiparticles
(holons) have been developed \cite{Kivelson-Rokhsar-Sethna1987} to describe
the high $T_c$ conductivity. Moreover, the coupling of massless
fermions to gauge fields in the $O(4)$ ($SU(2)$) non-linear sigma
models is confining. 
Assuming violation of 
$SU(2)$ to $U(1)$  as a deconfinement one gets the so called triple 
quantum electrodynamics (QED$_3$ model) \cite{Vafek-Tesanovic-Franz2002,Lee-Herbut}. To make 
a deconfined state stable, the number 
of matter fields should be large enough 
\cite{Hermele-Senthil-Fisher-Lee-Nagaosa-Wen}.
The zero-energy Majorana bound state (MBS) is associated with the non-abelian excitation.
The  vortex leads to the MBS, for example, in a superconductor $p_x+ i p_y$
\cite{Read-Green2000PhysRev,Ivanov2001PhysRev,SternOppenMariani2004PhysRev,StoneChung2006PhysRev}.

Thus, to  describe nodal-line Weyl semimetals a new theory is required, which would predict a phenomenon similar to
deconfinement, and would be characterized by a sufficiently large number $ N $ of  gauge fields.

Another difficulty is that the Fermi arcs are not closed,
because of that, the Fermi pockets are not formed.
From the Luttinger theorem it follows that the area of the Fermi
surface is the same as that of free fermions, i.e. it is
determined by the total density of electrons in the unit cell
\cite{Abrikosov-Gorkov-Dzyaloshinskii1965}. Violation 
of the Luttinger theorem in  Landau--Fermi liquid theory leads to a
non-conservation of the total electric charge.
Therefore, if the Fermi pockets are not formed, then the violation 
of the Luttinger theorem becomes unavoidable obstacle in utilizing
 Dirac particle physics \cite{Oshikawa2000,Paramekanti-Vishwanath2004} to describe
strongly correlated systems. 
The Majorana 
representation allows  any fermionic system,
either fermion number conserving or not, to be treated on equal
footing
\cite{Jaffe-Pedrocchi2015Annales-Henri-Poincare,Bender-Mannheim2010,Wei-Congjun-Li-Zhang-Xiang2016}.
The nodal lines restore 
the Fermi pockets and surround 
drumhead-like states \cite{PhysRevX6-2016Muechler,PhysRevB93-2016Hung-Liu-Vanderbit};
the last gives hope for the construction of a field theory 
of  Dirac materials provided one understood the 
origin and mechanisms of the decay 
of these lines.

Complex magnetic dynamics is developed  in double
perovskite compounds Ba$_2$YMoO$_6$, Ba$_2Me$OsO$_6$($Me$ = Li,~Na)
despite of perfect cancelation of  spin and angular momentum contributions at cubic symmetry
\cite{Cussen-Lynham-Rogers2006,Aharen-Magnetic-properties2010,Vries-Mclaughlin-Bos2010,
Carlo-Triplet2011,Steele-Low-moment-magnetism2011}. This testifies 
that SOC can effectively augment the Hubbard correlations effects    in Mott-–Hubbard physics
\cite{Covalency-and-vibronic-couplings2016,Minimal-ingredients2016}.

So,  quantum statistics of many-body systems with a particle-hole symmetry should be 
a non-abelian one, the  absence of which is the main obstacle in investigation of Majorana-like states.

 In the paper we  develop 
a quantum non-abelian statistical approach to (pseudo)Majorana fermionic systems
with  calculus of quasi-relativistic currents
and analyze emergent Majorana-like features of quantum charge transport 
in the Dirac materials.

We demonstrate that natural background to describe all types of Dirac materials is
in accounting of relativistic
exchange interactions, which destroy pseudo antiferromagnetic order in the Majorana basis.
In Section 2 we construct a transformation which produces one-to-one map of quasiparticle
states (hole (particle
)) with the negative (positive) energy  in one of two trigonal sublattices to states (particle 
(hole)) of the positive (negative) energy  in other trigonal sublattice of the hexagonal lattice.
Utilizing this transformation we
find  the equations of motion for a (pseudo)real braiding Majorana quasiparticle (electrically charged exciton)
on a 2D hexagonal lattice. In Section 3 we develop a relativistic theory of the
secondary-quantized field with a number of flavors $N=3$ on a hexagonal lattice within the quasi-relativistic
Dirac--Hartree--Fock self-consistent  field approximation \cite
{we-Kazan,we-arxiv2013,myNPCS2013,myNPCS17-2014,myarXiv1401-6880v1-2014,myJModPhys2014,myNATO2015,
NPCS18-2015GrushevskayaKrylovGaisyonokSerow,myNPCS18-2015,myTaylorFrancis2016,myIntJModPhys2016}.
In the section we also demonstrate that the relativistic exchange leads to dynamically gapped Fermi arcs in
topological semimetals.
We use the $\vec k \cdot \vec p$ perturbation theory and maximally-localized Wannier functions,
which  provide  accurate characterization of  points of
interest in the Brillouin zone (BZ) in terms of a relatively small number of parameters
with the first-principles accuracy and  linear-scaling computational costs.
Opposite to known non-relativistic $\vec k \cdot \vec p$ approaches 
\cite{k-p_method2009Springer-Verlag,LuttingerPhyRev1955,DresselhausKipKittel1955PhysRev,Kormanyos2D2015,
Yo-SLeeNardelliMarzari2005ArXiv,Mostofi2008ComputPhysCommun,Marzari-Mostofi2012RevModPhys,Xu-Yu-Fang-Dai-Weng2017PhysRev95},
we propose a quasi-relativistic chiral band structure theory. Calculating  Majorana bands in Section~3, we
neglect the Majorana dynamical mass. However, the presence of heavy and light carriers
is considered as a mass correction to complex conductivity in Section~4.
In Section 5  splitting zero-conductance peaks,
longitudinal magnetoresistance, minimal direct current (dc) conductivity, negative differential conductivity,
appearance of Majorana resonance and anti-resonance pairs 
in frequency dependence of dielectric permeability and other phenomena 
of Majorana braiding in charge transport 
in topological SMs are predicted within a  non-abelian quantum statistical approach developed in Section 4.

\section{Pseudo Majorana fermion model}
Equation of motion for a Majorana bispinor
$(\psi_{AB}^\dagger, (\psi^*_{BA})^\dagger)$ in a
 monoatomic hexagonal layer
(monolayer), comprised of two trigonal  sublattices $A,\ B$ reads \cite{myNPCS18-2015,mySymmetry2016}:
\begin{eqnarray}
\left[\vec \sigma_{2D}^{BA}\cdot \vec p_{AB} -c^{-1} \widetilde  {\Sigma_{AB}\Sigma_{BA}}
\right]\left|
\psi_{AB}\right\rangle =
i {\partial \over \partial t}\left| \psi^*_{BA}\right \rangle \label{Majorana-bispinor01}, \\
\left[\vec \sigma_{2D}^{AB}\cdot \vec p\,^*_{BA}
-c^{-1}\left( \widetilde {\Sigma_{BA}\Sigma_{AB}}\right)^*\right]
\left|\psi^*_{BA}\right\rangle  = - i {\partial \over \partial t}\left|
\psi_{AB}\right \rangle \label{Majorana-bispinor02}.
\end{eqnarray}
Here, the sublattice wave functions $\left|\psi_{AB}\right \rangle$ and $\left| \psi^*_{BA}\right \rangle$
relate to each other as follows:
\begin{eqnarray}
\left| \psi^*_{BA}\right \rangle \propto \left(\Sigma_{rel}^{x}\right)_{BA} \left|\psi_{AB}\right \rangle
\label{Majorana-bispinor01-1}, \\
\left|\psi_{AB}\right \rangle \propto \left( \Sigma_{rel}^{x}\right)_{AB}\left| \psi^*_{BA}\right \rangle
\label{Majorana-bispinor02-1};
\end{eqnarray}
$\left( \Sigma_{rel}^{x}\right)_{AB}\equiv \Sigma_{AB}, \
\left(
\Sigma_{rel}^{x}\right)_{BA}\equiv \Sigma_{BA}$ are
the relativistic exchange interaction operators for the trigonal sublattices
$A,\ B$ of the hexagonal lattice;
the dynamic mass operator terms $\widetilde {\Sigma_{BA(AB)} \Sigma_{AB(BA)}}$ are defined as
\begin{equation}
\widetilde
 {\Sigma_{BA(AB)} \Sigma_{AB(BA)}} =
\left(i\Sigma_{rel}^{x}\right)_{BA(AB)}   \left( i\Sigma_{rel}^{x}\right)_{AB(BA)} ;
\label{mass-operator}
\end{equation}
a transformed  2D vector $\vec \sigma _{2D}^{AB}$ of the Pauli matrices
and  a transformed  2D momentum $\vec p_{BA(AB)}$ are  introduced  as
\begin{eqnarray}
\vec \sigma _{2D}^{BA(AB)} = \left(\Sigma_{rel}^{x}\right)_{BA(AB)}
\vec \sigma\, \left(\Sigma_{rel}^{x}\right)_{BA(AB)}^{-1}
\label{transformedPauli-matrixes},
\\
\vec p_{BA(AB)}= \left(\Sigma_{rel}^{x}\right)_{BA(AB)} \vec p\, \left(
\Sigma_{rel}^{x}\right)_{BA(AB)}^{-1};
\label{transformedPauli-momentum}
\end{eqnarray}
$\vec \sigma$ is the 2D vector of the Pauli matrices: $\vec \sigma =\{\sigma_1,\ \sigma_2\}$;
$\vec p$ is the 2D momentum operator, $c$ is the speed of light.
One can see, that when neglecting
the  operator \eqref{mass-operator},
 Eqs.~\eqref{Majorana-bispinor01}, \eqref{Majorana-bispinor02} are equations of motion
for a Majorana-like massless particle.

The system of Eqs.~\eqref{Majorana-bispinor01}, \eqref{Majorana-bispinor02} can be approximated by 
a Dirac-like equation in the following way. 
Let us rewrite, 
for example, \eqref{Majorana-bispinor01} for the steady state 
\begin{eqnarray}
\left[\vec \sigma_{2D}^{BA}\cdot \vec p_{AB} -c^{-1} \widetilde {\Sigma_{AB}\Sigma_{BA}}
\right]\left|
\psi_{AB}\right\rangle = E_{qu}\left| \psi^*_{BA}\right \rangle \label{Majorana-bispinor01-2}.
\end{eqnarray}
According to \eqref{transformedPauli-matrixes} and \eqref{transformedPauli-momentum},
the bispinor component $\left|\psi_{AB}\right\rangle$
can be obtained as $\left| \psi_{AB}\right\rangle= \Sigma_{AB}\left|\tilde \psi_A
\right\rangle$. Due to  \eqref{Majorana-bispinor01-1}
$\left|\tilde \psi_A \right\rangle$ also defines  the component of the Majorana spinor
$\left| \psi^*_{BA}\right \rangle $ as
$\left| \psi^*_{BA}\right \rangle \propto \left(\Sigma_{rel}^{x}\right)_{BA}
\left|\tilde \psi_A \right\rangle$. Hence
\begin{eqnarray}
\Sigma_{BA(AB)}^2\propto \Sigma_{BA(AB)}. \label{braiding-condition}
\end{eqnarray}
Owing to the condition \eqref{braiding-condition} and taking into account that in accord with 
\eqref{Majorana-bispinor02-1} the operator $\Sigma_{AB}$ can be considered as
a Fermi velocity operator 
$\hat v_F$ in \eqref{Majorana-bispinor01-2}, the following expansion  holds up to normalization constant
$\left<0\right|\hat v_F \left|0 \right>$:
\begin{eqnarray}
\left| \psi_{AB}\right\rangle \propto \Sigma_{AB} {\Sigma_{BA}\over
\left<0\right|\hat v_F \left|0 \right>} \left|
\psi_{AB}\right\rangle 
={1\over \left<0\right|\hat v_F \left|0
\right>}\left\{
 \Sigma_{BA}\left| \psi_{AB}\right\rangle +
\left[\Sigma_{AB}, \Sigma_{BA}\right]\left|
\psi_{AB}\right\rangle\right\} 
\approx \left\{ 1  + {\left(\Delta \Sigma + \left[\Sigma_{AB}, \Sigma_{BA}\right]\right)
\over \left<0\right|\hat v_F \left|0 \right>}\right\} \left| \psi_{AB}\right\rangle \nonumber \\
\label{permutations1}
\end{eqnarray}
where $\left[\cdot, \cdot \right]$ denotes the
commutator, $\Delta \Sigma = \Sigma_{BA} - \Sigma_{AB} $.
Substituting (\ref{Majorana-bispinor02-1}, \ref{permutations1}) into the right-hand side of the equation 
\eqref{Majorana-bispinor01-2}, one gets the basic Dirac-like equation in the following form:
\begin{eqnarray}
\left[ \vec \sigma_{2D}^{BA}\cdot \vec p_{AB}-c^{-1} \widetilde {\Sigma_{AB}\Sigma_{BA}}
\right]
\left| \psi_{AB}\right\rangle  = \tilde E\left\{ 1
+\left(\Delta
\Sigma + \left[\Sigma_{AB}, \Sigma_{BA}\right]\right)/
\left<0\right|\hat v_F \left|0 \right> \right\}\left|
\psi_{AB}\right \rangle \label{Majorana-bispinor1}
\end{eqnarray}
where $\tilde E=E/\left<0\right|\hat v_F \left|0 \right>$.

\section{Gauge field theory of 2D Dirac materials }
The quasi-relativistic Dirac--Hartree--Fock exchange interaction $\left(\Sigma_{rel}^{x}\right)_{AB(BA)}$ in
a tight-binding approximation for the  system of equations
\eqref{Majorana-bispinor01}, \eqref{Majorana-bispinor02} reads (see
equations (\ref{Sigma-AB01}, \ref{Sigma-BA01}) of the
Supplementary Information)
\cite{NPCS18-2015GrushevskayaKrylovGaisyonokSerow,myNPCS18-2015,myTaylorFrancis2016}:
\begin{eqnarray}
 \left(\Sigma_{rel}^{x}\right)_{AB}
 ={1\over \sqrt{2}(2\pi)^{3}}
e^{-\imath (\theta_{k_{A}}-\theta_{K_B})}
\sum_{i=1}^{3} \exp\{\imath [\vec K^i_{A} - \vec q_i ] \cdot \vec
\delta_i\} \int  V(\vec r) d {\vec r}
\nonumber \\
 \times \left(
\begin{array}{cc}
\sqrt{2}\psi_{\mbox{\small p}_z} (\vec r )
\psi^*_{\mbox{\small p}_z, - \vec \delta_i} (\vec r )
 &
\psi_{\mbox{\small p}_z} (\vec r ) [\psi^*_{\mbox{\small
p}_z}(\vec r)
+ \psi^*_{\mbox{\small p}_z, - \vec \delta_i}(\vec r)]\\
\psi^*_{\mbox{\small p}_z, - \vec \delta_i} (\vec r)
[\psi_{\mbox{\small p}_z, \vec \delta_i}(\vec r)+
\psi_{\mbox{\small p}_z}(\vec r)] & {[\psi_{\mbox{\small
p}_z, \vec \delta_i}(\vec r)+ \psi_{\mbox{\small p}_z}(\vec
r)] [\psi^*_{\mbox{\small p}_z}(\vec r)
+\psi^*_{\mbox{\small p}_z, - \vec \delta_i}(\vec r)] \over
\sqrt{2}}
\end{array}
\right)
, \label{Sigma-AB3}
\end{eqnarray}
\begin{eqnarray}
\left( \Sigma_{rel}^{x}\right)_{BA}
= {1\over \sqrt{2}(2\pi)^{3}} e^{-\imath
(\theta_{K_A}-\theta_{K_B})}
\sum_{i=1}^{3} \exp\{\imath [\vec K^i_{A} - \vec q_i ] \cdot \vec
\delta_i\}  \int  V(\vec r) d {\vec r}
\nonumber \\
 \times \left(
\begin{array}{cc}
{[\psi_{\mbox{\small p}_z, \vec \delta_i}(\vec r)+
\psi_{\mbox{\small p}_z}(\vec r)] [\psi^*_{\mbox{\small
p}_z}(\vec r) +\psi^*_{\mbox{\small p}_z, - \vec
\delta_i}(\vec r)] \over \sqrt{2}}
 &
- \psi^*_{\mbox{\small p}_z, - \vec \delta_i} (\vec r )
[\psi_{\mbox{\small p}_z, \vec \delta_i}(\vec r)+
\psi_{\mbox{\small p}_z}(\vec r)]
\\
- \psi_{\mbox{\small p}_z} (\vec r ) [\psi^*_{\mbox{\small
p}_z}(\vec r) + \psi^*_{\mbox{\small p}_z, - \vec
\delta_i}(\vec r)] & \sqrt{2}\psi_{\mbox{\small p}_z} (\vec
r ) \psi^*_{\mbox{\small p}_z, - \vec \delta_i} (\vec r)
\end{array}
\right)
.
 \label{Sigma-BA3}
\end{eqnarray}
Here the origin of the reference frame is located at a given site on the lattice $A$($B$),
 $V(\vec r)$ is the  Coulomb potential, $\psi_{\mbox{\small p}_z} (\vec r)$ is
the wave function of p$_z$-electron, and designation
$\psi_{\mbox{\small p}_z,\ \pm \vec \delta_i}(\vec r)$
for atomic orbital  of p$_z$-electron with
radius-vector $\vec r\pm \vec \delta_i$ in the neighbors lattice sites 
$\vec \delta_i$, nearest to the reference site 
is  introduced in the following way: 
$\psi_{\mbox{\small p}_z,\ \pm \vec \delta_i}(\vec r_{2D})
=\psi_{\mbox{\small p}_z}(\vec r\pm \vec \delta_i)$;
 $\vec r\pm \vec \delta_i$ is the p$_z$-electron radius-vector,
$\vec K_{A}$ ($\vec K_{B}$) is the Dirac point  (valley) $\vec K$($\vec K'$) in the Brillouin zone.

The wave functions are defined up to  a phase multiplier.
Let us denote the phases of the wave functions 
$\psi_{\mbox{\small p}_z}(\vec r)$ and $\psi_{\mbox{\small p}_z, \pm \vec \delta_k}(\vec r)$,
$k=1,\ 2,\ 3$ as $\alpha_{ 0}$ and $\alpha_{\pm, k}$, $k=1,\ 2,\ 3$ respectively.
A set of these phases is a four-dimensional (4D) phase
$\alpha ^\mu _{\pm}$, $\mu=0,\ \ldots,\ 3$, whose components
play a role  of the space-time components of a lattice gauge
field. The 4D-phases  $\alpha ^\mu_{\pm}$, $\mu=0,\ \ldots,\ 3$
enter to the matrix elements $\Sigma_{ij,AB(BA)}$, $i,j=1,2$
in \eqref{Sigma-AB3} and \eqref{Sigma-BA3} in the following way:
$\Sigma_{ij,AB(BA)}$ include bilinear on
$\psi$, $\psi^*$  combinations of wave functions so that
4D-phase  $\alpha ^\mu_{\pm}$ enter into
(\ref{Sigma-AB3}--\ref{Sigma-BA3}) in the form
%

\begin{eqnarray}
|\psi| |\psi _{\pm \vec
\delta_k}| \exp\left\{\imath \left( \alpha_0 - \alpha_{\pm,
k}\right)\right\}
=|\psi| |\psi _{\pm \vec
\delta_k}| \exp\left\{\imath \Delta \alpha _{\pm, k} \right\}
. \label{phase_variation}
\end{eqnarray}

%

Therefore, an effective number 
$N$ of flavors in our gauge field theory is equal to 3. 

\begin{figure}[hbt]
\begin{center}
\hspace{-1cm} (a)\hspace{8cm} (b)\\
\includegraphics[width=3.cm,height=1.4cm,angle=0]{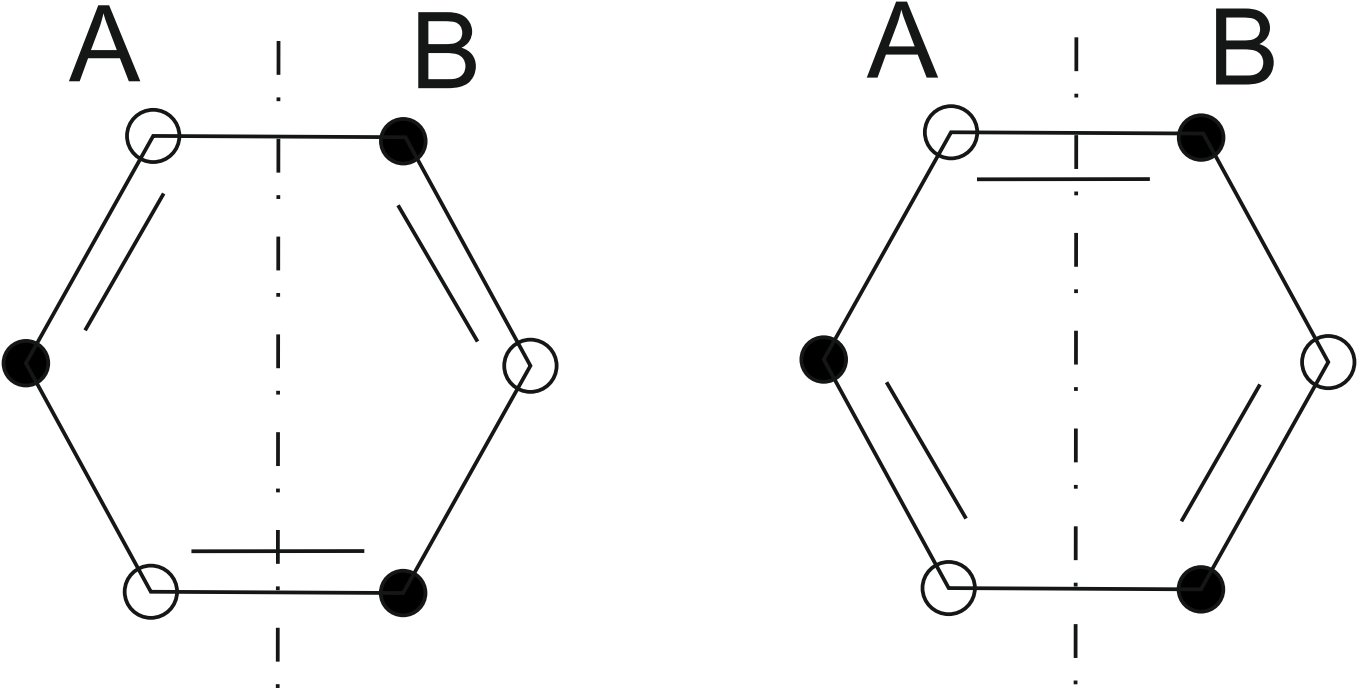}\hspace{3cm}
\includegraphics[width=5.cm,height=1.4cm,angle=0]{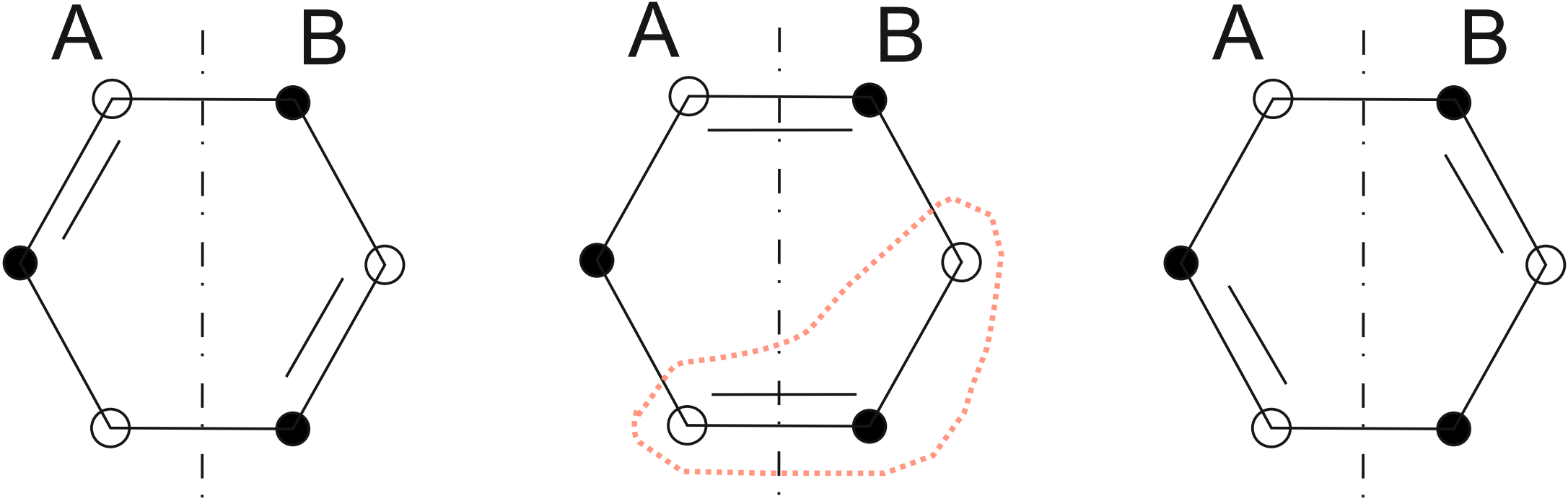}\\
 (c) \\
 \includegraphics[width=3.0cm]{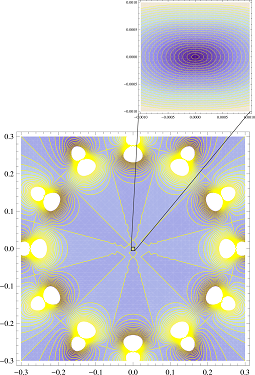} \hspace{0.2cm}
\includegraphics[width=3.0cm]{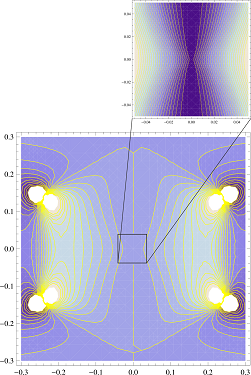} \hspace{0.2cm}
\includegraphics[width=3.0cm]{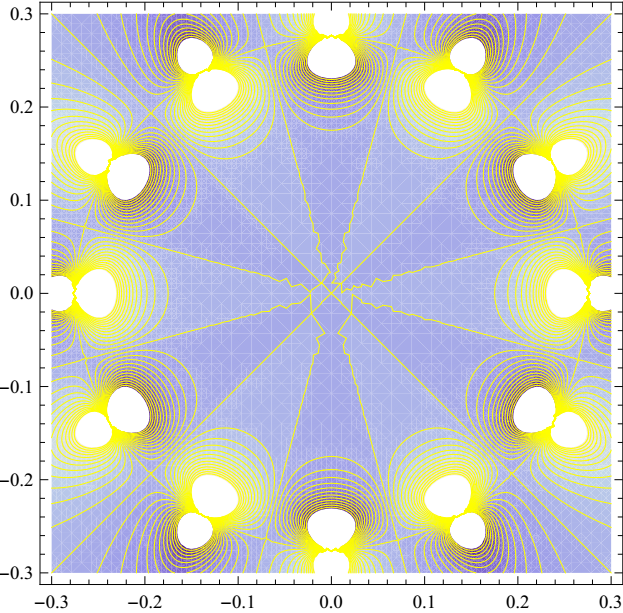}\\
(d)  \\
\includegraphics[width=8.cm,height=4.cm,angle=0]{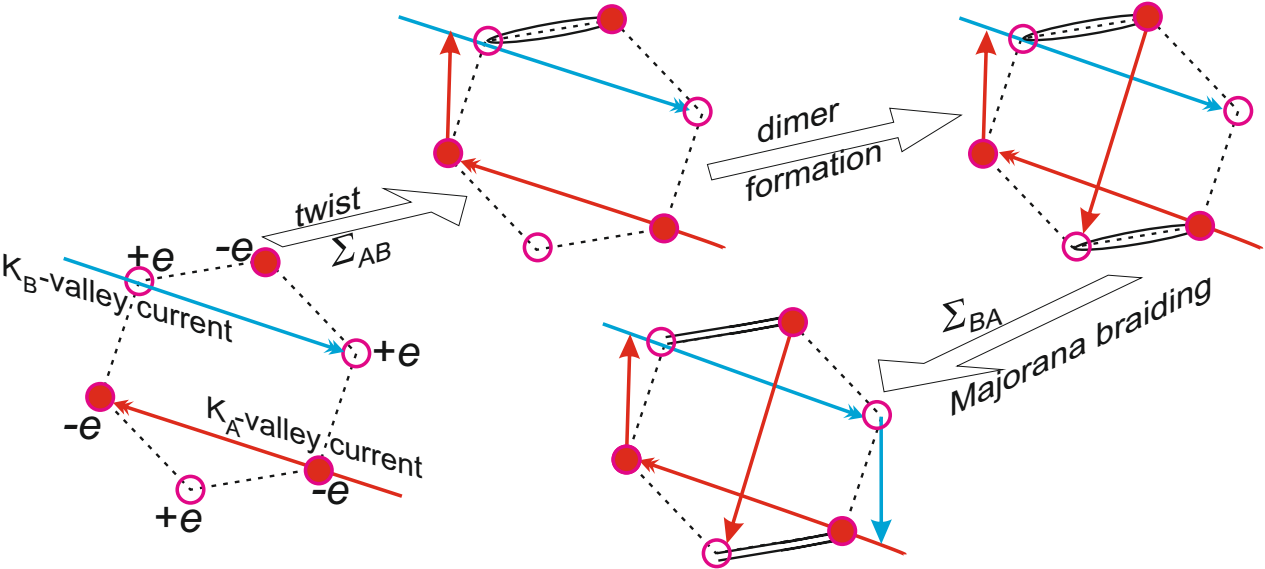}
\end{center}
\caption{Models $N=2$ (a), $N=3$ (b).
p$_z$-electrons belonging to triangular sublattices $A$ and $B$ are designated through solid
and unfilled circles respectively.
An electrically charged
exciton is designated by dashed line contour in figure (b).
%
(c) Action of exchange operators  on 
the particle (hole) Hamiltonians $H_{AB(BA)}$ in
\eqref{Majorana-bispinor1-first-approximation}
\cite{NPCS18-2015GrushevskayaKrylovGaisyonokSerow}: (left) Contour
plots of sum of original Dirac bands, (middle) of single action of
$\left( \Sigma_{rel}^{x}\right)_{BA}$; (right) of action of
exchange operators product $\left( \Sigma_{rel}^{x}\right)_{AB}
\left( \Sigma_{rel}^{x}\right)_{BA}$ restoring the system.
Inset in figure (c, left) demonstrates that the  steady state point in the Dirac point 
$K(K')$ is of center type.
Inset in figure (c, middle) shows an occurrence of a saddle point.
(d) Schematic diagram for  braiding Majorana fermions through twisted valley currents, dimer formation
and the exchange $\Sigma_{AB(BA)}$. Hole and electron currents are shown in blue and red respectively.}
\label{color-models} 
\end{figure}

A model of the Dirac material with two or three flavors as a model
with two or three dimer electron configurations is represented
schematically 
in fig.~\ref{color-models}a,b. Electrons of the first model in
 fig.~\ref{color-models}a  are paired $\pi(p_z)$-electrons. Among 
p$_z$-electrons of the second model in fig.~\ref{color-models}b there exist two 
unpaired electrons.  The first model with   $\pi (p_z)$-electrons
has been proposed   
in \cite{Wallace}  for graphite. 
Part of electrons in the second model 
are unpaired ones, and respectively dimers are formed  as in
the Anderson RVB picture. The weak exchange 
for the second 
model 
leads to a gapless band structure
as it has been demonstrated in \cite{MySolidState,allMyDokladyNANBelarus}, and
respectively to metallicity. 
The strong quantum exchange for the first 
model 
leads to the appearance of the gap which has been experimentally  observed for
graphene bilayers \cite{Origin2008NatMat}.

In the model 
 $N=2$ the valley currents are absent
 due to the fact that all 
p$_z$-electrons are pared.   
Contrary to this, valley currents exist for 
$N=3$ \cite{MyJNPCS2017Vol20}.
In contrast 
to the RVB picture, the stochastic staggered spin flows or identical to them
d-density wave states \cite{Chakravarty-Laughlin-Morr-Nayak2002} are absent for our model.
Their place, as we show further, is occupied by the 
quantum staggered valley currents of the electric charge carriers with a precessing spin similar
to a spin vortex. The spin precession 
is due to SOC-coupling. 
Experimentally, 
such a staggered orbital order (a staggered quadruple ordered phase
with distinct orbital polarization on two-sublattices)
without lattice distortions has been found 
in \cite{staggered orbital order}.


To account 
translational symmetry, we introduce the phase multiplier $\exp\left\{\imath \Delta \alpha _{\pm, i} \right\}$
to the wave function at site $\pm\delta_i$, $i=1,2,3$ in the following form:
\begin{equation}
\label{c-alpha}
\exp\left(\pm \imath c_\pm 
(q)({\vec q}\cdot{\vec
\delta_i})\right).
\end{equation}
To eliminate arbitrariness in the choice of phase factors
$ c_ \pm $ in \eqref {c-alpha}, one needs gauge condition for the phase fields. 

\subsection{First-order approximation}
Let us consider the case of small 
wave numbers $q$, $q\to 0$. Then, in the first-order approximation, a gauge condition can be chosen as follows.
The phases 
$\alpha_0$ and $\alpha_{\pm, k}$, $k=1,\ 2,\ 3$ of the wave functions 
$\psi_{\mbox{\small p}_z}(\vec r_{2D})$ and $\psi_{\mbox{\small
p}_z, \pm \vec \delta_k}(\vec r_{2D})$, $k=1,\ 2,\ 3$ respectively
 are the same for 
p$_z$-electrons 
in the expressions (\ref{Sigma-AB3}, \ref{Sigma-BA3})  due to \eqref{c-alpha}.
By virtue of the arbitrariness in the choice
of phases at $ q \to 0 $,
the phases in the first-order approximation were chosen the same for all lattice sites.

At  power series expansion on 
a small parameter ${1\over \left<0\right|\hat v_F \left|0 \right>}$ in the right hand side of 
\eqref{Majorana-bispinor1}, one can neglect the second term 
for small 
wave numbers $q$. Then accounting the fact that 
the  mass term $\widetilde {\Sigma_{AB(BA)}\Sigma_{BA(AB)}}$ vanishes in the Dirac point 
one gets:
\begin{eqnarray}
\vec \sigma_{2D}^{BA}\cdot \vec p_{AB}  \left|
\psi_{AB}\right\rangle = { E^{(0)}\over \left<0\right|\hat v_F \left|0 \right>  }
\left|\psi_{AB}\right \rangle .\label{Majorana-bispinor1-first-approximation}
\end{eqnarray}
In this case 
the exchange interaction terms 
(\ref{Sigma-AB3}, \ref{Sigma-BA3}) 
are given by the matrixes with real integrands. 
It turns out that with such a phases choice for 
$\psi_{\mbox{\small p}_z}(\vec r_{2D})$ and $\psi_{\mbox{\small
p}_z, \pm \vec \delta_k}(\vec r_{2D})$, we obtain an imaginary part for the energy
$E^{(0)}$ in  \eqref{Majorana-bispinor1-first-approximation} without the mass term
\cite{myNPCS18-2015,myIntJModPhys2016}. The deviations of the first-order approximation  from the massless Dirac fermion model
\cite{Semenoff1984} are of the order of $|q|^4$ \cite{myJModPhys2014}.

Under the action 
of $\Sigma_{BA(AB)}$ calculated in the first-order approximation,
 the Dirac point as a center of circumference trajectories
is transformed into a hyperbolic point of a saddle type and subsequent action
of $\Sigma_{AB(BA)}$ restores neutral-stability state of the center type 
in fig.~\ref{color-models}c. Therefore,
the exchange operator $\Sigma_{AB(BA)}$ plays the role of a braiding operator.
 Braiding scheme through the formation of a dimer is shown
in fig.~\ref{color-models}d.

\subsection{Second-order approximation}
In the second-order approximation the relative phases 
$\Delta \alpha _{\pm, j}$ in $j$-th primitive cell 
are
  different for different cells. 
Substituting the relative phases \eqref{c-alpha} 
of particles 
and holes 
into \eqref{Sigma-AB3} one gets the exchange interaction operator $\Sigma_{AB}$ in the second order approximation:
\begin{eqnarray}
 \Sigma_{AB} ={1\over \sqrt{2}(2\pi)^{3}}
e^{-\imath (\theta_{k_{A}}-\theta_{K_B})}
 \left(
\begin{array}{cc}
\Sigma_{11}^{AB} &\Sigma_{12}^{AB}\\
\Sigma_{21}^{AB}& \Sigma_{22}^{AB}
\end{array}
\right),  \label{Sigma-AB3-second-approximation}\\%
\Sigma_{11}^{AB}= I_{11}\sqrt{2} \left\{\sum_j  e^{-\imath
c_-
(\vec q)(\vec q  \cdot \vec \delta_j)} \exp\{\imath [\vec K^j_{A} - \vec q ] \cdot \vec \delta_j\} \right\},
 \label{Sigma-AB11-second-approximation} \\
\Sigma_{12}^{AB}=   \left\{\sum_j \left( I_{12} +  I_{11}
 e^{-\imath c_-
 (\vec q)(\vec q  \cdot \vec \delta_j)} \right)
\exp\{\imath [\vec K^j_{A} - \vec q ] \cdot \vec \delta_j\}
\right\} ,  \label{Sigma-AB12-second-approximation}\\
\Sigma_{21}^{AB}=   \left\{\sum_j \left(I_{21} e^{\imath
(c_+
(\vec q) - c_-
(\vec q))(\vec q  \cdot \vec \delta_j)} +  I_{11} e^{-\imath   c_-
(\vec q)(\vec q \cdot \vec \delta_j)} \right) \exp\{\imath [\vec K^j_{A} - \vec q]
\cdot \vec \delta_j\}
\right\},  \label{Sigma-AB21-second-approximation}\\
\Sigma_{22}^{AB}= {1 \over \sqrt{2}}   \left\{\sum_j \left( I_{22}
e^{\imath c_+
(\vec q) (\vec q  \cdot \vec \delta_j)}
 +
I_{12}+ I_{21} e^{\imath (c_+
(\vec q) - c_-
(\vec q))(\vec q  \cdot \vec \delta_j)}
+I_{11} e^{- \imath c_-
(\vec q))(\vec q \cdot \vec \delta_j)} \right) \exp\{\imath [\vec K^j_{A} - \vec q
] \cdot \vec \delta_j \}
\right\},  \label{Sigma-AB22-second-approximation}
\end{eqnarray}
\begin{eqnarray}
 I_{11} = \int  V(\vec r)  \psi^{(0)}_{\mbox{\small p}_z}{\psi^*}^{(0)}_{\mbox{\small p}_z-\vec \delta_j}
 \ d\vec r , \ \
 I_{12} = \int  V(\vec r)  \psi^{(0)}_{\mbox{\small p}_z}{\psi^*}^{(0)}_{\mbox{\small p}_z}
 \ d\vec r ,
\label{Sigma-AB3-second-approximation_Int12}\\
 I_{21} = \int  V(\vec r)  \psi_{\mbox{\small p}_z+\vec \delta_j}{\psi^*}_{\mbox{\small p}_z-\vec \delta_j}
 \ d\vec r ,\ \
 I_{22} = \int  V(\vec r)  \psi_{\mbox{\small p}_z+\vec \delta_j}{\psi^*}_{\mbox{\small p}_z}
 \ d\vec r \ ;
\label{Sigma-AB3-second-approximation_Int12}
\end{eqnarray}
and similar formulas for $\Sigma_{BA}$.
%
Now, neglecting 
the mass term, we can find the solution of 
the equation \eqref{Majorana-bispinor1} by the successive approximation technique 
as:
\begin{eqnarray}
\vec \sigma_{2D}^{BA}(\alpha_{\pm, i})\cdot \vec p_{AB}  \left|
\psi_{AB}(\alpha_{\pm, i})\right\rangle +{E^{(0)} \left(\Delta
\Sigma + \left[\Sigma_{AB}, \Sigma_{BA}\right]\right)\over
\left<0|\hat v_F|0\right>^2 }\left| \psi_{AB}(\alpha_{\pm,
i})\right \rangle =
{E^{(1)} \over \hat v_F}\left| \psi_{AB}(\alpha_{\pm, i})\right \rangle .\nonumber \\
\label{variational-Majorana-bispinor}
\end{eqnarray}
Here $\Delta \Sigma,\  \Sigma_{AB}, \Sigma_{BA}$ are determined by the expressions
(\ref{Sigma-AB3-second-approximation}--\ref{Sigma-AB3-second-approximation_Int12}).

Eigenvalues $E^{(1)}_i, \ i=1,\ 2$ of \eqref{variational-Majorana-bispinor}  are functions of $c_\pm
$.
The gauge condition 
is imposed as a  requirement on the absence of imaginary parts in eigenvalues
$E^{(1)}_i, \ i=1,\ 2$ of \eqref{variational-Majorana-bispinor}. This condition can be written as a
system of two equations of the form
\begin{equation}\label{sys}
\Im m(E^{(1)}_{i})=0, \ i=1,\ 2.
\end{equation}
Direct solution of this system turns out to be unstable for some
specific points in the momentum space. Instead, for every point in the
momentum space we use a minimization procedure with the price
function $f(c_+
, c_-
)=\left|\Im m \
E^{(1)}_{1}\right|+\left|\Im m \ E^{(1)}_{2}\right|$. Its absolute
minimum evidently coincides with the solution of the system (\ref{sys}).

\begin{figure}[hbt]
\hspace{-1cm} (a)\hspace{5cm} (b)\hspace{5cm} (c)\\
\includegraphics[width=6.cm,height=4.6cm,angle=0]{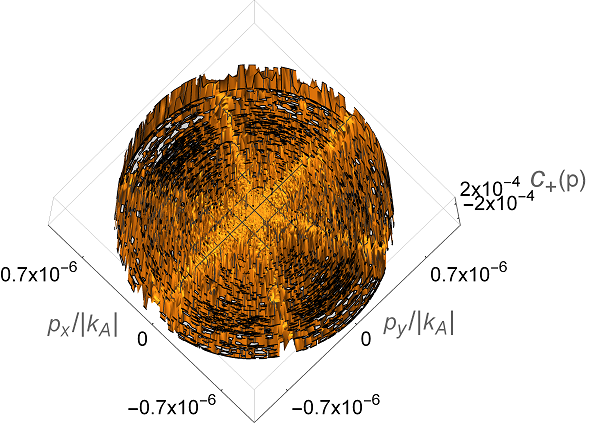} 
\includegraphics[width=5.cm,height=4.6cm,angle=0]{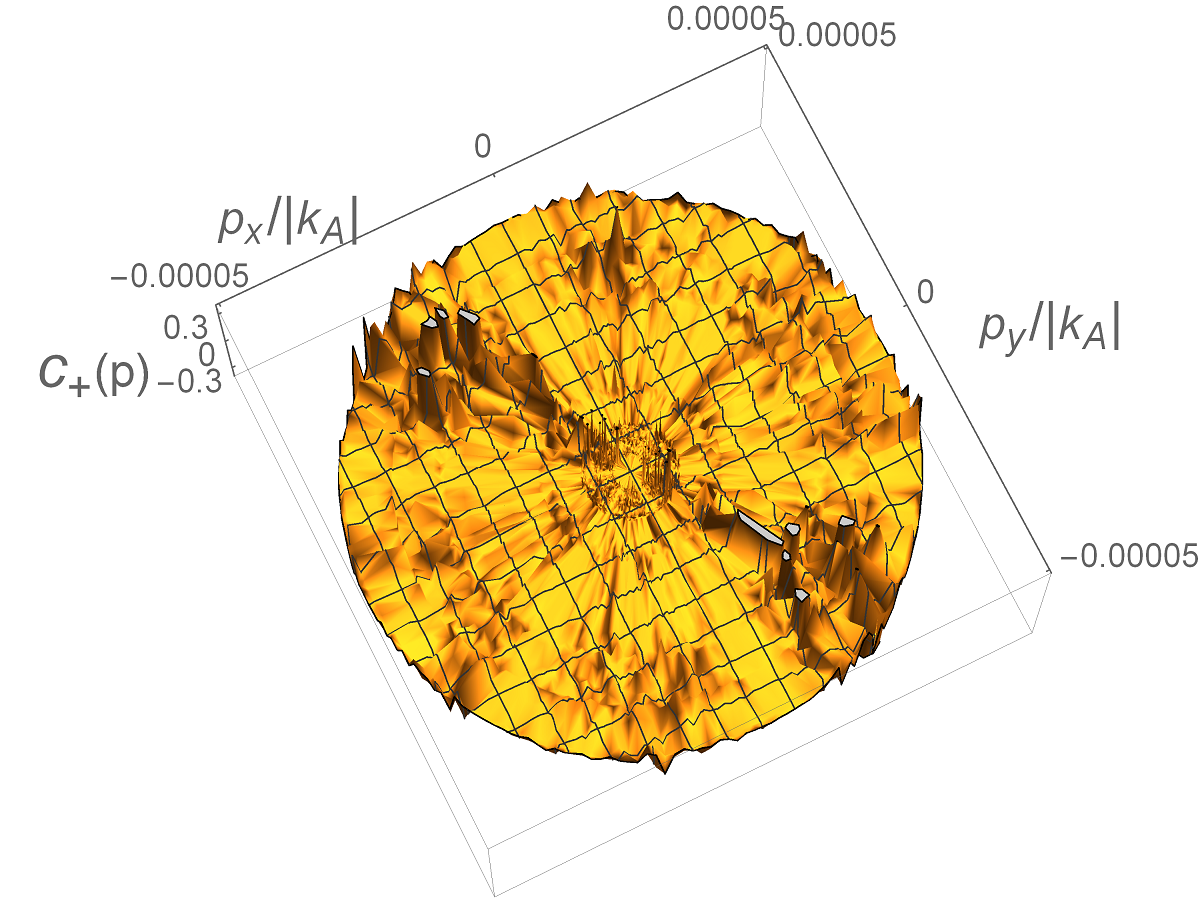}
\includegraphics[width=5.cm,height=4.6cm,angle=0]{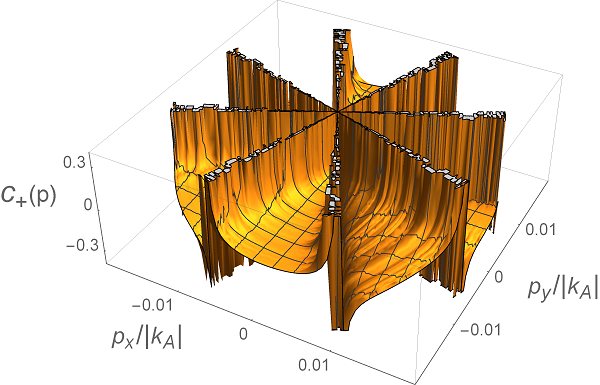}\\
\hspace{-1cm} (d)\hspace{5cm} (e)\hspace{5cm} (f)\\
\includegraphics[width=6.cm,height=4.6cm,angle=0]{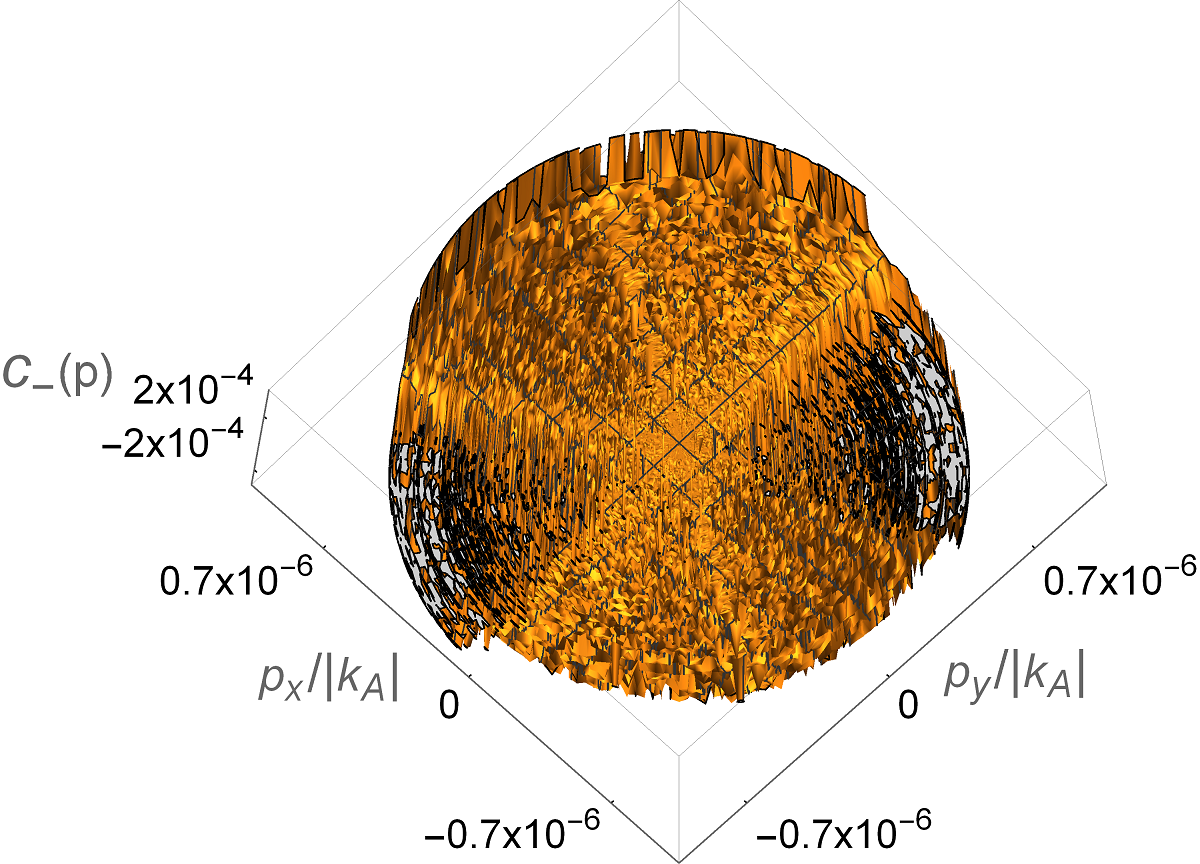}
\includegraphics[width=5.cm,height=4.6cm,angle=0]{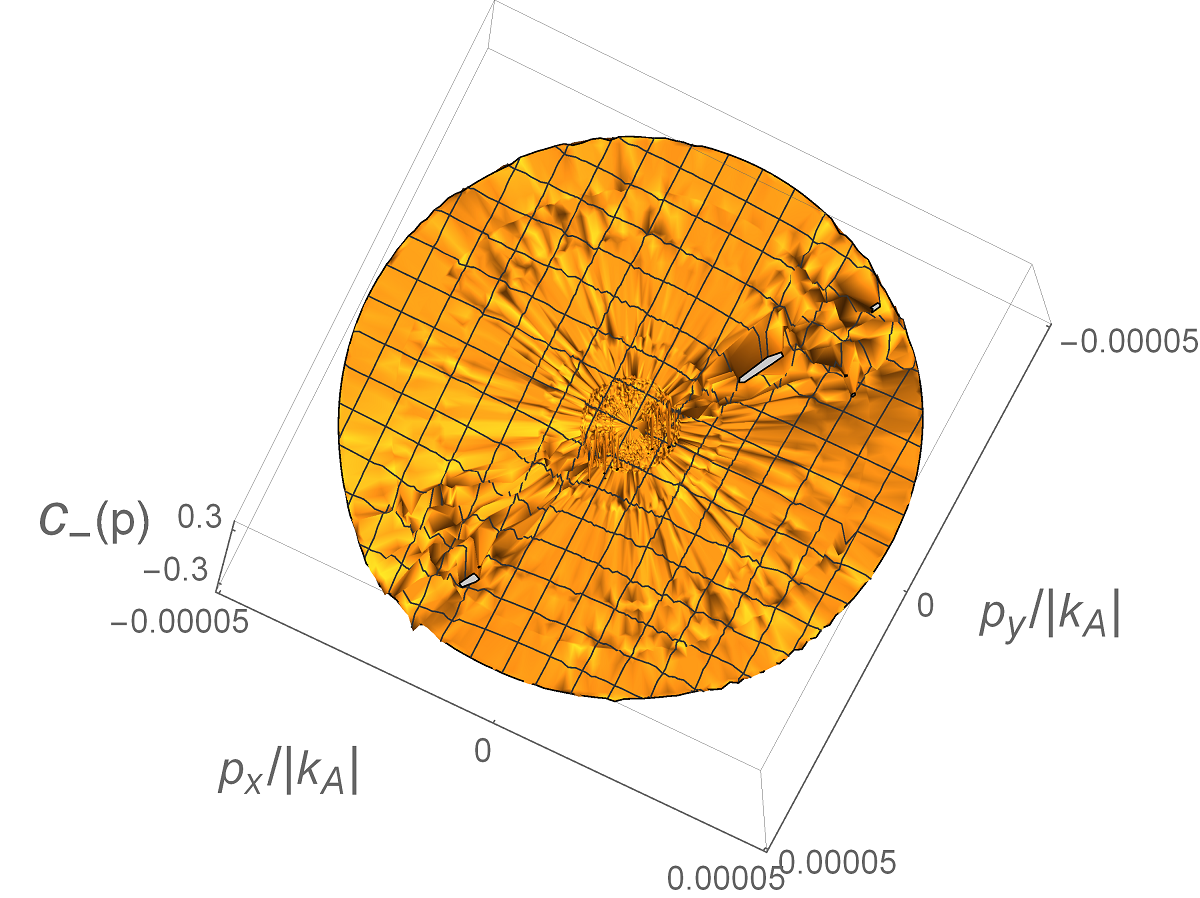}
\includegraphics[width=5.cm,height=4.6cm,angle=0]{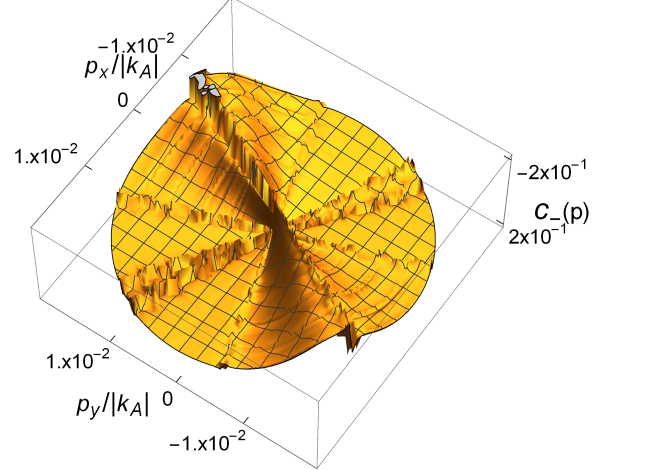}\\
\hspace{-1cm} (g)\hspace{5cm} (h)\\
\includegraphics[width=6.5cm,height=4.cm,angle=0]{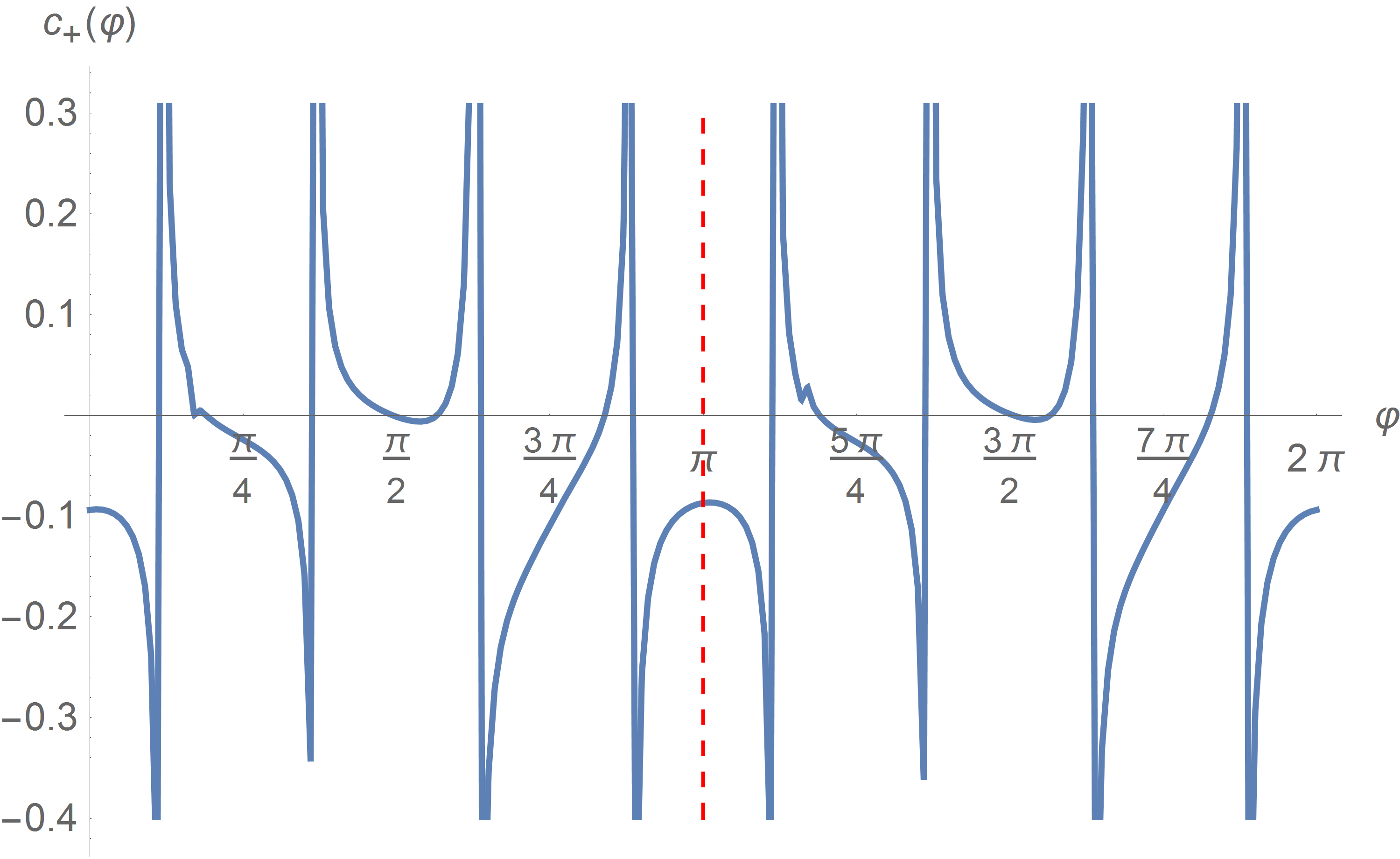} \hspace{1cm}
\includegraphics[width=6.5cm,height=4.cm,angle=0]{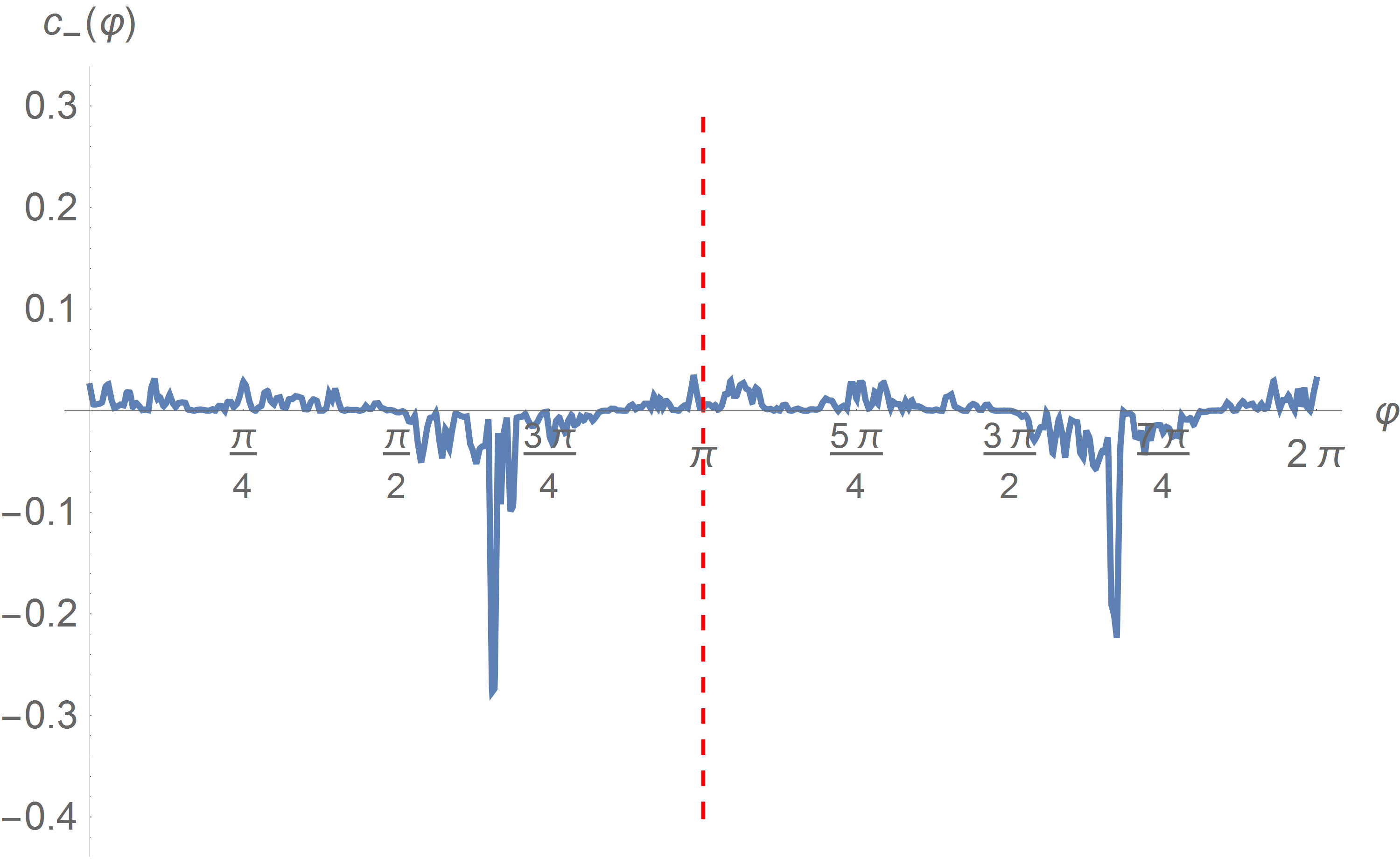}
\caption{Scenario 
of $\mathbb{Z}_2$ topological gapless-phase emerging through deconfinement. 
(a, d) Zero-value 
gauge field $\{c_+,c_-\}$ in the Dirac point $K(K')$ 
and its 
$\epsilon$-neighbourhood, 
$\epsilon \sim 2\cdot 10^{-8}$ and emergent hexagonal symmetry of the field in the energy range $ \sim
10^{-7}$. (b, e) Origin of 2D $\mathbb{Z}_2$-topological semimetal at energies $ \sim  5\cdot 10^{-7} - 10^{-5}$.
(c, f) Deconfinement of the  gauge field $\{c_+,c_-\}$  by SOC at the energies $ \sim 10^{-2}$.
Angle dependencies of phases $c_+$ and $c_-$ entering in the exchange
matrix at different excitation energies $E(q)$: $0.02$ (g) and
$5\cdot 10^{-5}$ (h) respectively.}  \label{gauge-field-C+}
\end{figure}

The phase factors $  c_{\pm} $  \eqref{c-alpha} entering \eqref{variational-Majorana-bispinor}
  periodically change their values on the polar angle $ \phi $ with the period
$\pi$ in fig.~\ref{gauge-field-C+}, and hence our model describes
2D $\mathbb{Z}_2$-topological insulators
\cite{Zak-phase1989,Short-Course-on-Topological-Insulators,Wilczek-Zee1984PRL,Topological-Dirac-nodal-lines2017NatCom}.
The gauge fields $  c_{\pm} $ hold hexagonal symmetry near 
the Dirac point $K(K')$ 
 and are rotated on $60^\circ$ with respect to each other in figs.~\ref{gauge-field-C+}a,~d.
At high momenta the gauge field $c_+$  changes symmetry to octagonal one 
in $\vec q$-space as figs.~\ref{gauge-field-C+}c,~b demonstrate. The behaviour of the phases 
$\{c_+,c_-\}$ in figs.~\ref{gauge-field-C+}a,~d is the same 
and hence they describe the same gauge field. 
The gauge field fluctuates strongly and is characterized by the hexagonal symmetry near the Dirac point 
$K(K')$ due to phase entanglement $\{c_+,c_-\}$.
A core of vortex is observed in figs.~\ref{gauge-field-C+}a,~d so
that  $c_+$, $c_-$ fluctuate
least of all at the boundary of six identical sectors of the circle.
With the increase of the excitation energy $ E (q) $, the value of the gauge field begins to increase only in one of
three pairs of sectors of the circle in figs.~\ref{gauge-field-C+}a, d.

With the increasing $ E (q) $, the amplitude of the fluctuations decreases
and 2D $\mathbb{Z}_2$-topological phase is originated. The phases $\{c_+,c_-\}$ become two 
different gauge fields as it is shown in figs.~\ref{gauge-field-C+}c,~f; and, hence, the  gauge field  is
deconfined. The phase $c_+$ changes sharply its sign four times 
due to the eight-fold Dirac cone for the deconfined quasiparticle
state at large excitation energy $E(q)$ (high momentum $q$) in
fig.~\ref{gauge-field-C+}g. The phases $c_+$, $c_-$ start to fluctuate strongly
with the decrease of the 
value $E(q)$ as, for example,  $c_-(\phi)$ dependency  demonstrates in fig.~\ref{gauge-field-C+}h.

\subsection{Dichroism of the
Dirac bands, deconfinement, nodal lines and  drumhead surface states}

\begin{figure}[hbt]
\begin{center}
  (a) \hspace{5cm} (b) \hspace{5cm} (c)\\
\includegraphics[width=6.5cm,height=4.6cm,angle=0]{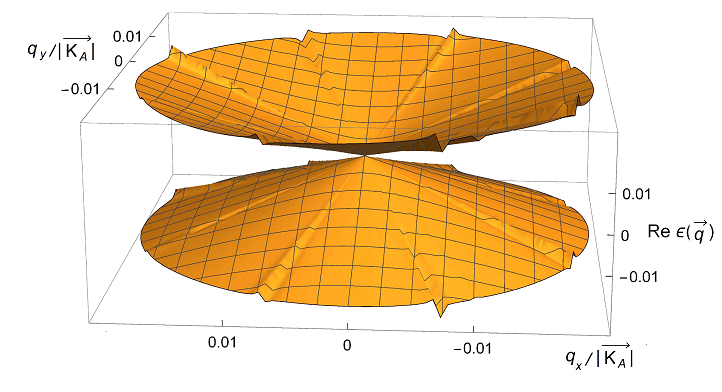} 
\includegraphics[width=5.5cm,height=4.6cm,angle=0]{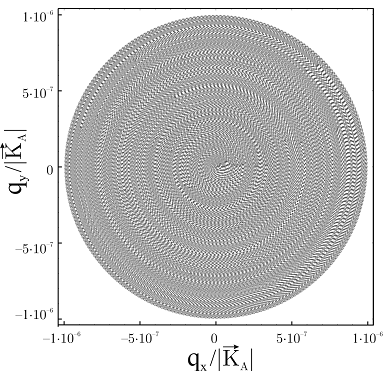}
\includegraphics[width=4.5cm,height=4.5cm,angle=0]{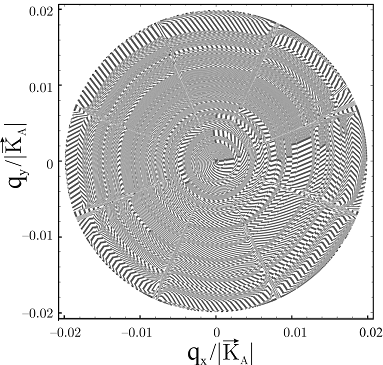} 
\\
 (d) \hspace{10cm} (e) \\
\includegraphics[width=4.5cm,height=4.5cm,angle=0]{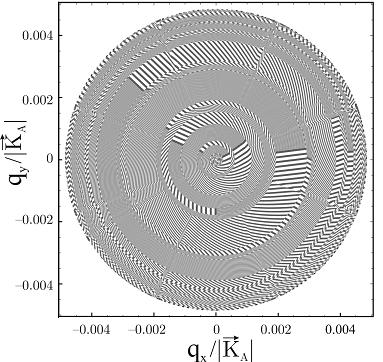}
\hspace{-0.4cm}
\includegraphics[width=3.4cm,height=2.3cm,angle=0]{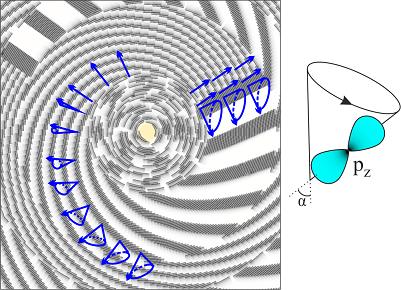}
\includegraphics[width=7.4cm,height=5.3cm,angle=0]{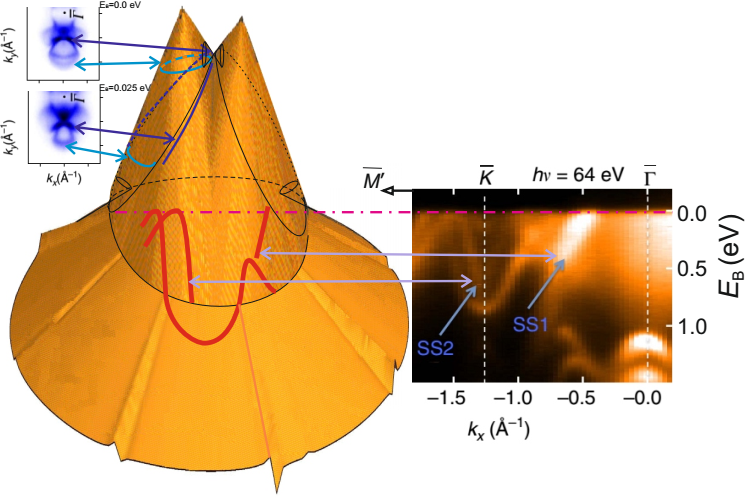}
\end{center}
\caption{Band structure 
with the Majorana  zero-energy modes in the Dirac cone apexes for the
semimetal model $N=3$ (a), their spin-orbit texture  of the scale: 
$q/|K_A|\sim 5\times 10^{-7}$ (b), 0.01 (c), 0.002 (d) in contour plots (b--d);
and a model vortex of the precessing orbitals 
in inset to figure~(d). The angle $\alpha$, $0\le \alpha\le \pi/2$ is the precession angle 
of p$_z$-orbital
. 
(e) Sketch of ARPES mappings  on a Majorana band structure
for an orthorombic T$_d$-MoTe$_2$ (blue and violet lines) and PbTaSe$_2$ (red lines).
Insets are (left) photoemission spectral intensity map of T$_d$-MoTe$_2$ at binding energy $E_B=0.0,\  0.025$~eV
\cite{Signature-of-type-II-Weyl-semimetal2017NatCom} and
(right) ARPES spectra of PbTaSe$_2$ taken along  $\overline{M}- \overline{K} - \overline{\Gamma} $ with 64-eV photons
\cite{NatCommunic7-2016Bian}.
 }
 \label{Vortices-band-structure}
\end{figure}

A helicoidal spin-valley-orbit texture of the Majorana bands
\eqref{variational-Majorana-bispinor} shown in fig.~\ref{Vortices-band-structure} is
originated in the $\sigma - p_z$ coupling between the orbital momentum $p_z$ of $\pi$-electron
and the $\pi$-electron spin $\sigma $ oriented along two directions: tangent 
$\vec n_{\phi}$ and radial 
$\vec n_{r}$. One can see that the Dirac point hosts such a 
"defect", as 
a core of vortex in the in fig.~\ref{Vortices-band-structure}b-d. This topological defect is  a Majorana zero-energy mode.
Multiple vortices structure of these bands is visualized in the form of 
concentric  circles of varying degrees of helicoidality and different widths.
 The spin-valley-orbit texture varies in both space and time. Dichroism  
 of ARPES spectra is a manifestation 
of the  $\pi$-electron-orbit pseudo-precession in inset to fig.~\ref{Vortices-band-structure}d. 
Since the pseudo-Majorana modes are simultaneously their 
antimodes, Majorana sinks 
(vortex cores) are simultaneously Majorana sources 
(antivortex anticores). Sinks and sources locating at the same place 
braid particles  and holes into Majorana fermions.

Deconfinement 
which splits the four-fold particle (hole) Dirac cone, should lead to the divergence of the locations of sinks and sources.
Meanwhile, a nodal ring $NL_D$, 
an image of which is  schematically  presented in fig.~\ref{Vortices-band-structure}e
, separates the region of 
eightfold 
splitting of the Dirac cone in fig.~\ref{Vortices-band-structure}a.
The left- and right-hand  $\pi$(p$_z$) electrons  possess relativistic
total angular momentum $J=3/2$  due to SOC after the deconfinement.
The emergent tilted Dirac cones which have been introduced earlier as tilted Dirac cone replicas in
\cite
{myNPCS2013
}
break 
the vortices in fig.~\ref{Vortices-band-structure}c,~d.
Vortex current lines pass from one vortex cores to others.
The connections of different vortices form an infinite number of
Fermi arcs which link divergent Majorana sinks and anti-sinks. 
Hence,  together with 
the nodal ring $NL_D$, four more pairs of 
Weil nodal lines are formed. 

Divergence of sinks and sources 
transforms 
the brainding Majorana excitations into another type of massless fermions in such a way that  their vacuum 
core and anticore  states  
become uncorrelated. 
Thus, before the deconfinement, the secondary quantized wave function always describes a fermionic state,
entangled with a particle
--hole pair, whereas after the deconfinement
 the non-correlated  vacuum spin up and down states lead to
$SU(2)\otimes SU(2)\cong O(4)$ symmetry for
the system \cite{MyLambert}.

For certain crystal groups describing nodal-line semimetals 
quasiparticles are eight-fold degenerate fermions with $SU(2)$ holonomy of
wave functions (Wilson loop \cite{Alexandradinata-2014,Alexandradinata-2016})\cite{PhysRevLett116-2016Wieder,Science353-2016Bradlyn}.
Thus, the Majorana particles decay into a continuous set of Weyl nodes 
and  a number of quasiparticle excitations in such  a decay channel  does not change.
Since the number of particles remains unchanged, 
the Luttinger theorem still holds. Indeed, the hole (particle) pockets  are observed 
in ARPES-spectra for nodal-line TIs \cite{PhysRevX6-2016Muechler,Nature527-2015Soluyanov}.
In the $q^4$ model \eqref{Majorana-bispinor1-first-approximation}
with dumping $\gamma$, nodal lines decay into a discrete set of Wey nodes
\cite{mySymmetry2016}.


Now, we can predict 
features of band structures, which should be observed 
in 
ARPES of topological SMs. Removing the degeneracy 
of  zero-energy Majorana mode leading also 
to divergence (unbraiding) of Majorana sinks and sources, 
yields as well to 
two additional nodal rings  $NL_1$ and $NL_2$ as it is shown in
fig.~\ref{Vortices-band-structure}e
. $NL_D,\ NL_i$, $i=1,2$ form a contour of drumhead-like states dispersing inwards with
respect to $K$. The three nodal rings and the  drumhead-like surface states in ARPES spectrum shown
in  fig.~\ref{Vortices-band-structure}e 
are features of  ARPES spectra in the right inset to this figure for superconductor PbTaSe$_2$ \cite{NatCommunic7-2016Bian}.
At the place of touching 
of $NL_1$ and $NL_2$ Majorana sink and source form 
a Dirac-like band, orthogonal 
to the origin Dirac band as it is shown in fig.~\ref{Vortices-band-structure}e.
Such a band structure shown in left inset to fig.~\ref{Vortices-band-structure}e is a feature of
type-II topological Weyl SMs as WTe$_2$, MoTe$_2$
\cite{JETPLetters103-2016Zyuzin,PhysRevX6-2016Muechler,Signature-of-type-II-Weyl-semimetal2017NatCom,Nature527-2015Soluyanov}.

So, the spin valley-currents coupling, which  is small on the energy, turns out to be topologically protected by emergent vortices.

\subsection{ Majorana dynamical mass operator and  chiral anomaly}

\begin{figure}[hbt]
\begin{center}
 (a)\hspace{4cm} (b)\hspace{4cm} (c) \hspace{4cm} (d)
\\
\includegraphics[width=4.cm,height=2.4cm,angle=0]{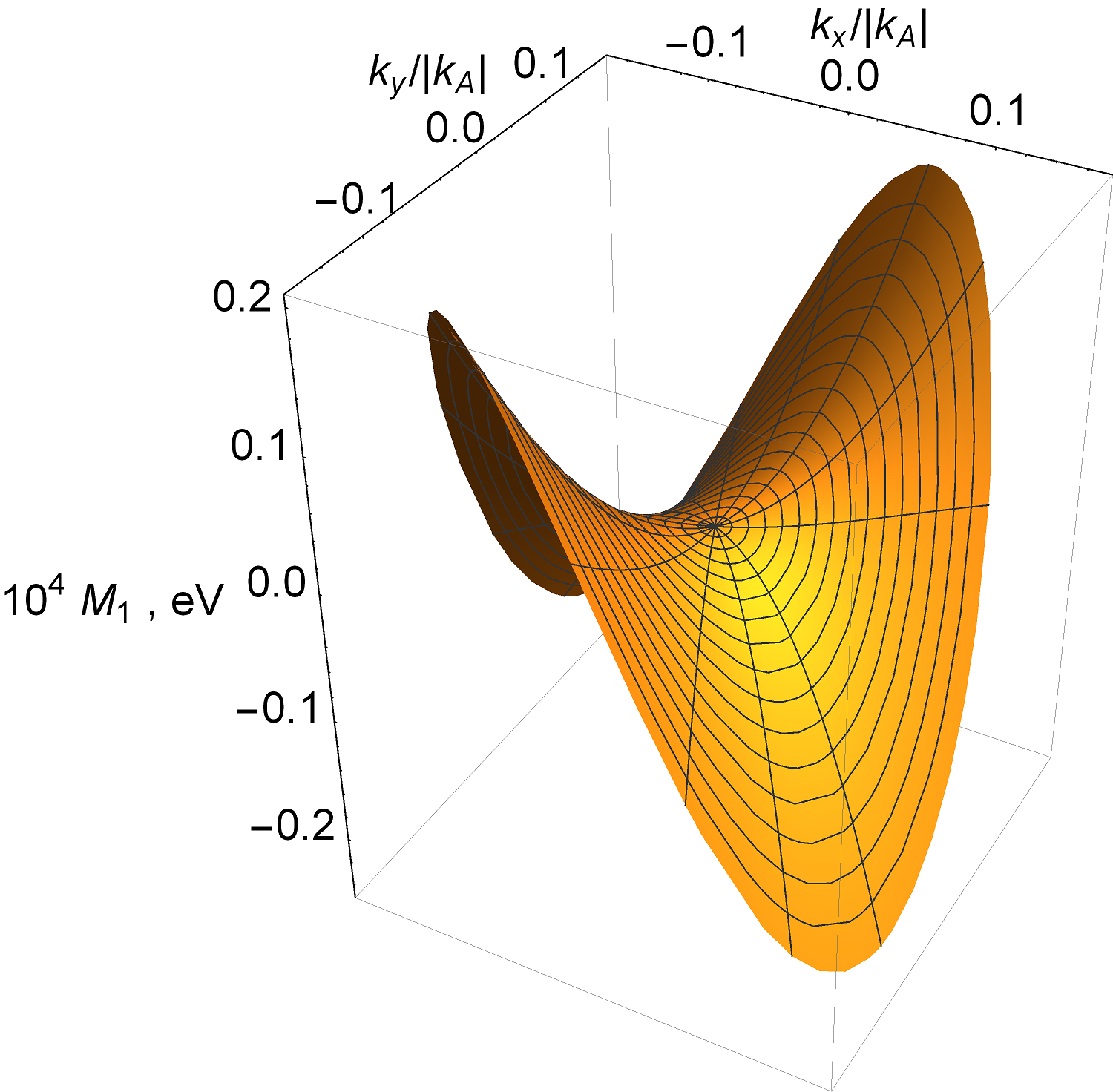}
\includegraphics[width=4.cm,height=2.4cm,angle=0]{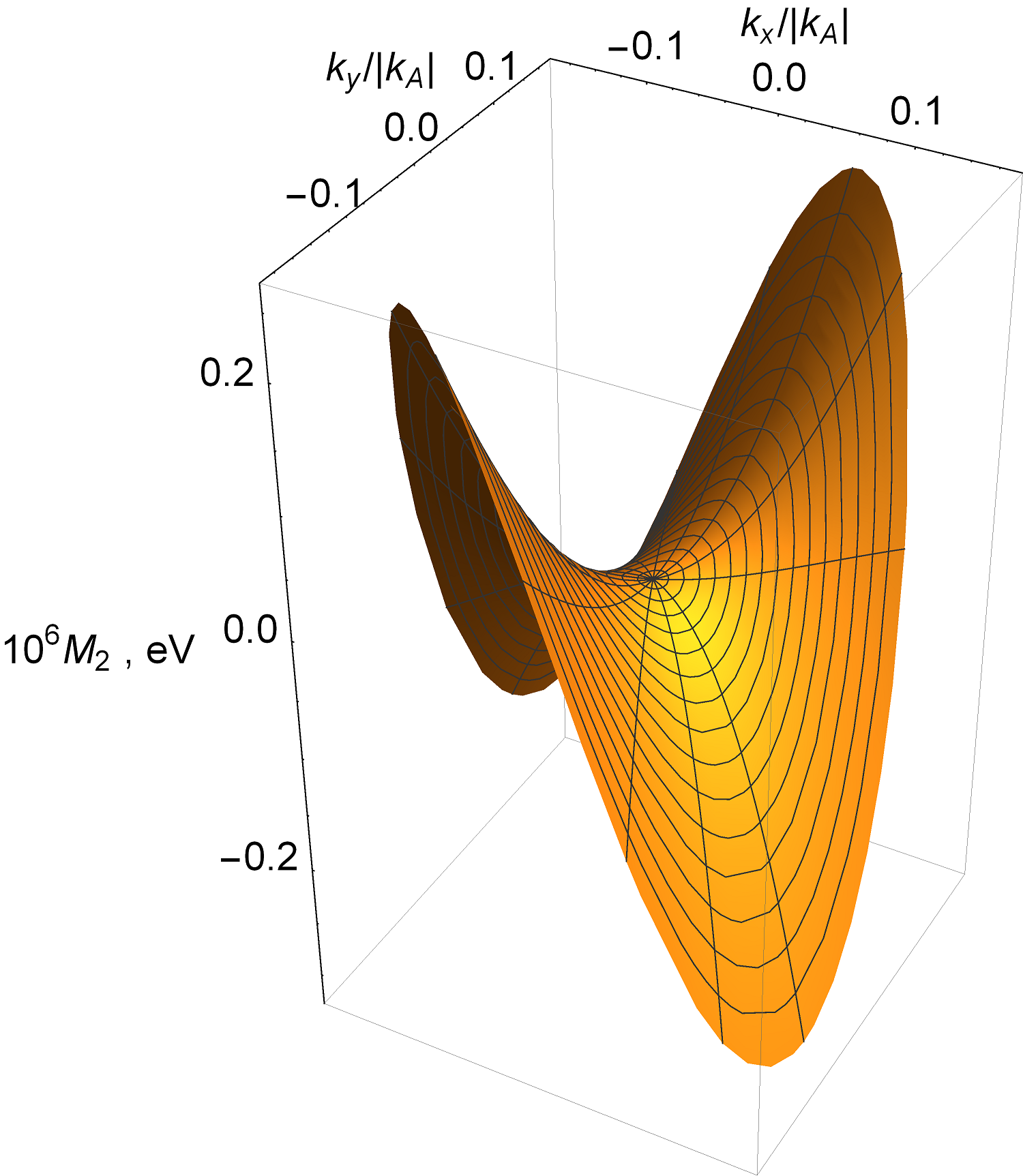}
\includegraphics[width=4.cm,height=2.4cm,angle=0]{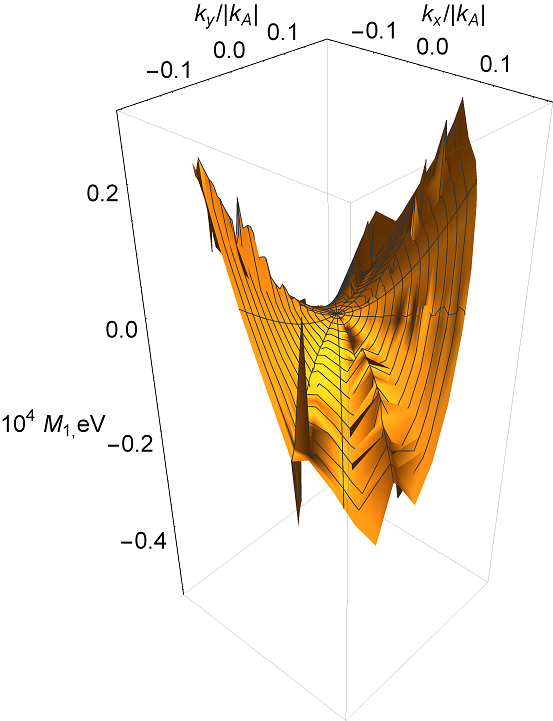} \hspace{0.5cm}
\includegraphics[width=4.cm,height=2.4cm,angle=0]{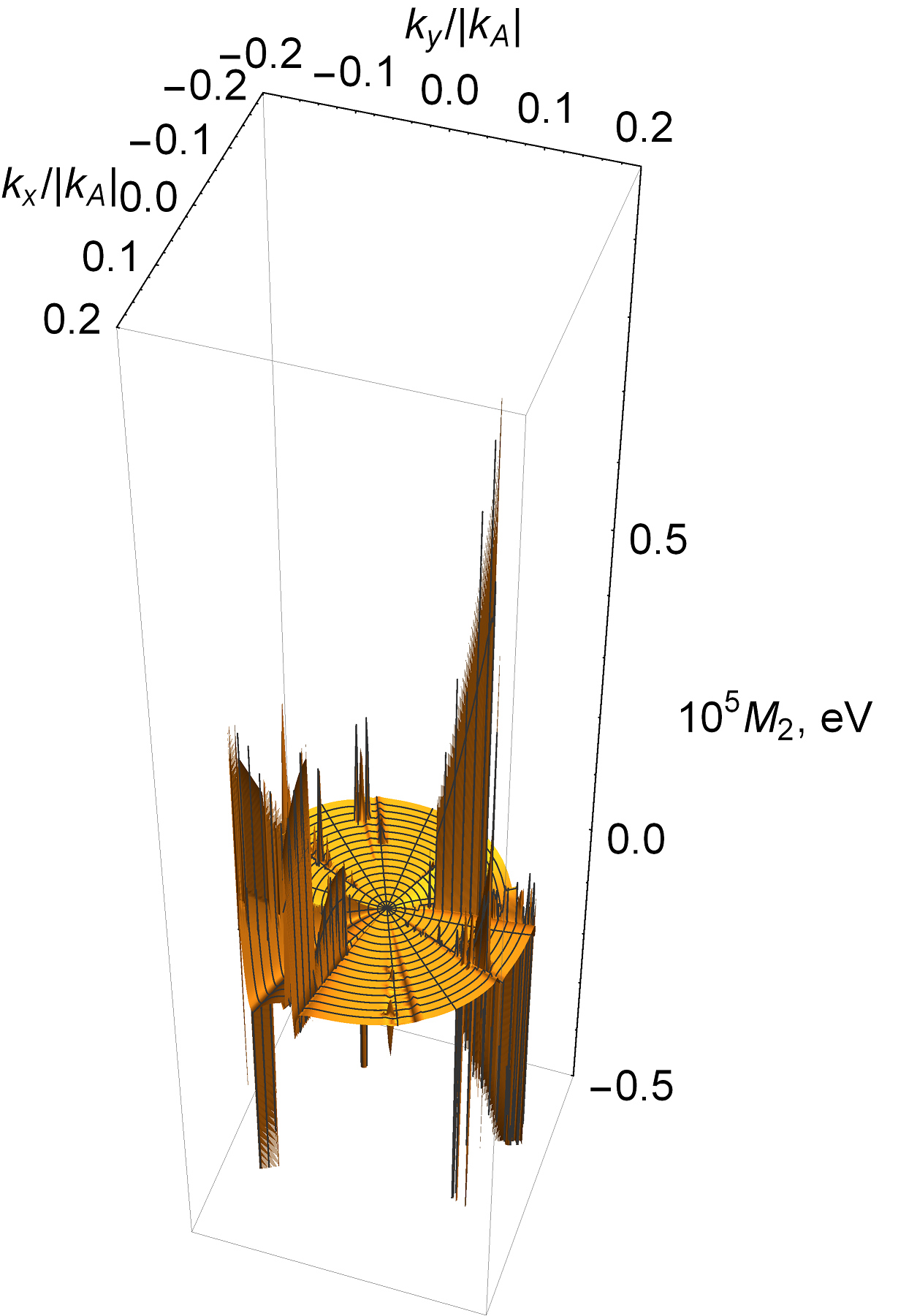}
\end{center}
\caption{Eigenvalues 
of the mass operator  in approximations of zero 
(a, b) and non-zero 
(c, d) gauge fields. (a, c) 
-  for particles 
(holes), (b, d) 
- for holes (particles).} \label{Mass-eigen-values}
\end{figure}

Eigenvalues 
of the mass operator $\widetilde {\Sigma_{AB}\Sigma_{BA}}$ \eqref{mass-operator} are represented 
in fig.~\ref{Mass-eigen-values}. 
These values are the dynamic masses of  particle 
and hole components of the Majorana state. Since the eigenvalues are equal to zero
in the Dirac points $K, \ K' $ 
Majorana zero energy modes exist 
in our model. 

Let us prove that the chirality 
of zero-energy modes  is preserved in the Dirac points. 
Within the approximation of zero gauge-phases (the first-order approximation) 
$c_{\pm}(\vec k)$, and respectively zero-valued  gauge fields, 
the eigenvalues 
of the mass operator differ by two orders of magnitude from each other outside of the Dirac
point, as one can see from the comparison of particle and hole masses 
in fig.~\ref{Mass-eigen-values}a,b. Since the mass operator is not diagonal in the energy representation of the massless
Hamiltonian $ H_0 $, the masses of Majorana fermions are obtained by mixing  particle and hole states.
 The density of 2D-states (DOS) holds the van Hove singularity
\cite{Bassani1975}
, because DOS is divergent in the Dirac point.
Since a hyperbolic point (saddle) is a feature of the dependence of mass  on
momentum in fig.~\ref{Mass-eigen-values}a,~b this singularity remains after including the mass term.
Therefore, 
particle and hole densities are concentrated at the energy $ E = 0 $,
that leads to 
particle--
hole annihilation. Since the 
mass term is alternating in one point, the chirality is preserved 
for the Majorana zero-energy modes only. 
In the first-order approximation there is no such a neighborhood 
of the Dirac point, where 
the mass term takes zero values 
and hence the Dirac bands are not chiral everewhere except of 
the Dirac points. Thus,  dichroism 
of the bands can not be described 
in the zero-gauge-field approximation. In what follows we show that dichroism
can be observed in the second-order approximation.

The comparison of particle and hole masses 
in fig.~\ref{Mass-eigen-values}a,~c demonstrates that in the second-order
approximation 
of  non-zero gauge fields 
$c_{\pm}(\vec q)$,    qualitatively the same momentum dependence remains
for one of the mass operator eigenvalues.
The dependence of the  mass-term second eigenvalue 
 upon the momentum  in fig.~\ref{Mass-eigen-values}d exhibits a singular alternating behaviour  
 and gets a zero values in the vicinity of the  
Dirac point. It means that  
chirality is preserved not only in the Dirac point but
in the Dirac band as well. 
Meanwhile,  the dichroism  is observed outside the Dirac point but only one of two right- or left-hand
Majorana modes   remains chiral.
The appearance of the mass term peak in fig.~\ref{Mass-eigen-values}d
leads to divergence 
of DOS at some another value of the energy 
$E$, that makes the particle and hole densities
spatially separated from to each other.
As a result, there exist non-zero energy  Majorana modes in the model $N=3$ without particle--
hole annihilation. 

Values of 
spectral weight (spectral function) 
$A(q, \omega)$ are much higher for a heavy particle than for
the light one. 
Accordingly, the dielectric permeability of the system of 
heavy particles is very small compared with those of the system of light ones
\cite{Kraft-Ropke}.
Hence, low-intensive 
Fermi arcs of the heavy particle (hole) components of Majorana states augment much more intensive
Fermi arcs of the light ones up to Fermi particle(hole) 
pockets.
Therefore, the Luttinger theorem is not violated 
in our approach. Moreover, the examples of the narrow-gap chalcogenide Pb$Me$, $Me =$ Te, Se, S being graphene
three-dimensional analogue \cite{Comput-Mater-Sci2005,Falkovsky2008PhysUspekhi} and of doped
Mott insulator La$_{2-x}$Sr$_x$TiO$_3$ \cite{Tokura1993PRL} appear to take the heavy mass route.

So, the construction and simulation of the mass operator 
in the topological SM theory point to the main difference of the chiral lattice theory of 
particle--hole pairs from the theory of Majorana fermions. 
A Majorana mass term $\cal{L}_M $ for left-hand 
$\psi _L$ and right-hand 
$\psi_R$ Majorana wave functions 
reads
${\cal L}_M \propto \left( \psi_L \right)^T\gamma_0 \psi_L +\left( \psi ^*_L \right)^T\gamma_0 \psi ^*_L$
owing to $\psi_R=\psi ^*_L$ and, hence, chirality of the Majorana theory is broken everywhere except of
zero-energy states \cite{PhysicsEMajorana-2014Wilczek}. Contrary to that 
in our Majorana-like theory,
the second eigenvalue of  the mass term equals to zero  in the vicinity 
of the Dirac point and therefore
the chirality breaks for only one constituent
of a particle-hole pair.  
The violation of the law of conservation of topological charge in the form
of weak imbalance in the number of topological
charges of the opposite signs is a feature of our theory.
It is called 
a chiral anomaly.

\section{Charge carrier transport  in the model $N=3$}

An equation with accounting of electron-photon interactions in the semimetal
can be obtained from 
\eqref{Majorana-bispinor1} 
 by ordinary use of the canonical momentum: 
\begin{eqnarray}
\left[ c \vec \sigma _{2D}^{BA} \cdot \left(\vec p_{AB} -{e\over
c}\vec A \right)-\widetilde {\Sigma_{AB}\Sigma_{BA}} \left(\vec
p_{AB}- e\vec A/c \right) \right] \widehat {\tilde \chi
^{\dagger}_{+\sigma_{_B} }} (\vec r)\left|0,-\sigma \right>
= c E_{qu}(p) \widehat {\tilde \chi
^{\dagger}_{+\sigma_{_B} }} (\vec r) \left|0,-\sigma \right> , 
\label{rel-from-pseudi-Dirac-whithout-mix3}\\
\widetilde {\Sigma_{AB}\Sigma_{BA}} \left(\vec p_{AB}- e\vec A/c
\right) =
\widetilde {\Sigma_{AB}\Sigma_{BA}}(
0)+ \sum_i \left. {d\widetilde {\Sigma_{AB} \Sigma_{BA}}\over d
p_i'} \right|_{p_i'=0} \left( p_i ^{AB}
-{e\over c} A_i \right) \nonumber \\
+ {1\over 2} \sum_{i,j} \left. {d^2\widetilde {\Sigma_{AB}
\Sigma_{BA}}\over d   p_i' d   p_j'} \right|_{p_i', \ p_j' =0}
\left( p_i ^{AB} -{e\over c} A_i \right)\, \left( p_j ^{AB}
-{e\over c} A_j \right) + \ldots . \label{mass-renormalization2}
\end{eqnarray}
In what follows, we omit the cumbersome designation 
"$AB$"\ if this does not lead to the lack of sense. 

Now, taking into account 
the equations \eqref{rel-from-pseudi-Dirac-whithout-mix3} and
\eqref{mass-renormalization2}), one can find the quasi-relativistic current 
\cite{Davydov} 
of charge carriers 
in SM as:
\begin{eqnarray}
j_i^{SM} =c^{-1}j_i, 
\nonumber \\
j_i=e \chi^{\dagger}_{+\sigma_{_B} } (x^+) v^i_{x^{+}x^-}
\chi_{+\sigma_{_B} } (x^-)
 -{e^2 A_i\over c\,
 \widetilde {\Sigma_{AB}\Sigma_{BA}}(\vec p_{AB}- e\vec A/c )}
 \chi_{+\sigma_{_B} }^\dagger (x^+) \chi_{+\sigma_{_B} }(x^-)
\nonumber \\
 + {e \hbar \over 2
 \widetilde {\Sigma_{AB}\Sigma_{BA}}(\vec p_{AB}- e\vec A/c ) }
\left[
 \vec \nabla \times \chi_{+\sigma_{_B} }^\dagger (x^+)\vec \sigma
 \chi_{+\sigma_{_B} }(x^-)
\right]_i . \label{graphene-quasirel-current}
\end{eqnarray}
Here
\begin{eqnarray}
x^{\pm}=x \pm\epsilon ,\ x=\{\vec r,\ t_0 \},\ t_0=0,\ \epsilon
\to 0 ;\label{current-limits}
\end{eqnarray}
$\vec v $ is the velocity operator determined by a derivative of
the Hamiltonian (\ref{rel-from-pseudi-Dirac-whithout-mix3}).
It worth to remark that in accord with
(\ref{current-limits}) the current is obtained from the quantity dependent
upon  two points
$x^{\pm}$, with subsequent performing the limit
$\epsilon \to 0$.
2D-rotor in the series expansion
(\ref{graphene-quasirel-current}) can be presented in the form
\begin{eqnarray}
\nabla \times \chi_{+\sigma_{_B} }^\dagger \vec \sigma
 \chi_{+\sigma_{_B} } \sim
 \sum_{i=1}^2\left [ \nabla \times \chi_{+\sigma_{_B} }^\dagger \vec \sigma
 \chi_{+\sigma_{_B} }\right]_i\vec e_i
 =
\left[ {\partial \over \partial y} \vec e_1 - {\partial \over
\partial x} \vec e_2 \right]\chi_{+\sigma_{_B} }^\dagger \sigma_z
 \chi_{+\sigma_{_B} }
 \label{curl-current}
\end{eqnarray}
where $\vec e_i$, $i=1,\, 2$ are unit vectors along the coordinate
axis directions. The substitution of
(\ref{curl-current}) into
(\ref{graphene-quasirel-current}) gives
\begin{equation}
\begin{split}
j_i= j_i^{Ohm}+j_i^{Zitterbew}+j_i^{spin-orbit},
\\
j_i^{Ohm}=e \chi^{\dagger}_{+\sigma_{_B} } (x^+) v^i_{x^+x^-}
\chi_{+\sigma_{_B} } (x^-),\\
j_i^{Zitterbew}= -{e^2 A_i\over c\,
 \widetilde {\Sigma_{AB}\Sigma_{BA}}(\vec p_{AB}- e\vec A/c )}
 \chi_{+\sigma_{_B} }^\dagger  \chi_{+\sigma_{_B} },
\\
j_{2(1)}^{spin-orbit} =
 (-1)^{1(2)} {\imath e \over 2 }
v^{1(2)}_{x^+ x^-}\chi_{+\sigma_{_B} }^\dagger  \sigma_z
 \chi_{+\sigma_{_B} }
 .
\end{split}
 \label{graphene-quasirel-current2}
\end{equation}
Terms
$j_i^{Ohm},\ j_i^{Zitterbew},\ j_i^{spin-orbit}$ in
(\ref{graphene-quasirel-current2}) describe
ohmic contribution which satisfies the Ohm law and
contributions of the polarization and magneto-electric effects respectively.


Now,  calculating  the currents (\ref{graphene-quasirel-current2}) one can find
the Ohmic conductivity (equation \ref{conduction2} in Supplementary Information),
the polarization and magneto-electric contributions to it
(equations \ref{Zitterbewegung_conduction} and \ref{spin-orbit-conduction}
in Supplementary Information) through the scalar product 
$(\cdot\, , \, \cdot) $ of the vectors 
$M  \vec v^i(p)$ and $N \vec v^i(p)$, $i=x,\ y$ as
\begin{eqnarray}
\sigma_{ii}^{Ohm}(\omega^{-+}, \ k)=
 {\imath e^2\bar{\beta}\over  (2\pi c)^2} \ \mbox{Tr}\
 \int
 \left(1-
  \widetilde {\Sigma_{AB}\Sigma_{BA}}(\vec p_{AB}- e\vec A/c )
   {d^2\widetilde {\Sigma_{AB} \Sigma_{BA}}\over d   p_i^{AB} d p_i^{AB}}
  (0)\right)
\left(
 M \vec v^i(p)
 \, , \,
N\vec v^i(p)
 \right)\,   d(\bar{\beta}\vec {p})   ,\ \ \
 \label{conduction3}\\
 \sigma_{ll}^{Zitterbew}(\omega^{-+}, \ k)={ \imath e^2\bar{\beta}  \over   (2\pi c)^2}
\mbox{Tr }\ \int {\widetilde {\Sigma_{AB}\Sigma_{BA}}(\vec p_{AB}-
e\vec A/c)\over 2}
 \sum_{i=1}^2
{d^2\widetilde {\Sigma_{AB} \Sigma_{BA}}\over d   p_i^{2}}
 \left(M \vec v^i(p) \, , \, N\vec v^i(p) \right) \,   d(\bar{\beta}\vec {p})   ,\ \ \
 \label{Zitterbewegung_conduction1}\\
 \sigma_{12(21)}^{spin-orbit}(\omega^{-+}, \ k )=
(-1)^{1(2)} {\imath \over 2 }{\imath e^2\bar{\beta }\over   (2\pi
c)^2} \mbox{Tr }\
 \int
 {\widetilde {\Sigma_{AB}\Sigma_{BA}}(\vec p_{AB}- e\vec A/c)
 }
{d^2\widetilde {\Sigma_{AB} \Sigma_{BA}}\over d p_{1(2)} d
p_{2(1)}} (0)
 \nonumber \\ \times
\left( M\vec v\, ^{1(2)}(p)
\, , \, N \vec v\, ^{1(2)}(p) \right)  \sigma_z   d(\bar{\beta}\vec {p}).
 \label{spin-orbit-conduction1}
\end{eqnarray}
where
matrices $M,\ N$ are given by the following expressions:
\begin{equation}
\begin{split}
M =
 {f[\beta ((H(p^+)-\mu)/\hbar)]
- f[\beta (H^\dagger (-p^-)-\mu/\hbar)]\over \beta ( z^{-+}) -
\beta (H(p^+)/\hbar) + \beta(H^\dagger (-p^-)/\hbar) },\
 N= {\delta\left(\hbar \omega^{-+}+ \mu^{-+}\right)\over  \hbar (z^{-+}
+\omega^e (\hat p^+) - \omega^h (-\hat p^-))\bar \beta }.
\end{split}
\label{operatorsM&N}
\end{equation}
Here $f$ is a Fermi -- Dirac distribution, $\beta$ is an inverse temperature.
In the diagonal Hamiltonian representation, the trace 
in (\ref{conduction3} -- \ref{spin-orbit-conduction1}) can be easily carried out, for example, as
\begin{eqnarray}
\left(\mbox{Tr}\ M \vec v\, ^x_{AB} \, , \, N\vec v\,
^y_{AB}\right)= \mbox{Tr}\ \left(M \vec v\, ^x_{AB}  \, , \, N
\vec v\,^y_{AB}\right)= (\vec e_x,\vec e_y) \sum_{i,k,l,m =1}^2
{v^\dagger}^{x, AB}_{ik} M^\dagger_{kl} N_{lm} v^{y, AB}_{mi}
 \nonumber \\
= (\vec e_x,\vec e_y) \sum_{i,\, k=1}^2 {v^\dagger}^{x, AB}_{ik}
M^\dagger_{kk}  N_{kk} v^{y, AB}_{ki} \label{trace1}
\end{eqnarray}
because matrices $M_{kl}, N_{kl}$ depending on the diagonal matrix $H_{AB}$ are diagonal ones.

Since there exists the change 
$-H(p)\to H^\dagger (-p)$, for every band 
$a$ ($a=1,\, 2$), it is possible to introduce Hamiltonians 
of a quasi-particle 
$ H_0^{a}$, $ \mathop{H_0^{a}}^\dagger$ with 
eigenvalues 
$E^e_{a}$, $E^h_{a}$  and respectively to quantize $M$ and $N$ (\ref{operatorsM&N}) as
\begin{eqnarray}
M ^\dagger = \left\{M^\dagger _{ab} \right\},\ M^\dagger _{ab}=
 {f[\bar{\beta} (H_0^{a}(p^+)-\mu)]
- f[\bar{\beta}( \mathop{H_0^{b}}^\dagger (-p^-)-\mu)]\over
\bar{\beta} \hbar z^{-+} - \bar{\beta}H_0^{a}(p^+) + \bar{\beta}
\mathop{H_0^{b}}^\dagger (-p^-) };
\label{matrix-quantization2}\\
N^\dagger= \left\{ N^\dagger_{ab}\right\},\ \ N^\dagger_{ab} = {\delta\left(\hbar
\omega^{-+}+ \mu^{-+}\right)\over (\hbar \omega + H_0^b(p^+) -
\mathop{H_0^a}^\dagger (-p^-) )\bar{\beta}} .
\label{matrix-quantization3}
\end{eqnarray}
As a result, the expression 
(\ref{trace1}) can be rewritten in the form 
\begin{eqnarray}
\left(\mbox{Tr}\ M v^x_{AB} \, , \, N v^y_{AB}\right) = (\vec
e_x,\vec e_y)\sum_{a=1}^2\left( v^{x, AB}_{aa} M^\dagger _{aa}
N_{aa} v^{y, AB}_{aa} + v^{x, AB}_{ab} M^\dagger _{ba} N_{ab}
v^{y, AB}_{ba} \right),\ a\neq b. \label{trace2}
\end{eqnarray}
After substitution of
(\ref{matrix-quantization2} -- \ref{trace2})
into
(\ref{conduction3}), we express, for example, the Ohmic contribution to conductivity as
\begin{eqnarray}
\sigma_{ij}^{Ohm}(\omega^{-+}, \ k)=
 {\imath e^2\bar \beta\over   (2\pi c)^2}
 \int (\vec e_i,\vec e_j) \left(1-
  \widetilde {\Sigma_{AB}\Sigma_{BA}}(\vec p_{AB}- e\vec A/c )
   {d^2\widetilde {\Sigma_{AB} \Sigma_{BA}}\over d   p_i^{AB} d p_j^{AB}}
  (0)\right)
\nonumber \\   \times
\sum_{a=1}^2\left\{ {v^\dagger}^{i}_{aa}(p,\ k)
{f[\bar{\beta} (H_0^{a}(p^+)-\mu)] - f[\bar{\beta}(
\mathop{H_0^{a}}^\dagger (-p^-)-\mu)]\over \bar{\beta} \hbar
z^{-+} - \bar{\beta}H_0^{a}(p^+) + \bar{\beta}
\mathop{H_0^{a}}^\dagger (-p^-) }
  v^{j}_{aa} (p,\ k)
  \right. \nonumber \\
\times
{1\over (\hbar \omega^{-+} + H_0^a(p^+) - \mathop{H_0^a}^\dagger
(-p^-) )\bar{\beta}}
+ {f[\bar{\beta} (H_0^{a}(p^-)-\mu)] - f[\bar{\beta}(
\mathop{H_0^{b}}^\dagger (-p^+)-\mu)]\over \bar{\beta} \hbar
z^{-+} - \bar{\beta}H_0^{a}(p^+) + \bar{\beta}
\mathop{H_0^{b}}^\dagger (-p^-) }
\nonumber \\
\left.\times
{{v^\dagger}^{i}_{ab} (p,\ k) v^{j}_{ba}(p,\ k)\over (\hbar \omega
^{-+}+ H_0^b(p^+)- \mathop{H_0^a}^\dagger (-p^-) )\bar{\beta}}
\right\}\ d(\bar{\beta}\vec {p})\delta\left(\hbar \omega^{-+}+
\mu^{-+}\right)
,\ a\neq b;\ \ a, b =1, 2.
 \label{conduction4}
\end{eqnarray}

\subsection{Approximation of the degenerate Dirac cone 
}

In the case of degeneration 
$E^e_{1,\, 2}(p)\approx \mp c v_F \hbar p$, а $E^h_{1,\, 2}(p)\approx \pm c v_F \hbar p$.
Since 
$E_a^e (p)= E_a^h (-p)$, using calculus in section II of Supplementary Information 
and neglecting 
the dynamic mass correction,  we get the intraband-transition contribution 
$\sigma_{aa,\, ij}^{Ohm}$ to conductivity due to the transitions 
in the same band: 
\begin{eqnarray}
\sigma_{aa,\, ij}^{Ohm}(\omega, \ k)=
 {\imath e^2\over c^2 (2\pi)^2} {\bar{\beta 
 }}
 \int  v^{i}_{aa}(p,\ k)
{f[\bar{\beta} (H_0^{a}(p^-)-\mu)] - f[\bar{\beta}(
\mathop{H_0^{a}}^\dagger (-p^+)-\mu)]\over \bar{\beta} \hbar z -
\bar{\beta}H_0^{a}(p^-) + \bar{\beta}  \mathop{H_0^{a}}^\dagger
(-p^+) }
  v^{j}_{aa} (p,\ k) \nonumber \\
\times
{1\over (\hbar \omega - \mathop{H_0^a}^\dagger (-p^+) +
H_0^a(p^-))\bar{\beta}} \ d(\bar{\beta}\vec {p})
=-
{\imath e^2\over c^2  (2\pi)^2} 
 \nonumber\\
\times\int
{\omega(k)v^{i}_{aa}(p,\ k)  v^{j}_{aa} (p,\ k)
\partial f[\bar{\beta} (E_{a}(p)-\mu)]/\partial E_a(p)
\over
 (\hbar z \hbar \omega -\omega(k)\hbar \omega
+ \hbar z \omega(k) - \omega^2(k))}\ d\vec {p}, \quad  i, j\in
\{x,y\}.
 \label{intra-zone-conduction}
\end{eqnarray}
Here $E_a(p)=E_a^e (p)=E_a^h (-p)$. Let us make the change 
$\omega(k)\to c\tilde \omega(k)$, and account for the existence of 
$\delta (\hbar(z^- - z^+)-\hbar z) $. Then calculating 
(\ref{intra-zone-conduction}) one  gets:
\begin{eqnarray}
\sigma_{aa,\, ij}^{Ohm}(\omega, \ k) =-
{\imath e^2\over  (2\pi)^2} 
 \nonumber\\
\times\int
{\tilde \omega(k)\tilde v^{i}_{aa}(p,\ k)  \tilde v^{j}_{aa} (p,\
k)
\partial f[(\epsilon_{a}(p)-\mu)/T]/\partial \epsilon_a(p)
\over (\hbar z_{12} \omega -c\tilde \omega(k) \hbar\omega + c
\hbar z   \tilde \omega(k) - c^2\tilde \omega^2(k) )}\ d\vec {p},
\quad  i, j\in \{x,y\}.
 \label{intra-zone-conduction1}
\end{eqnarray}
where  $\epsilon_a =E_a/c$, $\tilde v^{i}_{aa}=v^{i}_{aa}/c$,
$z_{12}=z_- -z_+$ is a photonic frequency.

Let us estimate the contribution to the conductivity of the interband transitions.
Multiplication on 
the fermionic frequency $\hbar \tilde \omega \, c^2 $ and division on 
the photonic frequency of interband transition 
$\hbar \tilde \omega_{12}= \epsilon_a(p^+)-\epsilon_b(p^-) $, $ \tilde
\omega_{12}= \omega_{12} /c $ are
possible due to 
$\delta (\hbar(z^- - z^+)-\hbar z) $. It allows to perform the following estimation
of interband contribution to the conductivity in
(\ref{conduction4}):
\begin{eqnarray}
\sigma_{ab,\, ij}^{Ohm}(\omega, \ k)=
 {\imath e^2\over c^2 (2\pi)^2} {\bar{\beta 
 }} 
 \int
 {f[\bar{\beta} (H_0^{a}(p^-)-\mu)]
- f[\bar{\beta}( \mathop{H_0^{b}}^\dagger (-p^+)-\mu)]\over
\bar{\beta} \hbar z - \bar{\beta}H_0^{a}(p^-) + \bar{\beta}
\mathop{H_0^{b}}^\dagger (-p^+) }
{v^{i}_{ab} (p,\ k) v^{j}_{ba}(p,\ k)\over (\hbar \omega -
\mathop{H_0^a}^\dagger (-p^+) + H_0^b(p^-))\bar{\beta}} \
d(\bar{\beta}\vec {p})
\nonumber \\
=
 {\imath e^2\hbar \over  (2\pi)^2}  \int
 \tilde \omega {f[ \bar{\beta}(H_0^{a}(p^-)-\mu)]
- f[\bar{\beta}( \mathop{H_0^{b}}^\dagger (-p^+)-\mu)]\over
  (\epsilon_a(p^+)-\epsilon_b(p^-))(\hbar\tilde z_{12} -
  \epsilon^e_{a}(p^-)
+    \epsilon^h_{b}  (-p^+) )}
{\tilde v^{i}_{ab} (p,\ k) \tilde  v^{j}_{ba}(p,\ k)\over (\hbar
\tilde \omega - \epsilon^h_{a}  (-p^+)  + \epsilon^e_{b}(p^-) ) }
\ d\vec {p}
,\ a< b .
 \label{inter-zone-conduction}
\end{eqnarray}
Here $\epsilon (p)= E(p)/c $. Since 
$E_a^h (p)-E_b^e (-p)= 2E(p)$, then 
(\ref{inter-zone-conduction}) can be transformed to the form 
\begin{eqnarray}
\sigma_{ab,\, ij}^{Ohm}(\omega, \ k)=
 {\imath e^2 \over  (2\pi)^2}  \int
\tilde \omega {f[ (\epsilon(p)-\mu)/T] - f[(
\epsilon(-p)-\mu)/T]\over
  2\epsilon (p)[(\hbar \tilde z_{12} )^2-  4 \epsilon^2 (p) ]}
\tilde v^{i}_{ab} (p,\ k) \tilde  v^{j}_{ba}(p,\ k)
\ d(\hbar\vec {p})
.
 \label{inter-zone-conduction1}
\end{eqnarray}
In the limit 
$\hbar z_{12}\to  \omega (k) +\imath \delta $, $\delta\to 0$ (we omit sign 
"$\ \widetilde{}\ $"\ 
), 
the expression 
(\ref{intra-zone-conduction1}) leads to 
\begin{eqnarray}
\sigma_{aa,\, ij}^{Ohm}(\omega, \ k) =-
{\imath e^2\over  \hbar (2\pi)^2} 
\int
{\omega(k)\tilde v^{i}_{aa}(p,\ k)  \tilde v^{j}_{aa} (p,\ k)
\partial f[(\epsilon_{a}(p)-\mu)/T]/\partial \epsilon_a(p)
\over ( \omega (k)+\imath \delta ) \, \hbar\omega  -
c^2\omega^2(k)}\ d(\hbar\vec {p}), \quad  i, j\in \{x,y\}.
 \label{intra-zone-conduction2}
\end{eqnarray}
If one neglects the small value of 
$\omega^2(k)$, in this limit the proposed estimation of conductivity is coincided  with that
in \cite{Falkovsky}.  

\section{Theory and experiment}

In this section 
we study  essential features of the electric charge transport
 by pseudo-Majorana  carriers in graphene
and compare the theoretical predictions with  experimental data.


\subsection{Braiding pseudo-Majorana modes  and topological skew currents}

Let us express  the massless ohmic contribution 
$\sigma_{ii}^{O}$ and the dynamical mass correction $\sigma_{ii}^{add}$ to it 
as
\begin{eqnarray}
\sigma_{ii}^{O}(\omega^{-+}, \ k) = {\imath \bar \beta  e^2
\over (2\pi c)^{2}}\mbox{Tr}
\left\{  
 \int d^2\vec p
\right.
\nonumber \\ \times
{v^\dagger }^i(p) \,
 {f[\beta ((H(p^+)-\mu)/\hbar)]
- f[\beta (H^\dagger (-p^-)-\mu/\hbar)]\over \beta ( z^{-+}) -
\beta (H(p^+)/\hbar) + \beta(H^\dagger (-p^-)/\hbar) }
\left.{\delta\left(\hbar \omega^{-+}+ \mu^{-+}\right)\over  \hbar
(z^{-+} +\omega^e (\hat p^+) - \omega^h (-\hat p^-)) }
v^i(p)\right\}, \ i=x,y
\label{conduction2-massless0}
\end{eqnarray}
 and
\begin{eqnarray}
\sigma_{ii}^{add}(\omega^{-+}, \ k) = {\imath \bar \beta  e^2
\over (2\pi c)^{2}}\mbox{Tr}
\left\{  
 \int d^2\vec p
 \left(
 -
  \widetilde {\Sigma_{AB}\Sigma_{BA}}(\vec p_{AB}- e\vec A/c )
   {d^2\widetilde {\Sigma_{AB} \Sigma_{BA}}\over d   p_i^{AB} d p_i^{AB}}
  (0)\right)\right.
\nonumber \\
\times {v^\dagger }^i(p) \,
 {f[\beta ((H(p^+)-\mu)/\hbar)]
- f[\beta (H^\dagger (-p^-)-\mu/\hbar)]\over \beta ( z^{-+}) -
\beta (H(p^+)/\hbar) + \beta(H^\dagger (-p^-)/\hbar) }
\left.{\delta\left(\hbar \omega^{-+}+ \mu^{-+}\right)\over  \hbar
(z^{-+} +\omega^e (\hat p^+) - \omega^h (-\hat p^-)) }
v^i(p)\right\} ,\ i=x,y;
\label{conduction2-topology-mass-correction-0}
\end{eqnarray}
respectively.

The polarization occurs because of the fact that 
ultrarelativistic massless quasiparticles can have hole-like states during its time evolution due to
uncertainty in the energy of these  particles \cite{Itzykson-Zuber2006Quantum-Field-Theory}.
The linear polarization current 
$j_l^{Zitterbew}= \sigma_{ll}^{Zitterbew}V$ is a displacement current. 
Here $V$ is a  voltage. Total conductivity $\sigma_{ii}$ 
in $i$-th direction can be obtained by addition of the polarization
contribution
$\sigma_{ii}^{Zitterbew}$ and the dynamical Ohmic mass correction
$\sigma_{ii}^{add}$ to $\sigma_{ii}^{O}$ 
\begin{equation}
\begin{split}
\sigma_{ii}=\sigma_{ii}^{O}+\sigma_{ii}^{tp}, \ \
\sigma_{ii}^{tp}\equiv \sigma_{ii}^{add}+\sigma_{ii}^{Zitterbew},\ \
\sigma_{xx}^{tp}= -\sigma_{yy}^{tp}
.
\label{conduction2-topology-mass-correction-1}
\end{split}
\end{equation}

\begin{figure}[hbt]
\begin{center}
\hspace{2cm}(a)\hspace{8cm} (b) 
\\
 \includegraphics[width=4.cm,height=4.cm,angle=0]{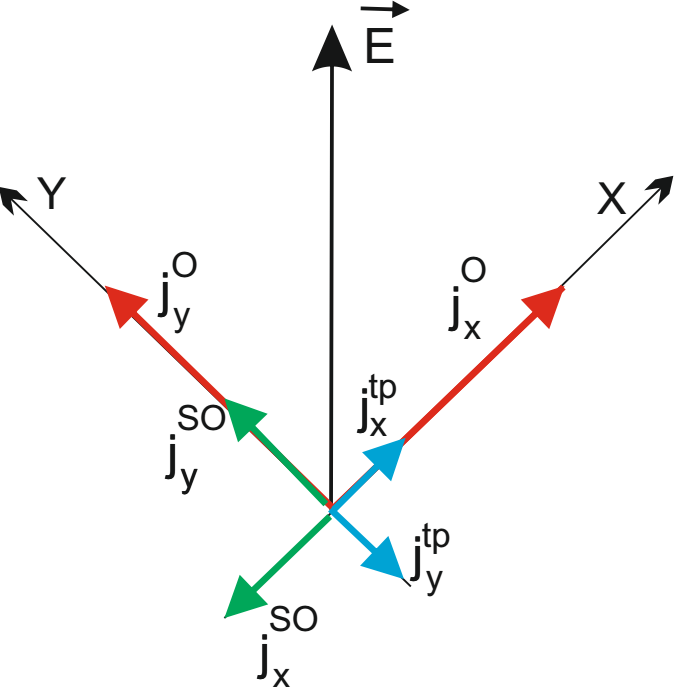}\hspace{0.8cm}
 \includegraphics[width=7.cm,height=4.cm,angle=0]{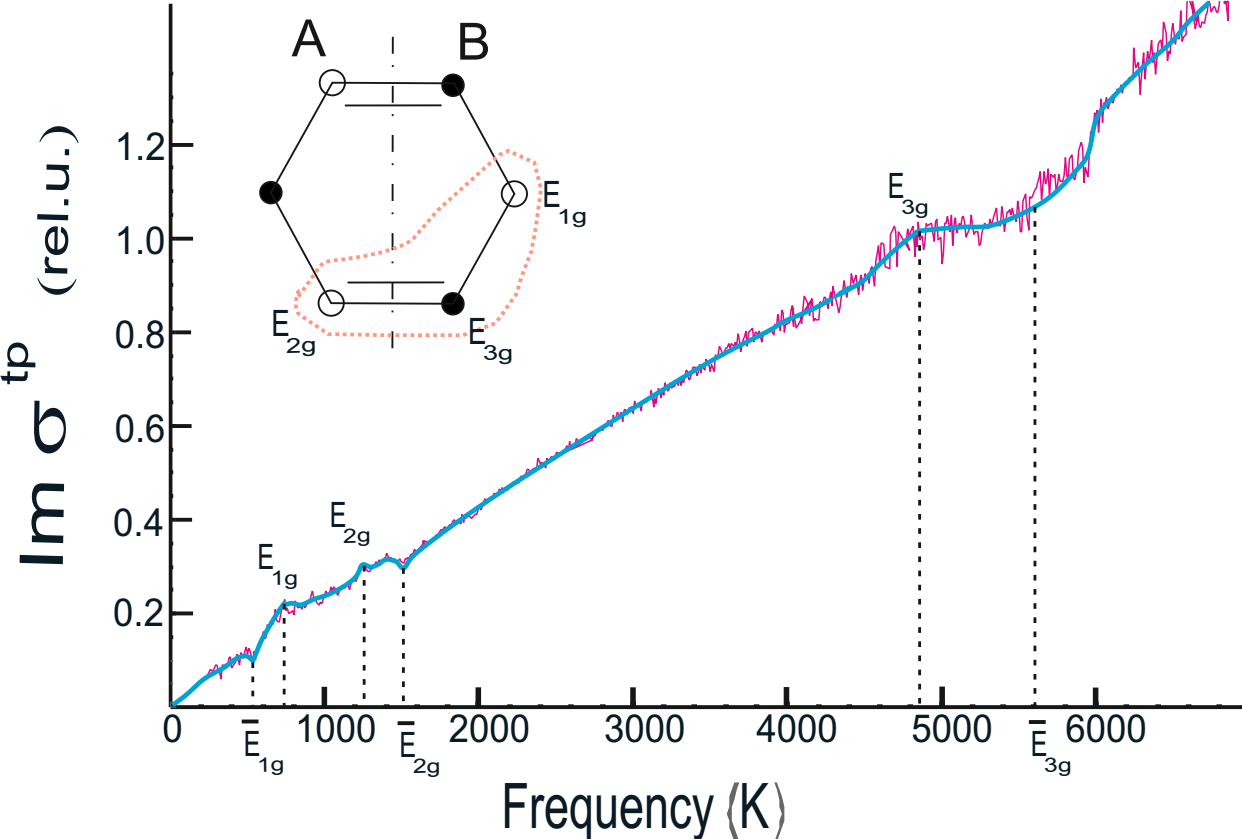}\\
\end{center}
\caption{(a) Sketch of currents in Dirac materials: $ j_{x(y)}^O$ is
a massless ohmic current along axis $X(Y)$, $  j_{x(y)}^{tp} $ is a sum of  the polarization
and dynamical-mass corrections to  $  j_{x(y)}^O $,
$j_{x(y)}^{SO}$ is a spin-orbit contribution to  $j_{x(y)}^O$.
$\vec E$ is an applied electric field. (b)
Frequency dependence of dielectric permeability
$\Im m \ \sigma_{xx}^{tp}(-\Im m \ \sigma_{yy}^{tp})$ (magenta curve) and
its fitting (blue curve) in rel.units $e^2/\hbar$ for the SM-model \eqref{variational-Majorana-bispinor}.
Spectrum consists of three Majorana resonances
(binding energies) $E_{1g},\ E_{2g},\ E_{3g}$ of negative (positive) charged exciton and
from three anti-resonances $\overline{E}_{1g},\ \overline{E}_{2g},\ \overline{E}_{3g}$
of positive (negative) charged exciton at temperature T=3K,
the higher chemical potential $\mu = 135$~K.
Inset shows a dimer configuration
with corresponding
binding (gap) energies.
}
\label{currents_sketch}
\end{figure}

Let us choose a reference frame so that hole 
$\vec j_{x}$ and 
electron 
$\vec j_{y}$ Ohmic massless currents are directed along Cartesian-coordinate axes 
$X,\ Y$ respectively. Due to independence of coordinate and
momentum spaces, the total current
$\vec J=\vec j_{x}+\vec j_{y}= \sigma ^{O}_{xx}\vec E_x+ \sigma ^{O}_{yy}\vec E_y $
is directed along 
an applied electrical field 
$\vec E= (\vec E_x,\ \vec E_y)$, as it is shown 
in fig.~\ref{currents_sketch}a. One can see that
the sums of the polarization and  dynamical mass corrections to the total  current are the same
but of different signs accordingly to \eqref{conduction2-topology-mass-correction-1}.
If the inverse symmetry of a semimetal is not broken,
then an angle $\arccos \left(\vec J\cdot \vec E/|J|/|E|\right)$ is equal to zero and respectively
currents
\begin{equation}
j_{x}^{tp}\equiv \sigma_{xx}^{tp}E_x, \
 j_{y}^{tp} \equiv\sigma_{yy}^{tp}E_y \label{chiral-anomaly}
\end{equation}
are mutually compensated along $\vec E$. 
The dependence of dielectric 
permeability 
$\Im m\ \sigma_{ii}^{tp}$ on frequency $\omega $  in fig.~\ref{currents_sketch}b holds 
three pairs of peak--antipeak. 
Accordingly, a three-particle excitation 
(negative (positive) charged exciton)  reveals itself as a state with 
three binding (gap) energies $E_{1g},\ E_{2g},\ E_{3g}$ 
(anti-binding energies 
$\overline{E}_{1g},\ \overline{E}_{2g},\ \overline{E}_{3g}$). These are Majorana resonances 
(antiresonances). The pairs resonance--antiresonance
are Majorana modes of three types (flavors), which correspond to three dimer configurations
\cite{MyJNPCS2017Vol20}. 
One of these configurations is shown in inset to 
fig.~\ref{currents_sketch}b.

Finding of Majorana resonances (antiresonances) proves that
a three-body scattering $S$-matrix is factorized into three two-body scattering S-matrices,
and, hence the Yang-Baxter equation  (YBE), which can be viewed as the factorization condition
\cite{Braid-Group-Knot-Theory}, 
holds.
Physical meaning of YBE is in such quantum entangling of two-body states in three-body one
that the three-particle excitations are obtained by entangling of
a some  particle-hole  pair with a particle (hole) \cite{IntJModPhys2014Ge-Yu}.
The quantum three-body entangling states are electrically charged excitons.
The last explains and proves the existence of charge transport 
in topological SM with equal number of electrons and holes as charge carriers. 

\begin{figure}[hbt]
\begin{center}
\hspace{2cm}(a)\hspace{8cm} (b) \\
 \includegraphics[width=7.cm,height=4.cm,angle=0]{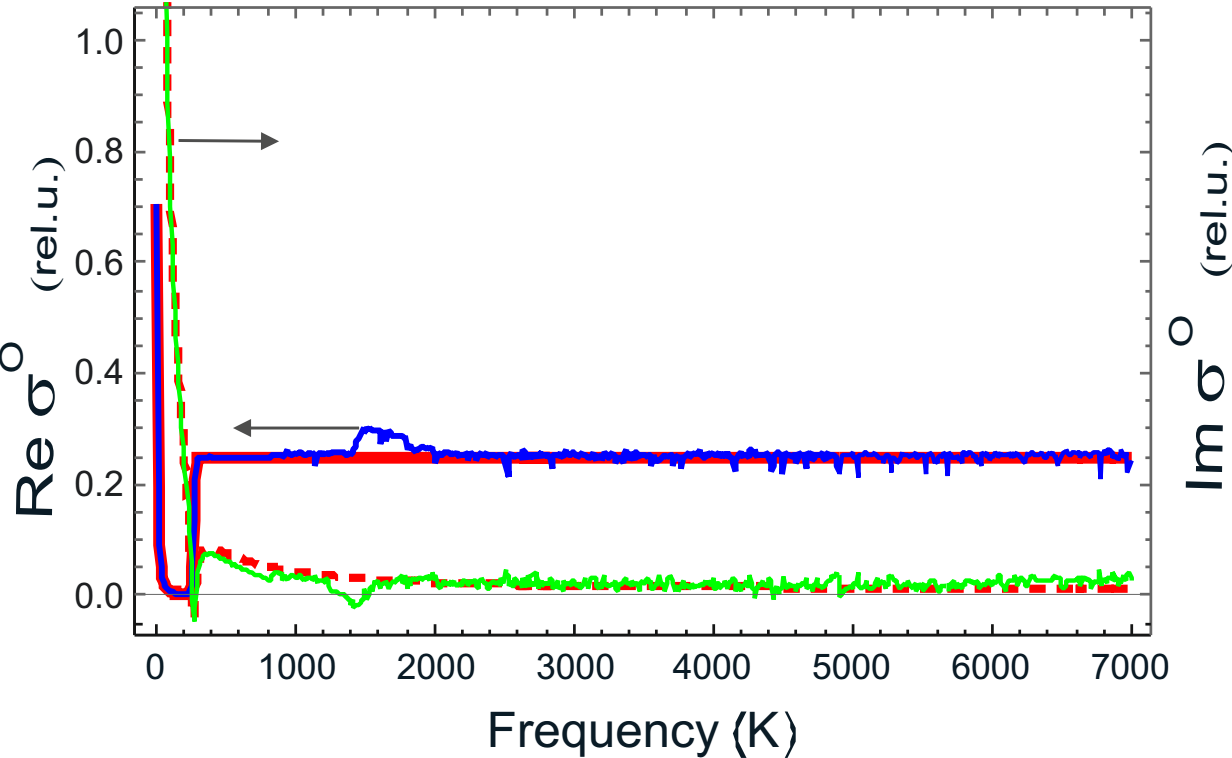}\hspace{1cm}
 \includegraphics[width=5.cm,height=4.cm,angle=0]{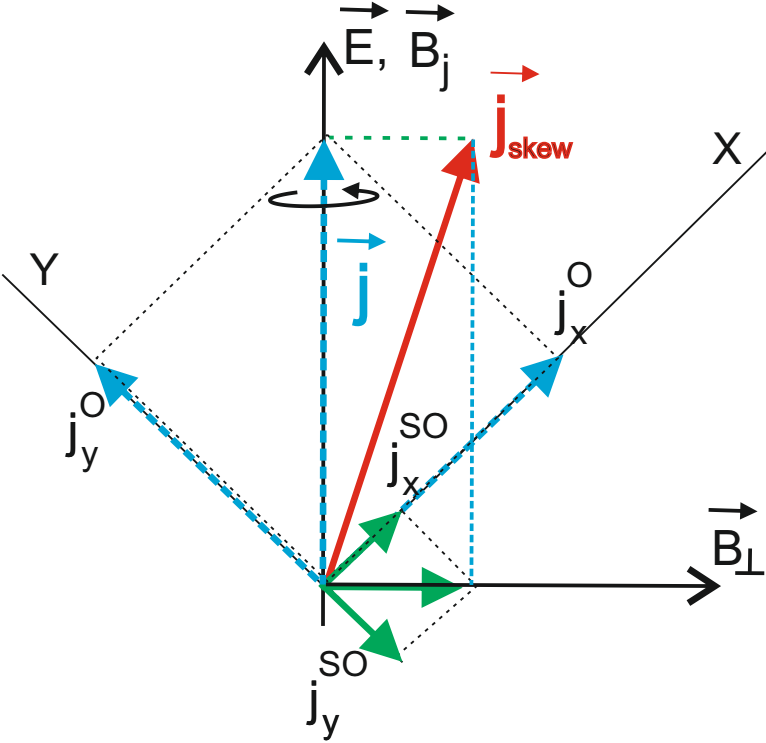}
\end{center}
\caption{
 (a) Frequency dependencies of  real (red and blue solid lines) and imaginary
(dashed red and solid green lines) parts of massless ohmic contribution 
$\sigma_{xx(yy)}^{O}$ to the conductivity
in rel.units $e^2/\hbar$ at very small wave number $q=10^{-8}|\vec K_A|$ for the Dirac massless fermion model
\cite{Falkovsky,Falkovsky2008PhysUspekhi}
(red solid and dashed lines) and for our model \eqref{variational-Majorana-bispinor}
 (blue and  green lines) , at temperature T=3K,
chemical potential $\mu =135$~K.  (b) Sketch of rotation of
the massless ohmic SM-currents $
j_{x}^O,\  j_{y}^O$ (dashed lines) as a result of addition with currents
$j_{2(1)}^{spin-orbit}$, generated by a magnetic field
${\vec B}_j$ of non-equilibrium spin
$\vec S $; $\vec j$ is a total massless ohmic current;
$\vec j^{skew}$ is a skew topological current. $j_{2(1)}^{spin-orbit}$ are depicted as $j_{x(y)}^{SO}$,
$\vec E$ is an applied electric field.
}
\label{topological-current}
\end{figure}

At higher values of chemical potential
$\mu$,
contrary to the massless Dirac fermion model of graphene conductivity with one region of negative values of
$\sigma_{ii}^{O}$ for a doped 
SM, in our non-abelian 
SM-model with 
pseudo-Majorana excitations there exist two regions with negative values
of dielectric permeability 
$\sigma_{ii}^{O}$ in fig.~\ref{topological-current}a.
One of these regions is stipulated by high value of chemical potential
$\mu$ due to doping, the second one is the region of plasma oscillations owing to the presence of Majorana modes.
%
Value of the optical conductivity 
$\sigma ^{opt}={e^2\over 4\hbar }$ for SM-model with Majorana model 
and the massless Dirac fermion model coincide in  the visible optical range and are equal to
$\sigma ^{opt}={e^2\over 4\hbar }$ in fig.~\ref{topological-current}a.

Since the precession 
of p$_z$-orbitals   holds in fig.~\ref{Vortices-band-structure}e,
the  total Ohmic massless 
current $\vec J=\vec j_x^O+ \vec j_y^O$  precesses
leading to emergence of  a magnetic field 
$\vec B_j$ of non-equilibrium spin 
$\vec S$ in fig.~\ref{topological-current}b in the absence of disordered influence of substrates on
$\vec S$. As a result of the non-equilibrium 
spin Hall effect (NSHE),
the currents  $J_{2(1)}^{spin-orbit}\sim  B_j^{1(2)}\sigma^{spin-orbit}_{12(21)} $,
$\vec B_j^{1(2)}\equiv \vec B_j^{x(y)}$  arise in the absence of an
external magnetic field 
$\vec B$. At the same time, 
a sum $\vec J_{1}^{spin-orbit} + \vec J_{2}^{spin-orbit}  $ is directed along
an external  magnetic field 
$\vec B_{\perp}$, which is orthogonal to the electric field applied to a sample
$\vec E$, as it is shown in fig.~\ref{topological-current}b, and, hence,  the non-equilibrium spin 
$\vec S$ is not revealed  in the ordinary Hall effect. 
The total current 
$\vec j^{skew} =\sum_{i=1}^2 \left(\vec j^O_{i}+\vec J_i^{spin-orbit}\right)$ is skewed,
since it flows at an angle
$\alpha$ to $\vec E$ in fig.~\ref{topological-current}b.

Thus, our model  qualitatively explains experimentally observed
skew topological currents in TIs \cite{Hsieh2009Science}
and aligned graphene/hBN superlattices \cite{Science346-2014Gorbachev} at zero magnetic field.

\subsection{Negative differential conductivity}

\begin{figure}[hbt]
\begin{center}
\hspace{-2cm}(a)\hspace{8cm} (b) \\
\hspace{-10cm}
\includegraphics[width=5.cm,height=4.cm,angle=0]{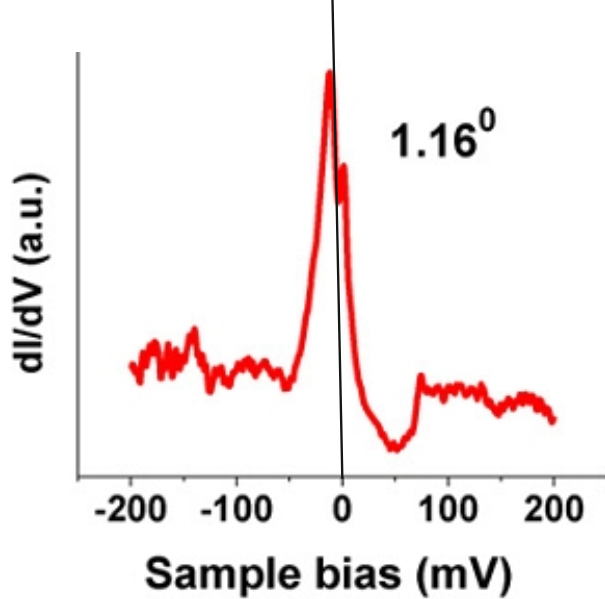}
 \vspace*{-3.5cm}\\
\hspace{-5.cm}
 \includegraphics[width=1.5cm,height=1.5cm,angle=0]{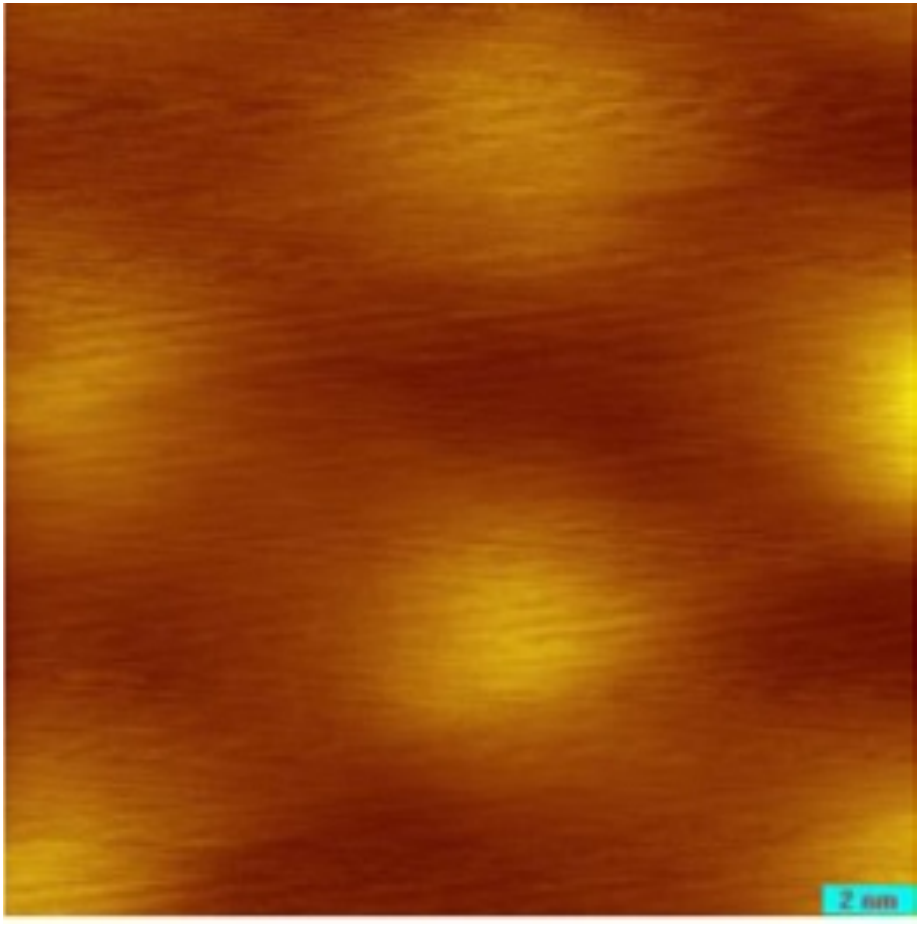}
\vspace*{-2.5cm}\\
\hspace{5cm}
\includegraphics[width=5.5cm,height=4.cm,angle=0]{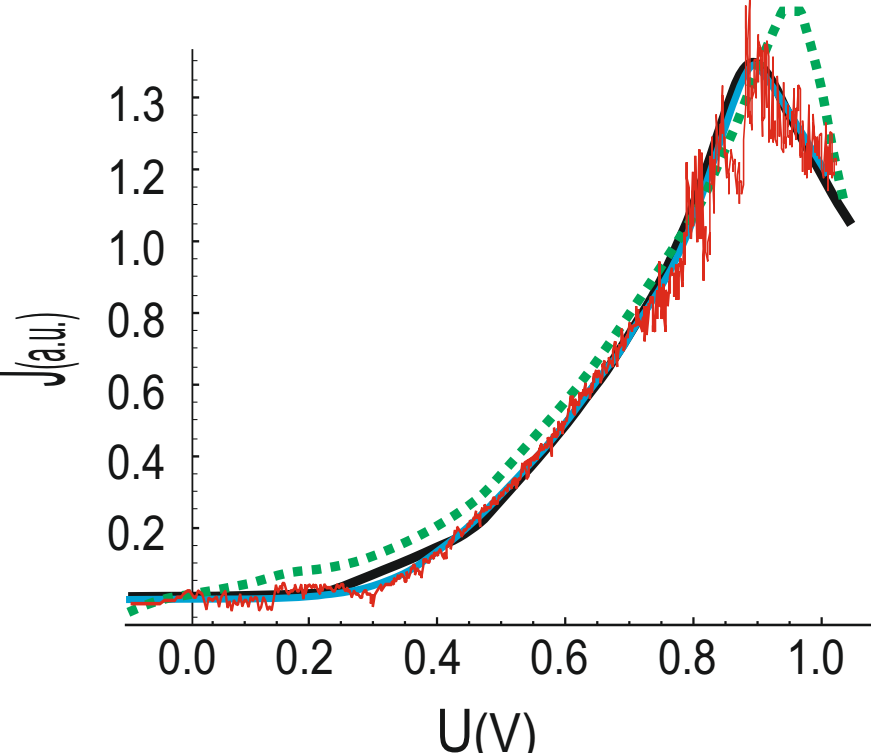}
\end{center}
\caption{ (a) Conductance $dI/dU$  as a function of voltage 
$U$ applied to 
twisted graphene placed on  graphite for a sample with  large Moir\'{e} pattern 
(inset, scale  2 nm) at $1.16^{\circ} $ misalignment angle
\cite{G_Li}. (b) Dependence of tunnel current $J$ on the bias voltage $U$ for heterostructure 
graphene/hBN/graphene: an experimental curve at $1.8^{\circ} $ misalignment angle
 (black solid curve), a theoretical one (green dashed curve) from
 \cite{Twist-controlled-resonant-tunnelling-in-graphene-boron-nitride-graphene2014},
 our theoretical simulation (red solid curve) of $J$ in the SM-model \eqref{variational-Majorana-bispinor}
 at temperature T=3K, chemical potential $\mu =135$~K with its fitting (blue solid curve).}
\label{twist-resonant-tunneling}
\end{figure}

Let graphene be disposed commensurately on the substrate,
for example, of hexagonal boron nitride (graphene/hBN/graphene) or
graphite  so that their hexagonal lattices practically coincide
(are rotated in respect to each other on a very small
misalignment angle
$\alpha_m < 2^{\circ}$)
 \cite{Zhi-Guo-Chen-ZhiwenShi,Pletikosic,Woods}.
In this case, the resonant influence of the substrate on graphene leads to appearance
of the interference  long period Moir\'e pattern on
STM- and AFM-images in fig.~\ref{twist-resonant-tunneling}a \cite{Andrei_Rep.Prog.Phys(2012),G_Li}. \
%
The heterostructures at $\alpha_m =0 $  are superlattices with center of inversion
when Dirac electron and hole cone of band
structures are degenerated. 
Accordingly, coincidence of 
Van Hove singularities of electron and hole  densities in hyperbolic Dirac points
leads to annihilation of electron-hole pairs, 
so that a
tunnel current is vanishing at 
small bias $U$. The last is observed as extrema 
of  conductance $dI/dU$ in fig.~\ref{twist-resonant-tunneling}a \cite{Andrei_Rep.Prog.Phys(2012),G_Li}.
$U$ is the electric
potential difference of the  field directed transversally to the surface of the heterostructure.
The polarization pair-production contribution $\sigma_{ii}^{Zitterbew}U$ and
the ohmic massless current $\sigma ^{O}_{ii}U 
$ define electron-hole contribution 
$I_{e-h}= (\sigma ^{O}_{ii} +\sigma_{ii}^{Zitterbew})U$ into the tunnel current 
$I_{e-h}$ through the alignment
graphene/hBN/graphene heterostructure that can  be negligibly small 
($I_{e-h}=0$) at small bias voltages $U 
$  due to recovering mirror symmetry.
Since the massless Ohmic and 
 the polarization contributions are  absent,  
the displacement current $J$ through the heterostructure is determined 
by the expression 
$J=\Re e\ \sigma_{ii}^{add} (U) U$ only. The bias current 
$J$ increases the heterostructure energy 
 on a value 
$\hbar\omega= C U 
^2/2 + A_0$, where $C$ is the electric capacitance 
of the heterostructure and $A_0$ is a constant. Then  the dependence 
$J=\Re e\ \sigma_{ii}^{add} (\omega) U 
$ in fig.~\ref{twist-resonant-tunneling}b can be found
with the condition 
$U\sim \sqrt{\hbar \omega}$. 
The current increase in  volt-ampere characteristics in fig.~\ref{twist-resonant-tunneling}b
is followed to its decrease in some range  of values of $U$ that is known as  
 a phenomenon of negative differential conductivity. 
Our theoretically predicted 
dependence $ J $ on $U$ 
is completely consistent with  the experimental data represented in
\cite{Twist-controlled-resonant-tunnelling-in-graphene-boron-nitride-graphene2014},
as fig.~\ref{twist-resonant-tunneling}b demonstrates.

\subsection{Longitudinal conductivity}
\begin{figure}[htbp]
(a) \hspace{8cm} (b) \\
\includegraphics[width=7cm,height=4cm]{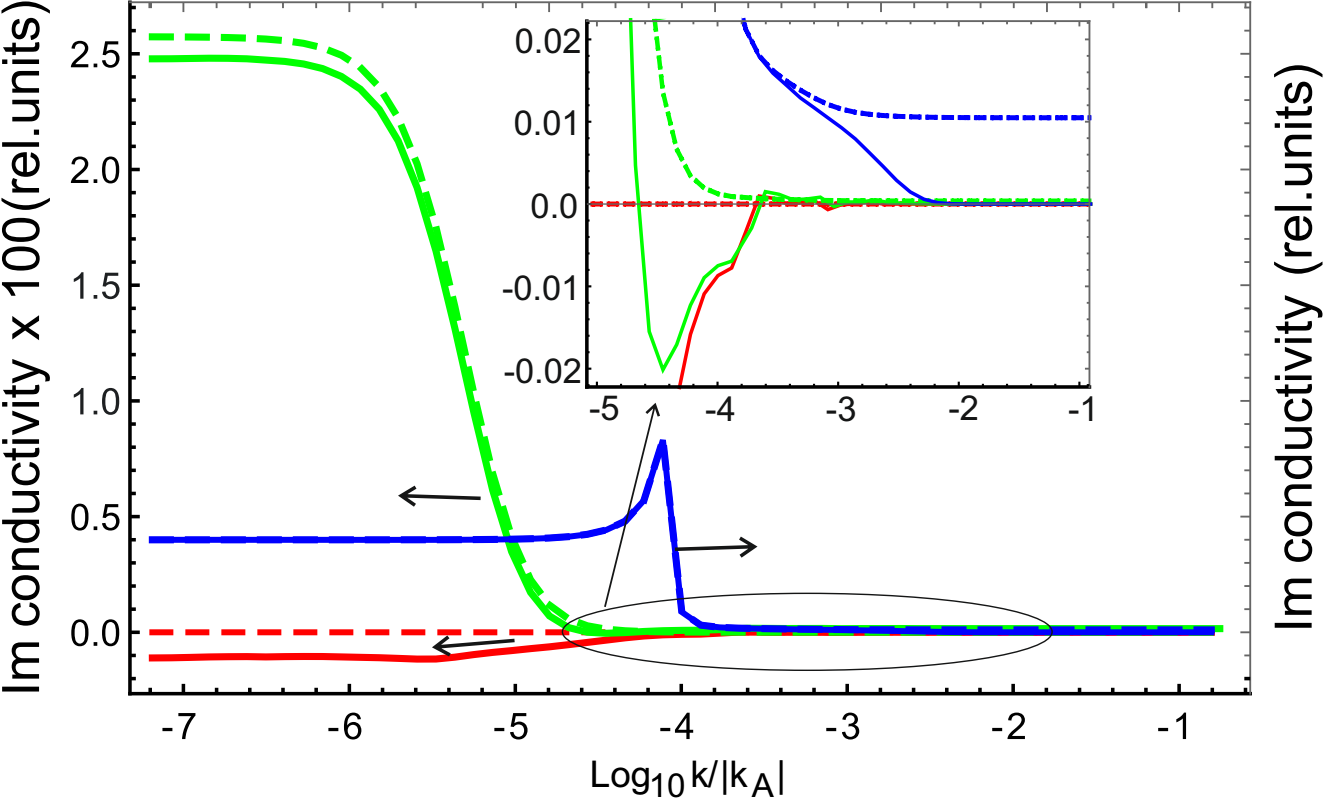} \hspace{0.5cm}
\includegraphics[width=7cm,height=4cm]{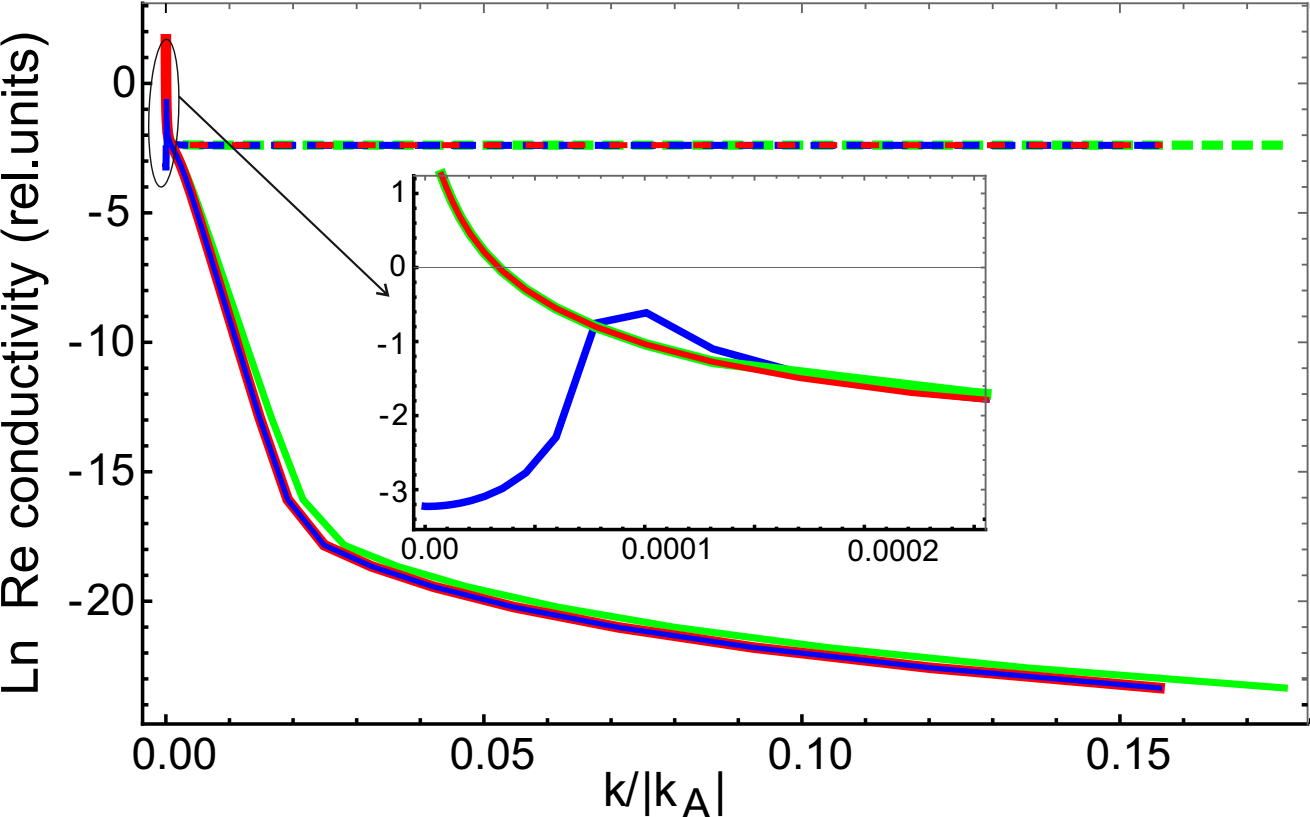}
\caption{Imaginary (a) and real (b) parts of longitudinal massless ohmic contribution 
$\sigma_{xx(yy)}^{O}$ to  conductivity vs  wave number $k$,
$\vec k = \vec p - \vec K_{A(B)}$
for our SM-model $N = 3$ \eqref{variational-Majorana-bispinor}
 (solid curves) and for the massless Dirac fermion model
 \cite{Falkovsky} (dashed curves),
at temperature 100 K and frequencies 13.3~K ($0.27
$~THz, blue color), $4\times 10^{-3}$~K ($83 
$~MHz, green color), $10^{-10}$~K ($2.08$~Hz, red color); chemical potential equals to 1~K. Inset to figure (a)
demonstrates the behaviour of $\Im m\ \sigma_{xx(yy)}^{O}$ in a neighborhood of small values;
 inset to figure (b)
demonstrates the behaviour of $\Re e\  \sigma_{xx(yy)}^{O}$ in
 the region of small  $k$ .   }
\label{dc-conductivity}
\end{figure}

For  attacking  the  so  called  "minimal  dc-conductivity"\ problem \cite{Ando2002,Ziegler}
it  is  important  to  obtain  the frequency dependence  of  longitudinal
dc-conductivity  for
frequencies  $\omega\to 0$ and  non-vanishing  wave  vectors $\vec k=\vec p/\hbar - \vec K_{A(B)}$.
The longitudinal   conductivity $\sigma_L(\omega,{\vec k})$ is defined through the conductivity tensor splitting into
longitudinal and transversal terms as \cite{Kraft-Ropke}
\begin{equation}
 \sigma_{ij}(\omega,{\vec k})=
 \left(\delta_{ij}-\frac{k_i k_j}{{\vec k}^2} \right)\sigma_{T}(\omega,{\vec k})+
\frac{k_i k_j}{{\vec k}^2}\sigma_{L}(\omega,{\vec k}) .
\end{equation}
Let us consider an influence of spatial dispersion on longitudinal
conductivity at low frequencies 
$\omega=10^{-10}$, $4\times 10^{-3}$, 13.3~K for graphene models $N=2,\ 3$ in fig.~\ref{color-models}.
According 
to the numerical results in fig.~\ref{dc-conductivity}, the dielectric permeability (imaginary part of complex conductivity
$\Im  m \ \sigma ^O_{ii}$ ) of non-doped graphene in the massless Dirac fermion model $N=2$ with spatial dispersion
is  positive at $\omega=4\times 10^{-3}$ and 13.3~K, but takes zero value 
at $\omega=10^{-10}$~K. In the model 
$N=3$ with spatial dispersion, the dielectric permeability
can gain zero and negative values  at all these frequencies.
Regions with zero and negative values of the dielectric permeability
are regions of plasmonic oscillations where the  complex frequencies
$\omega(k_{pl})+ i \gamma(k_{pl})$ satisfy the following equation  
\cite{Platzmann-Wolf1973,Mikhaylovski1977}:
\begin{equation}
\Im m \ \sigma_{ii}^{O}(k,z) - i \Re e \ \sigma_{ii}^{O}(k,z) = 0, \ z=\omega +i \gamma .
\label{plasmon}
\end{equation}
Expanding 
eq.~\eqref{plasmon} into series in terms of powers of $i\gamma$ in the vicinity of plasmonic resonance 
$\omega_{pl}$, $\Im m \ \sigma_{ii}^{O}(k_{pl},\omega(k_{pl}))=0$ gives the dumping constant
for the plasmons 
\begin{equation}
\gamma(k_{pl})=\left.{\Re e \ \sigma_{ii}^{O}(\omega(k)) \over
\frac{\partial \Im m \ \sigma_{ii}^{O}(\omega(k))}{\partial k}
\left(\frac{\partial \omega(k)}{\partial k}
\right)^{-1}}\right|_{k=k_{pl}} . \label{plasmon1}
\end{equation}
According 
to simulation presented in fig.~\ref{dc-conductivity}a,
$\Im m \ \sigma_{ii}^{O}(k,\omega)$ at $\omega =10^{-10}$~K is 
practically constant function 
in the model $N=2$, and, hence, the expression \eqref{plasmon1} diverges.
Therefore, contrary to 
the model $N=3$, the plasmon damping 
occurs instantly
in the model $N=2$, $\gamma^{N=2}\to \infty$. Long living plasmons 
in  the SM-model $N=3$  are able to provide 
screening in the electrophysical range.
The massless Dirac fermion model with instantly damped plasmon oscillations
does not allow to describe screening of an external electric field. 

Let us calculate 
the dc-conductivity $\sigma^{dc}$. We perform the inverse Fourier transformation
as
\begin{eqnarray}
\sigma^{dc}={1\over (2\pi)^2} \int {\sigma^O _{\|}}(\omega ,k) e^{i \vec k\cdot \vec r} d^2k
.
\label{dc-conductivity-estimation-0}
\end{eqnarray}
According to 
fig.~\ref{dc-conductivity}b, the Fourier image 
${\sigma^O _{\|}}(\omega ,k)$ of longitudinal conductivity 
at $\omega=10^{-10}$~K for the graphene model 
$N=3$  behaves itself approximately as 
the Dirac $\delta$-function, that is  equal zero everywhere except of
$k\to 0$, where $\sigma^O _{\|}(\omega ,0)=5.49e^2/\hbar$. 
Hence, the dc-conductivity \eqref{dc-conductivity-estimation-0} in the model $N=3$
gains non-zero value 
\begin{eqnarray}
\sigma^{dc,N=3}(\omega)={1\over (2\pi)^2} \int \sigma^O _{\|, N=3 }(\omega ,k) e^{i \vec k\cdot \vec r} d^2k
\approx {1\over 2\pi} \int \sigma^O _{\|N=3}(\omega ,0) \delta (|k|)e^{i \vec k\cdot \vec r} d k={5.49e^2\over h},
\ \omega=10^{-10}~K \label{dc-conductivity-estimation}
\end{eqnarray}
at temperature 100~K and chemical potential 1~K.
Contrary to this, the Fourier image 
of the dc-conductivity
in the graphene model $N=2$ takes a constant value
at large wave numbers $k$, hence, integrand in inverse Fourier transformation of
$\sigma^{dc,N=2}(\omega)$ in \eqref{dc-conductivity-estimation-0} is a highly oscillating function 
leading to zero value 
of the minimal dc-conductivity $\sigma^{dc,N=2}(\omega)$,  $\sigma^{dc,N=2}(\omega)=0$.

The minimal dc-conductivity of graphene
in devices with large area of graphene monolayer on SiO$_2$ turns out to be
$\sigma_{min} \sim 4 e^2/h$ \cite{Novoselov2004} at low temperatures ($ \sim 1.5 $~K).
The
minimal dc-conductivity of suspended graphene \cite{Bolotin,Du} and of graphene on boron nitride substrate
\cite{Dean} is $\sigma_{min}\sim 6 e^2/h$ at $T\sim 300 $~K
. Thus, our estimate
\eqref{dc-conductivity-estimation} is in a perfect agreement with experimental data.

\subsection{Chiral anomaly,  longitudinal magneto-conductivity, and splitting zero-bias conductance peaks }

\begin{figure}[hbt]
\begin{center}
\hspace{0cm}(a)\hspace{7cm} (b) \hspace{7cm} (c)\\
\includegraphics[width=5.cm,height=4.cm,angle=0]{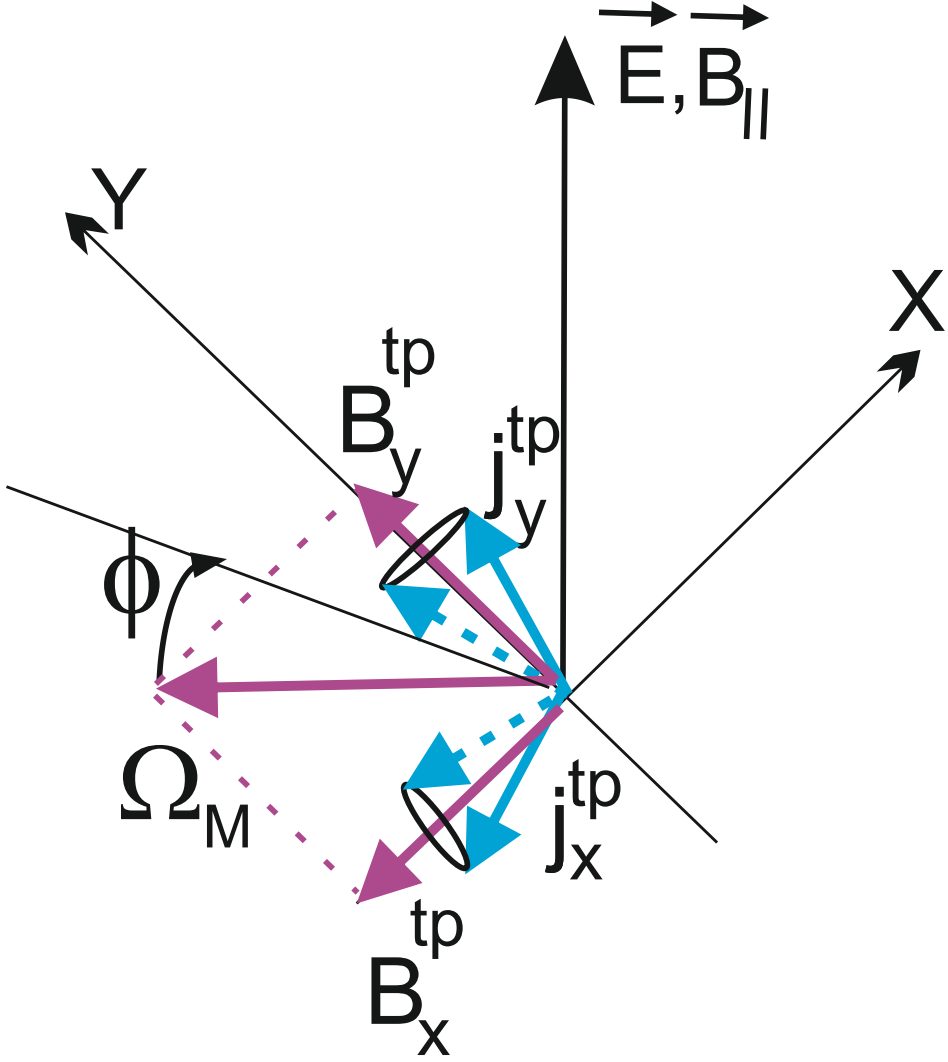}
 \includegraphics[width=5.cm,height=4.cm,angle=0]{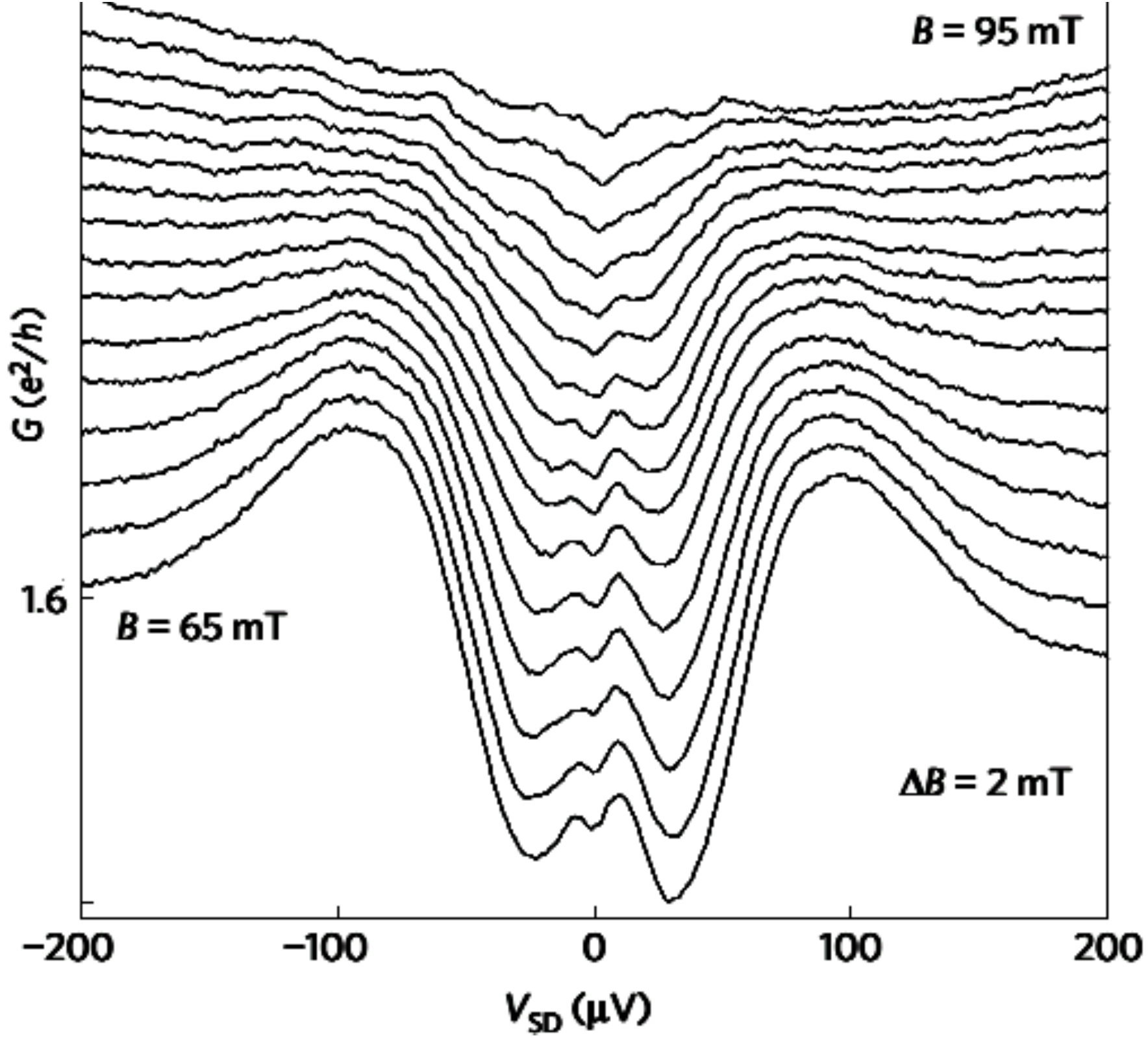}
 \includegraphics[width=5.cm,height=4.cm,angle=0]{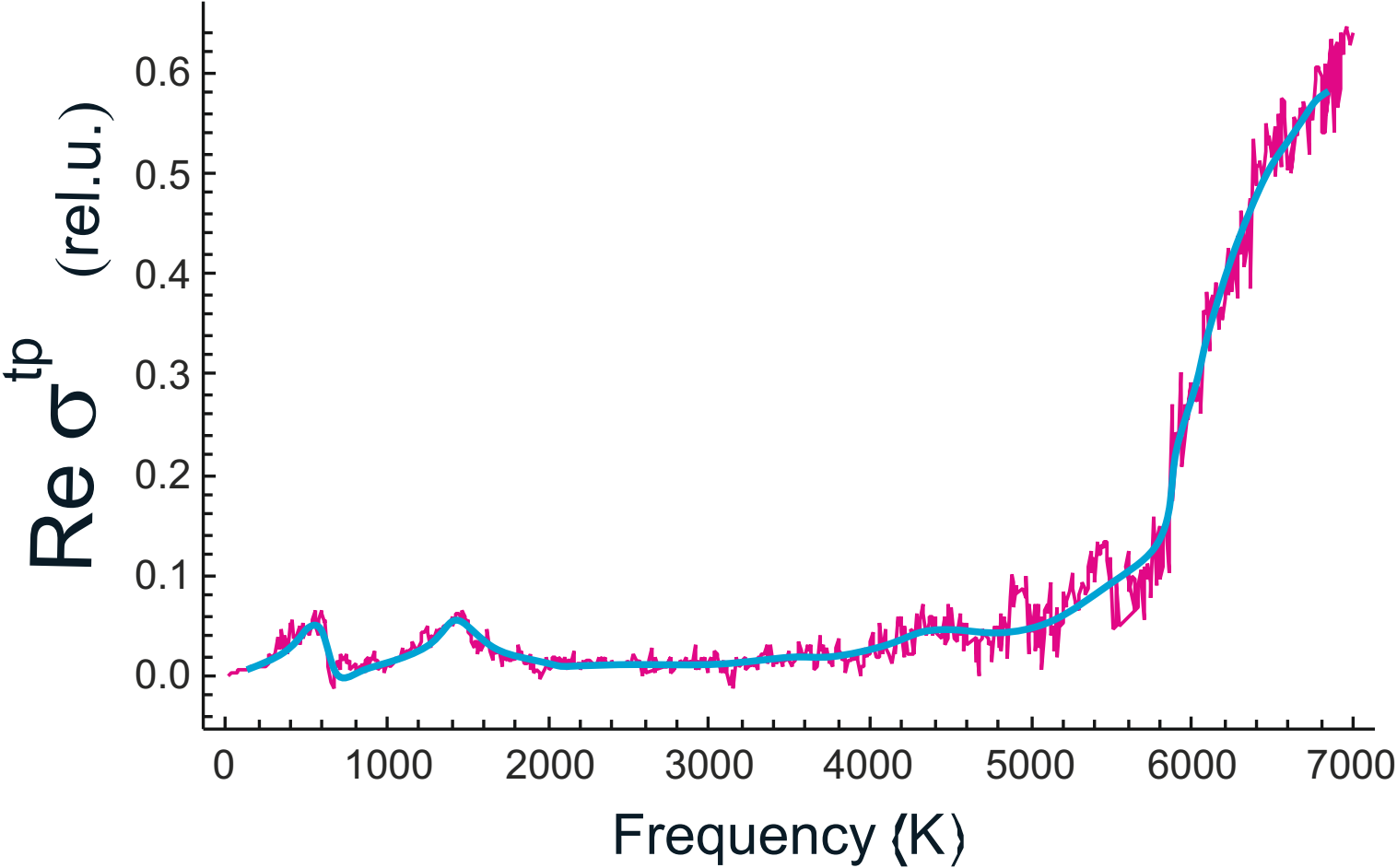}
\end{center}
\caption{ (a) Sketch of splitting zero-bias conductance peak:
$
{\vec B}^{tp}_{x(y)}$ is a magnetic field of precessing spins
of $\pi$(p$_z$)-orbitals,
$\phi$ is a turn angle
of the Majorana curvature $\vec \Omega_M$ in a magnetic field $\vec B_{\|}$,
parallel to
electric field $\vec E$.
(b) Experimental low-bias conductance  as a function of applied magnetic field parallel to  axis
of an induced one-dimensional topological superconductor
(a system composed of an aluminium superconductor in proximity to an indium arsenide nanowire)
at a higher chemical potential, from
\cite{Das2012Nat-Phys}. (c) Our theoretical simulation of SZBP by
the semimetal-model Hamiltonian \eqref{variational-Majorana-bispinor} as two low-frequency peaks in the
frequency dependence of real part of topological conductivity
$\sigma_{xx}^{tp}$
(magenta solid curve)  and its fitting (blue solid curve)
in rel.units $e^2/\hbar$ at temperature T=3K and
chemical potential $\mu = 135$~K.}
\label{SZBP-our-simulation}
\end{figure}

As it is sketched
in fig.~\ref{Vortices-band-structure}e, \ p$_z$-orbitals precess.
Accordingly, the topological currents  $j^{tp}_{x(y)}\propto \sigma_{xx(yy))}^{tp}$ precess 
as well, creating magnetic fields $\vec B_{x(y)}^{tp}$ in fig.~\ref{SZBP-our-simulation}a.
Let us denote the resulting magnetic field  
$\vec B_{x}^{tp} + \vec B_{y}^{tp}$ through $\vec \Omega _M$.

Generally, external electromagnetic fields 
$\left\{\vec A, \Phi\right\}$ reorient randomly the directions of magnetic fields
$B^{tp}_{x(y)}$,   disordering the vortex 
SM-lattice 
and, as a result, breaking topological currents 
$j^{tp}_{x(y)}\propto \sigma_{xx(yy))}^{tp}$. Let us place 
SM on superconducting substrate 
S. Then, nearly located and having equal topological charges vortexes
of SM and S repel each other. 
As a result, the vortex regions characterized by the definite sign 
$\pm$ of topological charge
appear 
in SM. 
If vortex lattices 
of SM and S are consistent, 
topological components 
$j^{tp}_{x(y)}$ of the current linked with 
$\vec B_{x(y)}^{tp} $ are rotated in an external magnetic field 
$\vec B_{\|}$ at conditions that $\vec B_{\|} \perp \vec \Omega _M$ and the electric field 
$\vec E$ and $\vec B_{\|}$ are parallel: 
$\vec E\| \vec B_{\|}$. Let us denote the distribution of the magnetic
fields inside a sample 
through $\left.\{\vec B_{i,SM}\}
\right|_{i=1}^N$, $N$ is a number of hexagons. 
$\vec \Omega _M$ is aligned along the direction of the effective magnetic field 
$\vec B_{eff}=\vec B_{\|}+ \sum_{i=1}^N \vec B_{i,SM} = \vec B_{\|}+\vec B_{SM}$
and hence,
is directed at an angle 
$\alpha $ to the electric field 
$\vec E$.
Meanwhile, the topological current 
$\vec J_{ZBP}$ (ZBP), called as 
a zero-bias peak,  emerges in the direction 
$\vec E$  as
\begin{equation}
 J_{ZBP} \propto \Omega _M \cos \alpha \label{JZBP}.
\end{equation}
Scheme of this phenomenon is demonstrated in 
fig.~\ref{SZBP-our-simulation}a.

If the magnetic field 
$\sum_{i=1}^N \vec B_{i,SM} $ is small compared with the magnitude
of the applied field 
$\vec B_{\|}$: $\vec B_{SM}\ll \vec B_{\|}$,
then  $\vec \Omega _M = \vec \Omega _{1,M} N$ trends to align along 
$\vec B_{\|}$ in the absence of 
the electric поля $\vec E$. Accordingly, the value 
$ J_{ZBP}$ trends to 
the quantum limit 
$\sum_{i=1}^N (\vec j^{tp}_{i,x}+ \vec j^{tp}_{i,y})$. 
The quantity 
$ \left|\vec j^{tp}_{1,x}+ \vec j^{tp}_{1,y}\right|$ is proportional to
the Majorana conductivity 
\cite{Goudarzi2017PhysicaE,Peng-Falko2015PRL}.

Now, let us consider the case 
$\sum_{i=1}^N \vec B_{i,SM} \gg \vec B_{\|}$. Then the 
following approximation for $\vec \Omega _M$ holds
\begin{equation}
\vec \Omega _M \approx \vec \Omega _{1,M}n_1 +  \Omega _{1,M}n_2 \vec B_{SM}/|B_{SM}| , \ \
n_1 \propto B_{\|},\ n_1+n_2=N.
\end{equation}
Here $\vec \Omega _{1,M}=\sum_{i=1}^2\vec B^{tp}_{1,i}$, $B^{tp}_{1,x(y)}$ are magnetic fields, 
generated by the topological currents
$j^{tp}_{1,x(y)}$ in one hexagon.
Then,
the current
$\vec J_{ZBP}$ \eqref{JZBP} is approximated by the following expression:
\begin{equation}
 J_{ZBP} \propto \Omega _{1,M} ( \cos \alpha ) B_{\|} \label{JZBP1}.
\end{equation}
Contrary to the ordinary Hall effect, the contributions 
$\sigma_{xx}^{tp}$ and $\sigma_{yy}^{tp}$
reveal themselves in the magnetic field 
$\vec B_{\|}$, directed along the electric field 
$\vec E$, when 
the current  orthogonal to $\vec E$, 
$\vec j^{tp}_{x}+ \vec j^{tp}_{y}$, $j^{tp}_{i}\propto \sigma_{ii}^{tp}$
is rotated in the field 
$\vec B_{\|}$  due to alignment of the spin of precessing 
$\pi($p$_z)$-orbital at the angle 
$\alpha$ to 
$\vec B_{\|}$, as it is shown in 
fig.~\ref{SZBP-our-simulation}a.
Hence, the longitudinal magneto-conductivity is observed.
Berry curvature 
$\vec \Omega_k$ leads to the change of the velocity 
of the charged particle from 
$\vec v$ in the field 
$\vec E$ to 
$\vec {\dot r}$ in the field 
$\vec B_{\|}$ and $\vec E$, $\vec E\, \| \, \vec B_{\|}$ as \cite{Lu-Shen2017Front-Phys,Yip2015Preprint}
\begin{equation}
\vec {\dot r} = \vec v +{e\over \hbar}\left( \vec \Omega_k \cdot \vec v\right)\vec B_{\|}, \ \vec v\| \vec E.
\label{JZBP2}
\end{equation}
Right hand sides 
of \eqref{JZBP1} and \eqref{JZBP2} are similar, 
but $\vec \Omega_M$ is a curvature of our Majorana model system.
Eq.~\eqref{JZBP1} describes 
negative magnetoresistance (NMR) 
that  represents the phenomenon of chiral anomaly 
at weak  magnetic fields parallel to electric ones \cite{H-ZhLuSh-QShen2017,Niemann2017}.

Bounding  the vortex lattice 
of SM to the vortex lattice of the S-substrate 
NMR manifests itself as 
a splitting zero-bias conductance peak (SZBP) in fig.~\ref{SZBP-our-simulation}b
at higher values of  chemical potential. The real part of the frequency dependence 
of $\sigma_{xx(yy)}^{tp}$ for higher chemical potential is calculated  based on
the formula 
\eqref{conduction2-topology-mass-correction-1}   and is represented 
in fig.~\ref{SZBP-our-simulation}c.  The  two low-frequency peaks similar 
to  SZBP  with the height of about 
$\sim 0.24 \sigma ^{opt}\sim 0.3768 e^2/h$
are observed in fig.~\ref{SZBP-our-simulation}c. For one-dimensional 
SM possessing strong SOC, the height of 
ZBP would be four times smaller 
$\sim 0.0942 e^2/h$. The last coincides with experimentally measured 
in \cite{Das2012Nat-Phys} ZBP ($\approx 0.1 e^2/h$) for such 1D topological
superconductor as  an indium arsenide nanowire on an aluminium superconductor substrate.
The effective magnetic field 
$\vec B_{eff}$ in a sample is determined 
through  the competition between vortex-vortex repulsion and Lorentz force in an external magnetic field. 
Therefore, SZBP disappears in  strong external magnetic fields 
$\vec B$, as it is shown in fig.~\ref{SZBP-our-simulation}b.

Magnetic-field-induced ZBP and its splitting, as
expected for  zero-energy Majorana state,  were  observed in
\cite{Mourik2012Science,Deng2012Preprint}.
Magnetic pinning of vortices in S/1D-ferromagnetic heterostructures
\cite{DiGiorgio2016ScientificReports}
\cite{Oppen-Peng-Pientka-Oxford2017}
, hybrid S/ferromagnetic TI \cite{Koren2011PhysRevB}
, and graphene/S
\cite{Dirks2011NaturePhysics}
structures  is called
a proximity effect
\cite{Buzdin2005RevModPhys,LiangFu-Kane2008PRL}.
Formation of superconductor-vortex clusters on topological defects
in  1D-ferromagnetic materials is possible
\cite{DiGiorgio2016ScientificReports}.
Magnetic anisotropy
of all these heterostructures manifests itself in an external magnetic
field in the form
of
quantized zero-bias conductance peak
\cite{Goudarzi2017PhysicaE,Peng-Falko2015PRL,Das2012Nat-Phys,Yang2011}.


\section{Discussion and Conclusion}
%
Finalizing our finding, a ${\mathbb{Z}}_2$ topological semimetal  model with a number of internal degrees
of freedom (flavors) $ N = 3 $ has been proposed, with three-body  excitations as charge carriers.
A novel Majorana-like quantum-field approach allows to calculate  low-frequency
conductivity with accounting of the polarization and magnetoelectric effects.
Using this approach, 
braiding Majorana 
particles have been found
and interpreted as dimer configurations 
in the RVB-picture. The developed 
non-abelian quantum statistics predicts  
 three flavor pairs Majorana resonance--antiresonance.

Dichroism 
of Dirac bands for the ${\mathbb{Z}}_2$ SM-model 
is provided by the vortex structure 
of the Dirac bands due to existence 
of zero-energy Majorana modes and chiral braiding Majorana modes.  It has been shown that 
the deconfinement of the Majorana-like modes leads to the appearance of 
a set of gapless Fermi arcs.

Chiral-anomaly revealed in charge transport in the form of 
zero-bias conductance peak and its splitting  is a specific feature 
of the SM-model $N=3$ with skew topological currents  in  zero magnetic fields.
This splitting peak is stipulated by non-zero curvature of the
Majorana model. 

In the model $N=3$ there exists a mechanism for
dynamical reduction of spatial dispersion of
states. This dynamic reduction provides a non-zero value of  minimal dc-conductivity
of graphene.  Plasmonic oscillations 
in the massless Dirac fermion model occur in electrophysic frequency range 
$\omega \lesssim 2$~Hz, but practically dump instantly. 
Contrary to this, the plasmonic oscillations 
in the Majorana-like massless fermion model have a finite dumping rate and exist both in
optical and electrophysic frequency ranges. 

Significant advantage 
of  the proposed Majorana-like 
fermion theory augment by a mixing mass term 
over the massless Dirac fermion  and Weyl SM-models consists in the achieved
qualitative and quantitative consistency  with experimentally observable properties 
of topological semimetals.
The ${\mathbb{Z}}_2$ topological SM-model with the flavors number  $N = 3$
 can be considered as an effective tools to
discover and investigate new Dirac materials and to develop
new devices for quantum computing.


{\bf Acknowledgments.}
This work has been supported in part by the State Scientific Program of
Fundamental Researches "Convergence-2020" of Belarus.

\newpage
\renewcommand{\theequation}{\Alph{section}.\arabic{equation}}
\setcounter{equation}{0}
\setcounter{section}{19}

\section*{\huge Supplementary Information }


\setcounter{section}{0}
\setcounter{figure}{0}
\renewcommand{\thefigure}{\Alph{section}.\arabic{figure}}
\section{Band-structure simulation details}
\setcounter{section}{19}

Bispinor wave functions of quasiparticles  can be represented through  free Dirac fields
of $\pi$~(p$_{z}$)-electrons:
\begin{eqnarray}
\left(
\begin{array}{c}
\widehat {\chi ^{\dagger}_{-\sigma_{_{A(B)}} }} (\vec r)
\left|0,-\sigma \right>\\
\widehat {\chi ^{\dagger}_{\sigma_{_{B(A)}} }} (\vec r)
\left|0,\sigma \right>
\end{array}
\right) ={e^{-\imath (\vec K_{A(B)} -\vec q_{_{A(B)}})\cdot \vec r}
\over \sqrt{2}}
\left(
\begin{array}{c}
\exp\{- \imath \theta_{k_{A(B)}}\}\phi_1\\
\exp\{-\imath \theta_{k_{A(B)}}\}\phi_2\\
-\exp\{\imath \theta_{k_{B(A)}}\}\phi_2\\
\exp\{\imath \theta_{k_{B(A)}}\}\phi_1
\end{array}
\right)
\equiv \left(
\begin{array}{c}
\chi_{-\sigma_{A(B)}}\\
\chi_{+\sigma_{B(A)}}
\end{array}
\right),\ \ \label{pi-electronA_down}
\end{eqnarray}
\begin{eqnarray}
\left(
\begin{array}{c}
\widehat {\chi ^{\dagger}_{+\sigma_{_{A(B)}} }} (\vec r)
\left|0,\sigma \right>\\
\widehat {\chi ^{\dagger}_{-\sigma_{_{B(A)}} }} (\vec r)
\left|0,-\sigma \right>
\end{array}
\right) ={e^{-\imath (\vec K_{A(B)} -\vec q_{_{A(B)}})\cdot \vec r}
\over \sqrt{2}} 
\left(
\begin{array}{c}
-\exp\{\imath \theta_{k_{A(B)}}\}\phi_2\\
\exp\{\imath \theta_{k_{A(B)}}\}\phi_1\\
\exp\{- \imath \theta_{k_{B(A)}}\}\phi_1\\
\exp\{- \imath \theta_{k_{B(A)}}\}\phi_2
\end{array}
\right)
\equiv \left(
\begin{array}{c}
\chi_{+\sigma_{A(B)}}\\
\chi_{-\sigma_{B(A)}}
\end{array}
\right) \label{pi-electronA_up}
\end{eqnarray}
\noindent where $ \theta_{k_{B(A)}}$ are phases for free fermion fields \cite{Peskin-Schroeder},
\begin{eqnarray}
\phi_i=
{1\over (2\pi)^{3/2}\sqrt{N/2}}\sum_{\vec R^{A(B)}_{l}}
\exp\{\imath [\vec K_{A(B)} - \vec q_{_{A(B)}} ]
[\vec R^{A(B)}_{l}-\vec r]\}
\psi_{\{n_i\}}(\vec r -\vec R^{A(B)}_{l})\ \ \
\ \ \ \ \ \label{Bloch-function-reduc}
\end{eqnarray}
is a Bloch function; maximally-localized Wannier functions
$\psi_{\{n_i\}}$, $i=1,2$ are defined as
\begin{equation}
\psi_{\{n_2\}}=c_1\psi_{\mbox{\small p}_z}(\vec r\pm \vec \delta_i)+
c_2\psi_{\mbox{\small p}_z}(\vec r),\ \sum_{i=1}^2 c_i^2=1;\\
\psi_{\{n_1\}}= \psi_{\mbox{\small p}_z}(\vec r);
\end{equation}
$\vec \delta_i$ is a distance to nearest 
$i-$th site $\vec  \delta_i$, $i=1,2,3$; definition of wave vector $\vec q_{_{A(B)}}$ uses the reference frame origin located in
the Dirac point $K_{A(B)}$, $\psi_{\mbox{\small p}_z}(\vec r)$ is
the p$_z$-wave function of a hydrogen-like atom with an "effective charge" \ 1
centered at the graphene lattice sites.

An exchange interaction term $\left(\Sigma_{rel}^{x}\right)_{AB}$ is determined as
\cite{myNPCS2013,myNPCS17-2014,NPCS18-2015GrushevskayaKrylovGaisyonokSerow}
\begin{eqnarray}
&\Sigma_{rel}^{x}\left(
\begin{array}{c}
\widehat {\chi } ^{\dagger}_{_{-\sigma_{_A}} }(\vec r) \\
\widehat {\chi }^\dagger _{\sigma_{_B}}(\vec r)
\end{array}
\right)\left|0,-\sigma \right> \left|0,\sigma \right>
=
 \left(
\begin{array}{cc}
0& \left( \Sigma_{rel}^{x}\right)_{AB}
\\
\left( \Sigma_{rel}^{x}\right)_{BA} & 0
\end{array}
\right)
\left(
\begin{array}{c}
\widehat {\chi }^{\dagger}_{-\sigma_{_A} } (\vec r) \\
\widehat {\chi} ^\dagger _{\sigma_{_B}}(\vec r)
\end{array}
\right)\left|0,-\sigma \right> \left|0,\sigma \right> \label{exchange}
, \\[3mm]
&\left( \Sigma_{rel}^{x}\right)_{AB}
\widehat {\chi }^\dagger _{\sigma_{_B}}(\vec r)\left|0,\sigma \right>
=
\sum_{i=1}^{N_v\,N}\int { d \vec r_i}
\widehat {\chi }^\dagger _{\sigma_i{^B}}(\vec r)\left|0,\sigma \right>
\langle 0,-\sigma_i|{\widehat \chi}^\dag_{-\sigma_i^A} (\vec r_i)
V(\vec r_i -\vec r)
{\widehat \chi}_{-\sigma_B}(\vec r_i)|0,-\sigma_{i'}\rangle ,
\label{Sigma-AB}
\\[3mm]
& \left( \Sigma_{rel}^{x}\right)_{BA}
\widehat {\chi }^{\dagger}_{_{-\sigma_{_A}} } (\vec r)
\left|0,-\sigma \right>
=\sum_{i'=1}^{N_v\,N}\int { d \vec r_{i'}}
\widehat {\chi }^{\dagger}_{_{-\sigma_{i'}^A} } (\vec r)
\left|0,-\sigma \right>
\langle 0,\sigma_{i'}|{\widehat \chi}^\dag_{\sigma_{i'}^B} (\vec r_{i'})
V(\vec r_{i'} -\vec r)
{\widehat \chi}_{_{\sigma_A}}(\vec r_{i'})|0,\sigma_i\rangle.
\label{Sigma-BA}
\end{eqnarray}
In the approximation of free 
$\pi$-electrons, all wave functions 
entering the expression (\ref{Sigma-AB}, \ref{Sigma-BA}) are
\begin{equation}
\begin{split}
\widehat {\chi} ^{\dagger}_{-\sigma_{i'}{^A} } (\vec r)
\equiv \widehat {\chi ^{\dagger}_{-\sigma_{_A} }} (\vec r)
,\ \
\widehat {\chi }^\dagger _{\sigma_i{^B}}(\vec r) \equiv
\widehat {\chi ^\dagger _{\sigma_{_B}}}(\vec r) \mbox{ for }
\forall \ i,\ i'.
\end{split}
\label{avarage-bispinor1}
\end{equation}
Then, in this approximation we can write matrices 
$\left( \Sigma_{rel}^{x}\right)_{AB}\approx \Sigma_{AB}$ and $ \left( \Sigma_{rel}^{x}\right)_{BA} \approx \Sigma_{BA}$
without self-action as 
\begin{eqnarray}
&\left( \Sigma_{rel}^{x}\right)_{AB}
\widehat {\chi ^{\dagger}_{+\sigma_{_B} }} (\vec r)
\left|0,\sigma \right> \approx   \Sigma_{AB}\ \chi_{\sigma_B}
=
\sum_{i=1}^{N_v N-1} \int d\vec r_i
\nonumber \\
&\times  V(\vec r_i -\vec r)
\left[
\chi_{-\sigma_A} (\vec r_i)\cdot
\chi_{-\sigma_B}^*( \vec r_i)
\right]\chi_{+\sigma_B} (\vec r)
=
{1\over 2^{3/2}} \sum_{i=1}^{N_v N-1} \int d\vec r_i V(\vec r_i -\vec r)
 \nonumber \\
& \times
\left[
\begin{array}{cc}
e^{-\imath  \theta_{K_A}} \phi_1 (\vec r_i)
e^{\imath  \theta_{K_B}} \phi^*_1(\vec r_i)
&
e^{-\imath  \theta_{K_A}} \phi_1 (\vec r_i)
e^{\imath  \theta_{K_B}} \phi^*_2 (\vec r_i)\\
e^{-\imath  \theta_{K_A}} \phi_2 (\vec r_i)
e^{\imath  \theta_{K_B}} \phi^*_1 (\vec r_i)&
e^{-\imath  \theta_{K_A}}\phi_2(\vec r_i)
e^{\imath  \theta_{K_B}} \phi^*_2 (\vec r_i)
\end{array}
\right]
\left[
\begin{array}{c}
- e^{-\imath [(\vec K_A -\vec q_A)\cdot \vec r - \theta_{K_B}]}
\phi_2 (\vec r) \\
e^{-\imath [(\vec K_A -\vec q_A)\cdot \vec r - \theta_{K_B}]}
\phi_1(\vec r)
\end{array}
\right], 
&\label{Sigma-AB1}
\end{eqnarray}
\begin{eqnarray}
&\left( \Sigma_{rel}^{x}\right)_{BA}
\widehat {\chi ^{\dagger}_{-\sigma_{_A} }} (\vec r)
\left|0,-\sigma \right> \approx  \Sigma_{BA}\ \chi_{-\sigma_A}
=
\sum_{i=1}^{N_v N-1} \int d\vec r_i  \nonumber \\
&\times V(\vec r_i -\vec r)
\left[
\chi_{+\sigma_B} (\vec r_i)\cdot
\chi_{+\sigma_A}^*( \vec r_i)
\right]\chi_{-\sigma_A} (\vec r)
=
{1\over 2^{3/2}}\sum_{i=1}^{N_v N-1} \int d\vec r_i V(\vec r_i -\vec r)
 \nonumber \\
 & \times
\left[
\begin{array}{cc}
-
e^{\imath  \theta_{K_B}} \phi_2 (\vec r_i)
(-1)
e^{-\imath  \theta_{K_A}}\phi^*_2(\vec r_i)
& -
e^{\imath  \theta_{K_B}}\phi_2 (\vec r_i)
e^{-\imath \theta_{K_A}} \phi^*_1(\vec r_i)\\
e^{\imath \theta_{K_B}}\phi_1 (\vec r_i)(-1)
e^{-\imath  \theta_{K_A}}\phi^*_2(\vec r_i)
&
e^{\imath  \theta_{K_B}} \phi_1(\vec r_i)
e^{-\imath  \theta_{K_A}} \phi^*_1(\vec r_i)
\end{array}
\right]
\left[
\begin{array}{c}
e^{-\imath [(\vec K_A -\vec q_A)\cdot \vec r+  \theta_{K_A}]}
\phi_1(\vec r) \\
e^{-\imath [(\vec K_A -\vec q_A)\cdot \vec r+  \theta_{K_A}]}
\phi_2(\vec r)
\end{array}
\right]. \label{Sigma-BA1}
\end{eqnarray}
It follows from the expressions (\ref{Sigma-AB1}, \ref{Sigma-BA1}) that 
the matrices 
$\Sigma_{AB}$ and $\Sigma_{BA}$ have the form 
\begin{eqnarray}
\Sigma_{AB}
 =
{1\over 2} e^{\{-\imath (\theta_{K_A}-\theta_{K_B})\}}
\sum_{i=1}^{N_v N -1} \int d\vec r_i V(\vec r_i -\vec r)
\left[
\begin{array}{cc}
\phi_1 (\vec r_i)\phi^*_1(\vec r_i) &
\phi_1 (\vec r_i) \phi^*_2 (\vec r_i)\\
\phi_2 (\vec r_i) \phi^*_1 (\vec r_i)&
\phi_2(\vec r_i) \phi^*_2 (\vec r_i)
\end{array}
\right]
, \label{Sigma-AB01}\\
\Sigma_{BA}
=
 {1\over 2} e^{\{-\imath (\theta_{K_A}-\theta_{K_B})\}}
\sum_{i=1}^{N_v N -1} \int d\vec r_i V(\vec r_i -\vec r)
\left[
\begin{array}{cc}
\phi_2 (\vec r_i)\phi^*_2(\vec r_i) &
- \phi_2 (\vec r_i) \phi^*_1(\vec r_i)\\
- \phi_1 (\vec r_i) \phi^*_2(\vec r_i)
&
\phi_1(\vec r_i) \phi^*_1(\vec r_i)
\end{array}
\right]
. \label{Sigma-BA01}
\end{eqnarray}
The quasirelativistic Dirac--Hartree--Fock exchange interaction (\ref{Sigma-AB01}, \ref{Sigma-BA01}) in
tight-binding approximation are determined by Eqs.~(\ref{Sigma-AB3}, \ref{Sigma-BA3}) of the main text.
In the direction perpendicular to the SM layer, the integration has
been limited to a distance equals two C--C bond lengths,
the phases have been chosen as $\theta_{K_A}=\theta_{K_B}$.


Simulations of the band structure have been performed
for two variants of approximation to the exchange operator. The first one
already mentioned above is the series expansion of the exchange matrices on deviation
of the wave vector from the Dirac point up to
4th order in the length of wave vector $\vec q$. As it has been shown in
\cite{myJModPhys2014,myNATO2015,NPCS18-2015GrushevskayaKrylovGaisyonokSerow,myNPCS18-2015},
this approximation leads to small imaginary part for the
energy (spectral line width), this in fact not very bad as it effectively corresponds to a finite decay.
The approximation 
of the zero gauge-phases is valid 
in the Dirac point.

The second approximation is the use of the exchange interaction matrices calculated based on
$\pi$(p$_z$)-orbital wave functions with full exponents and with non-zero
gauge-phases (see the detail of the approximation in
\cite{NPCS18-2015GrushevskayaKrylovGaisyonokSerow,myNPCS18-2015,myIntJModPhys2016,mySymmetry2016}
). The last approximation holds real eigenenergies for all wave vectors.
The condition of the real values for the energy playing a role of
the gauge condition in the second approximation has been achieved by minimization (on respect
to a given phases set) of its imaginary part.
This results into
finite accuracy at the energy evaluation in situation
with non-zero phases. 

Numerical simulations of eigenenergies have been performed
for approximately $3500 \times 200$ 
grid points $(r,\ \phi)$ in 2D $\vec q$-space  for the purposes of
subsequent estimation of charge transport properties of the
system. For  simulations we use the spectral line width equals
to 1~K.

\begin{figure*}
(a) \hspace{8cm} (b) \\
\includegraphics[width=7cm,height=6cm]{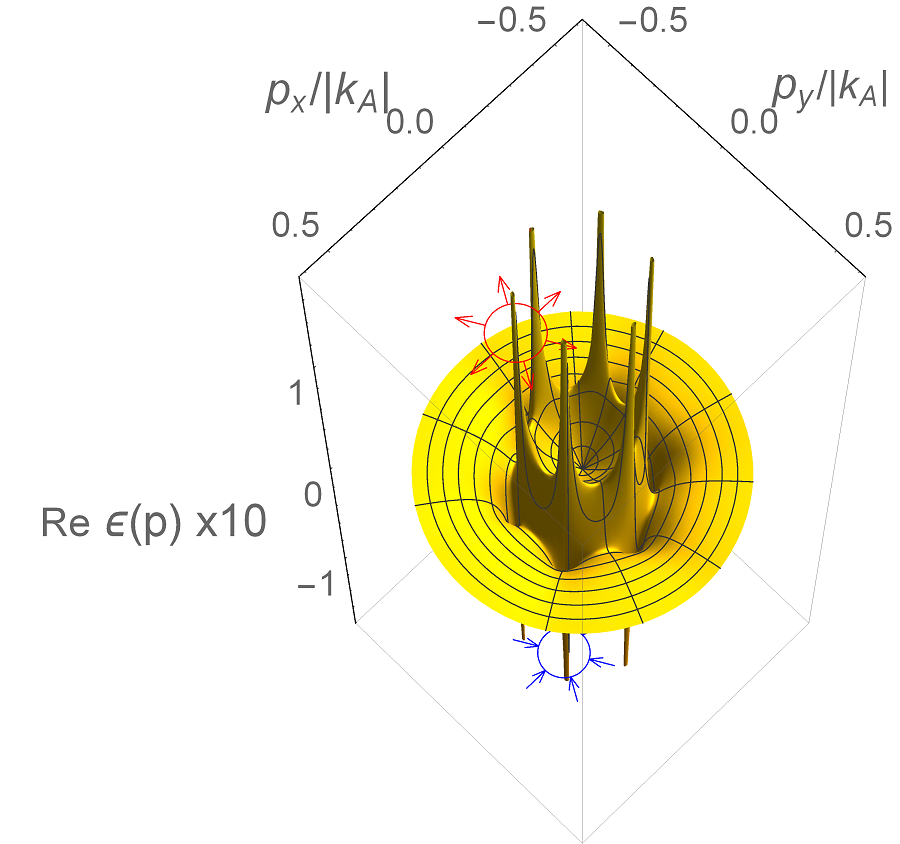}
\includegraphics[width=6cm,height=5cm]{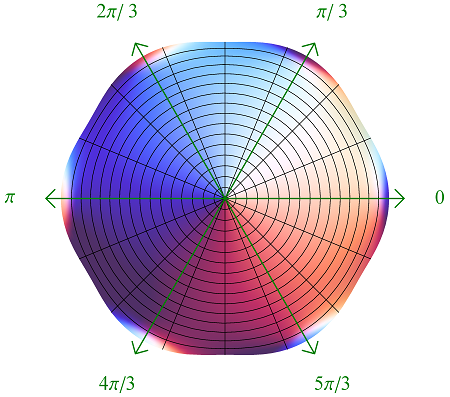}\\
(c) 
\includegraphics[width=8cm]{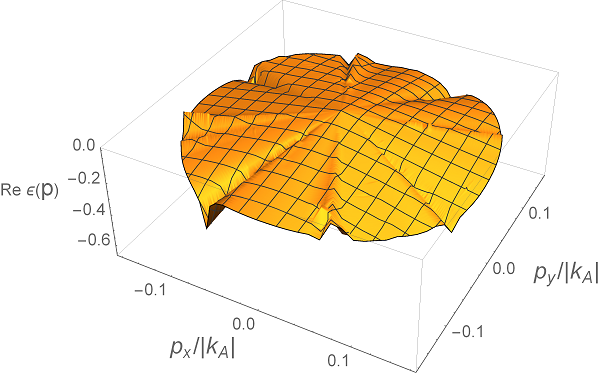}
\caption{Energy bands of 2D semi-metal calculated in two approximations:
(a,~b) in the first-order approximation with zero gauge phases in   4th order Taylor series expansion
 of exponential factors of the exchange operator for Pb monolayer (a) and graphene (b);
(c) in the second-order approximations
 with non-zero gauge phases in the exact expression for the
exchange for graphene.
The Dirac cone and its six tilted replicas in the Dirac point is presented in figure (b).
 One of six pairs Weyl-like nodes: source and sink is indicated in figure (a).
 }
\label{fig-band}
\end{figure*}

The calculated Majorana model bands $\epsilon$ in fig.~\ref{Vortices-band-structure}a and fig.~\ref{fig-band} have the form of 
cones near the Dirac point $K_{A(B)}$ at wave numbers $q<0.0002K_A $. Then the Dirac cone is splitted, and
 eight tilted sub-replicas emerge at large $q$.

We have calculated the Fermi velocity
$v_M(\vec q_F)$ for different
wave vectors
$\vec q_F= \{q_F, \phi\}$  by the formula
$v_M={\partial \epsilon \over \partial q_F} $.
According to the Table~\ref{majorana-velocity},   the Fermi velocity $v_M(\vec q_F)$
 decreases with a factor two 
in respect to its value
$v_M(0)$ in the Dirac point.

\begin{table}[htbp]
\caption{
%
The dependence of the charge carriers velocity  $v_M(\vec q_F)$
on  wave vector $\vec q_F=\{q_F,\ \phi\}$ at the Fermi level in polar coordinates
$ \{q_F,\ \phi\}$ for various directions shown
in Fig.~\ref{fig-band}b.}
\begin{tabular}{|c|c|c|}
\hline \hline $q_F/K_A$  & $\phi$, rad 
& $v_M(\vec q_F)/v_M(0)$
\\
\hline\hline
   0    &     for all   $\phi$      & 1\\ \hline
    0.005        &    0            & 0.9985\\
        &      $\pi$/2          &  0.9993\\
     &    $\pi$            & 0.9985\\ \hline
           & $\pi$/2       & 0.9968\\
  0.02      &      2$\pi$/3         & 0.9920 \\
        &  4$\pi$/3               & 0.9135\\\hline
&       2$\pi$/3       &  0.6148\\[-2mm]
\vspace*{-1mm}  0.1& &\\[-2mm]
      &     5$\pi$/3         &  0.5012\\
\hline \hline
\end{tabular} \label{majorana-velocity}
\end{table}


\setcounter{section}{1}

\section{
Quantum statistics of the model $N=3$}
\setcounter{section}{19}
\setcounter{equation}{12}

\subsection{Perturbation theory}

To find   the quasiparticle current \eqref{graphene-quasirel-current2} in main text,
we use the perturbation theory
\cite{Dyson,MyLambert}. 
Power series expansion
of (\ref{mass-renormalization2}) allows to rewrite the potential energy
$V^{SM}$ for interaction  of the secondary quantized
fermionic field
$\chi_{+\sigma_{_B}}(x)$ with an electromagnetic field
(Eq.~(\ref{rel-from-pseudi-Dirac-whithout-mix3}) in main text)
in the following form
\begin{eqnarray}
V^{SM}= \chi^{\dagger}_{+\sigma_{_B} } \left[-c \vec \sigma
_{2D}^{BA} \cdot{e\over c}\vec A -
\widetilde {\Sigma_{AB}\Sigma_{BA}}(
0)- \sum_i \left. {d\widetilde {\Sigma_{AB} \Sigma_{BA}}\over d
p_i'} \right|_{p_i'=0} \left( p_i ^{AB}
-{e\over c} A_i \right)\right. \nonumber \\
\left. - {1\over 2} \sum_{i,j} \left. {d^2\widetilde {\Sigma_{AB}
\Sigma_{BA}}\over d   p_i' d   p_j'} \right|_{p_i', \ p_j' =0}
\left( p_i ^{AB} -{e\over c} A_i \right)\, \left( p_j ^{AB}
-{e\over c} A_j \right) + \ldots\right]\chi_{+\sigma_{_B} }.
\label{interaction_V_graphene}
\end{eqnarray}
Taking into account that 
$c \vec \sigma _{2D}^{BA} $ is a quantum analog of the current vector \cite{Fock}  
$\vec v$, and expressing 
$\vec p^{AB}$ through  the dynamical mass 
$\widetilde {\Sigma_{AB}\Sigma_{BA}}(\vec p_{AB}- e\vec A/c )$, the operator 
$V^{SM}$ (\ref{interaction_V_graphene}) is rewritten as 
\begin{eqnarray}
V^{SM}= -\chi^{\dagger}_{+\sigma_{_B} } \left\{ \vec v
\cdot{e\over c}\vec A +
\widetilde {\Sigma_{AB}\Sigma_{BA}}(
\vec p_{AB}- e\vec A/c ) \right.
\nonumber \\
\times \left[ {\widetilde {\Sigma_{AB}\Sigma_{BA}}(0)\over
\widetilde {\Sigma_{AB}\Sigma_{BA}}(\vec p_{AB}- e\vec A/c )} +
\sum_i
  \left.
        {d\widetilde {\Sigma_{AB} \Sigma_{BA}}\over d   p_i'}
  \right|_{p_i'=0}
  \left( v_i
     -{e\over c} \tilde A_i
  \right)
\right.\nonumber \\
\left.
        + {\widetilde {\Sigma_{AB}\Sigma_{BA}}(\vec p_{AB}- e\vec A/c )
        \over 2} \sum_{i,j}
  \left.
      \left.
       {d^2\widetilde {\Sigma_{AB} \Sigma_{BA}}\over d   p_i' d   p_j'}
      \right|_{p_i', \ p_j' =0}
      \left( v_i-{e\over c} \tilde A_i \right)\, \left( v_j -{e\over c} \tilde A_j
      \right)
      + \ldots
  \right]
\right\}
\chi_{+\sigma_{_B} }, \nonumber \\
\label{interaction_V_graphene1}
\end{eqnarray}
where $\vec {\tilde A}= \vec A/ \widetilde
{\Sigma_{AB}\Sigma_{BA}}(\vec p_{AB}- e\vec A/c )$. In the interaction representation,
if account for the first-order terms on
$ A_i$ and terms quadratic on 
$v_i$ in the expansion of the evolution operator 
$\hat U(x'-x'')$  only,  the Ohmic current 
$e \chi^{\dagger}_{+\sigma_{_B} } (x')v^i_{x'x} \chi_{+\sigma_{_B} } (x)$ reads 
\begin{eqnarray}
\chi^{\dagger}_{+\sigma_{_B} } (x^+) v^l_{x^+x^-}
\chi_{+\sigma_{_B} } (x^-) =\hat U^{\dagger}(x^{+}-x')
\chi^{\dagger}_{+\sigma_{_B} } (x') v^l_{x'\, x^-}
\chi_{+\sigma_{_B} } (x'') \hat U(x'' - x^{-})
\nonumber \\
=
\left[1- {\imath 
}\int V^{SM}(x^{+}-x') dt' d\vec {r'} +\ldots \right]
\chi^{\dagger}_{+\sigma_{_B} } (x')v^l_{x'x^-} \chi_{+\sigma_{_B}
} (x^-) =\left\{1 \right.
   -
{(-\imath)}\int \int \chi^{\dagger}_{+\sigma_{_B} } (x^{+}-x')
\times \nonumber \\
\sum_i \left[
  {e\over c}  {v^\dagger }^i_{x^{+},\, \bar{x}}  A_i
\right.
+\widetilde {\Sigma_{AB}\Sigma_{BA}}(\vec p_{AB}- e\vec A/c )
   {d\widetilde {\Sigma_{AB} \Sigma_{BA}}\over d   p_i^{AB}}(0)
   \ {{v^\dagger }^i_{x^{+},\, \bar{x}} }
   -{e\over c}{d\widetilde {\Sigma_{AB} \Sigma_{BA}}\over d   p_i^{AB}}
(0) A_i \nonumber
\\
- \left.
  {e\widetilde {\Sigma_{AB}\Sigma_{BA}}(\vec p_{AB}- e\vec A/c )\over 2c}
  \sum_{j}
  {d^2\widetilde {\Sigma_{AB} \Sigma_{BA}}\over d   p_i^{AB} d   p_j^{AB}}
  (0)\
  \left(
        {v^\dagger }^i_{x^{+},\, \bar{x}}\, A_j+
        {v^\dagger }^j_{x^{+},\, \bar{x}}\, A_i
  \right)
\right] \nonumber
\\
\left. \times \chi_{+\sigma_{_B} } (\bar{x})
   d\bar{t}\, d\vec {\bar{x}}\, dt'\, d\vec {x}\,'
+ \ldots \right\} \chi^{\dagger}_{+\sigma_{_B} } (x')
 \chi_{+\sigma_{_B} } (x'')\hat U(x''-x') v^l_{x'x^-}. \
\label{green-function-representation}
\end{eqnarray}
Here 
$x=\{\vec x,\ t\}$ is the 4-vector of space coordinates 
$\vec x$ and time 
$t$. The current 
(\ref{green-function-representation}) is the two-point function and, therefore,
has to be expressed through
two-point $G_1(\vec r - \vec r\,',\, \vec{\bar r}- \vec{\bar r}\,', t-t')$ and four-point 
$G_2(x,\, x',\, \bar{x},\, \bar{x}')$ Green functions. 
Using the relation 
$U(\vec r - \vec r\,', t) = \imath G_1(\vec r - \vec r\,', t)$ between the evolution operator 
$U(\vec r - \vec r\,', t)$ and the one-particle Green function 
$\imath G_1^{\mu \nu} (x,\ x') = (\chi^{\dagger})^{\mu}(x)\chi^{\nu}(x')$
\cite{MyLambert}
, one can express the product 
of \ the field components 
$\chi^\mu,\ \chi^{\nu}$, entering in the form of multipliers 
in (\ref{green-function-representation}), through 
$G_1(\vec r - \vec r\,',\, \vec{\bar r}- \vec{\bar r}\,', t-t')$ as
$$
G_1(\vec r - \vec r\,',\, \vec{\bar r}- \vec{\bar r}\,', t-t')=
{1\over \imath ^2} U(\vec r - \vec r\,', t-t')U(\vec r\,' - \vec
{\bar r}, t)= {1\over \imath ^2} U(\vec r - \vec {\bar r}, t)
$$
and through 
$G_2(x,\, x',\, \bar{x},\, \bar{x}')$ as $ G_2(x,\, x',\,
\bar{x},\, \bar{x}') =(\imath)^2 \chi^{\dagger} (x)\chi^{\dagger}
(x')\chi (\bar{x}) \chi (\bar{x}'). $ Now, we take into account the linear in 
$A_i$ terms only  in
Eq.~(\ref{green-function-representation}), then after the standard procedure
of transformation of the operator product into a normal form 
\cite{Dyson},
one can rewrite 
$j_i^{Ohm}$ in (\ref{graphene-quasirel-current2}) through 
two-point 
$G_1(\vec r - \vec r\,',\, \vec{\bar r}- \vec{\bar
r}\,', t-t')$ and four-point 
$G_2(x,\, x',\, \bar{x},\, \bar{x}')$ Green functions: 
\begin{eqnarray}
j_l^{Ohm}=\sum_{\mu} e {\chi^\mu}^{\dagger}_{+\sigma_{_B} } (x^+)
v^l_{x^+x^-} \chi^\mu_{+\sigma_{_B} }(x^{-})
\nonumber \\
=\mbox{Tr} \left\{\chi^{\dagger}_{+\sigma_{_B} } (x')
 \chi_{+\sigma_{_B} } (x'') +\imath {-\imath e\over (\imath)^3}\int \int
\sum_i \left[ {e\over c}  {v^\dagger }^i_{x^{+},\, \bar{ x}}
 A_i
+\widetilde {\Sigma_{AB}\Sigma_{BA}} \left(\vec p_{AB}- {e\vec
A\over c }\right) {d\widetilde {\Sigma_{AB} \Sigma_{BA}}\over d
p_i^{AB}} (0) \ {{v^\dagger }^i_{x^{+},\, \bar{ x}} }\right.
\right. \nonumber
\\
- \left.
  {e\widetilde {\Sigma_{AB}\Sigma_{BA}}(\vec p_{AB}- e\vec A/c )\over 2c}
  \sum_{j}
  {d^2\widetilde {\Sigma_{AB} \Sigma_{BA}}\over d   p_i^{AB} d   p_j^{AB}}
  (0)\
  \left(
        {v^\dagger }^i_{x^{+},\, \bar{x}}\, A_j+
        {v^\dagger }^j_{x^{+},\, \bar{x}}\, A_i
  \right)
\right] \nonumber
\\
\times \left. G_2(x^{+},\, \bar{x},\, x',\, x'') G_1( x'',\ x')
d\bar{t}\, d^2 \vec {\bar{ x}}\, dt'\, d^2\vec {x}\,' dt''\,
d^2\vec {x''} + \ldots \right\} v^l_{x'x^{-}} .\ \ \
\label{green-function-representation1}
\end{eqnarray}
Due to the fact that the Green functions 
$G_1$, $G_2$  are symmetric at change 
$\vec v \to -\vec v$, the integrals with expressions proportional 
to $v_i$, vanish in 
(\ref{green-function-representation1}).
Interparticle Coulomb interaction renormalizes the charge carriers mass
\cite{DiracFieldTheory} in a way as it is shown 
in fig.~\ref{Mass-eigen-values}. Therefore,
 $ 
 {d\widetilde {\Sigma_{AB} \Sigma_{BA}}\over d   p_i^{AB}} \approx 0$ near 
the Majorana zero-energy state, and respectively
the third term in the equation (\ref{green-function-representation1}) is also equal to zero.

\subsection{Low-frequency limit of the  conductivity 
in the models $N=2$ and $N=3$ with spatial dispersion 
}

Let us analyze the behavior of the low-frequency conductivity
of the models $ N = 2 $ and $ N = 3 $ with spatial dispersion of charge carriers.
First, let us prove the following theorem 
\begin{theorem}
Low-frequency conductivity of the charge carriers system  with spatial dispersion for the $ N = 3 $
model, in contrast to the model $ N = 2 $, takes nonzero finite values at frequencies
$ \omega \ll 1 $.
\end{theorem}

\Proof
Self-energy operator 
$\hat\Sigma _{self}$ is expressed through the polarization operator
$\hat \Pi $ as \cite{Kraft-Ropke}:
\begin{equation}
\hat\Sigma _{self} (\omega,\ k) G_1= i\hbar V(k) \hat \Pi(\omega,\
k) ,
 \label{complex-conductivity}
\end{equation}
where 
$V(k)$ is the Coulomb interaction operator in the momentum representation,
$ G_1$ is the one-particle Green function.
In the neighborhood of the Dirac point,
the one-particle Green function 
of negatively (positively) charged carrier  is $ G_1={\hbar^{-1}\over (\omega \mp k )}$.
Let us use the Feynman diagram technique. 
The Feynman diagram 
for $\hat \Pi $ has the form 
$\hat \Pi $ = \includegraphics[width=0.9cm,height=0.9cm,angle=0]{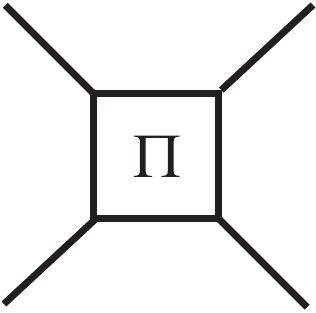},
The Feynman representation for 
$i\hbar V(k)$ is a wavy line. 
For holes, the left hand side of the expression (\ref{complex-conductivity})
is represented by the diagram 
$\hat\Sigma _{self} (\omega,\ k) G_1=$
\includegraphics[width=1.2cm,height=0.5cm,angle=0]{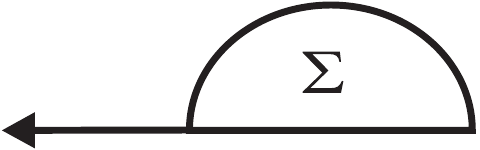},
whereas the right hand side of 
(\ref{complex-conductivity}) is given by the diagram 
$i\hbar V(k) \hat \Pi(\omega,\ k)=$
\includegraphics[width=1.cm,height=1.cm,angle=0]
{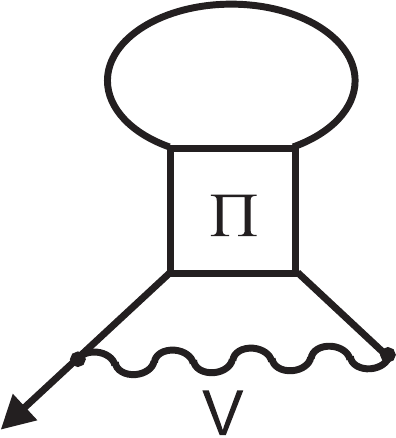}. Using the diagram
technique, the expansion of the expression
(\ref{complex-conductivity}) into  perturbation theory series is written  as
\begin{equation}
\includegraphics[width=1.2cm,height=0.5cm,angle=0]{hole-selfenergy-s.pdf}=
\includegraphics[width=1.1cm,height=1.cm,angle=0]
{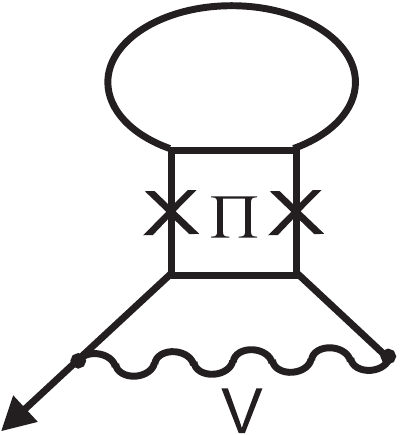}
+\includegraphics[width=1.1cm,height=1.cm,angle=0]
{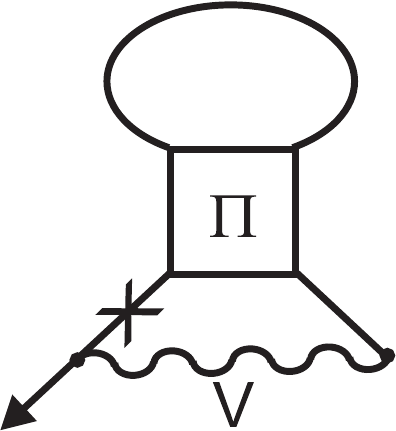} +
\includegraphics[width=1.1cm,height=1.cm,angle=0]
{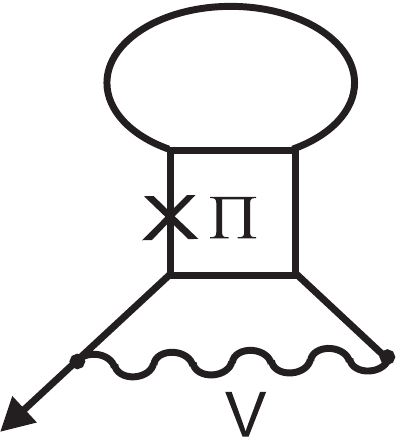}.
\label{feynman-diagram-series}
\end{equation}
Here, by the cross 
"$\times$"\ there have  been marked crossed out lines. 
The first and the second terms in the expansion 
(\ref{feynman-diagram-series}) are called the Hartree-Fock approximation for
$\Sigma^{XC}$, the rest terms are for correlation interaction 
(correlation energy), 
$
\hat\Sigma _{self} = \Sigma^{xc} + \Sigma^{c}.
$
Let us account for terms up to the second order in series on $V$
in 
(\ref{feynman-diagram-series}) as 
\begin{equation}
\includegraphics[width=1.2cm,height=0.5cm,angle=0]{hole-selfenergy-s.pdf}\approx
\includegraphics[width=1.2cm,height=0.4cm,angle=0]{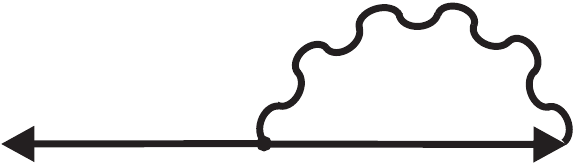}
+\includegraphics[width=0.7cm,height=1.1cm,angle=0]{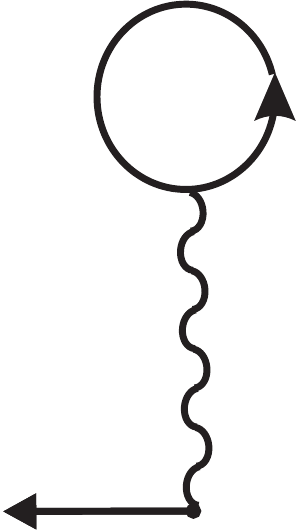}
+
\includegraphics[width=1.4cm,height=1.3cm,angle=0]{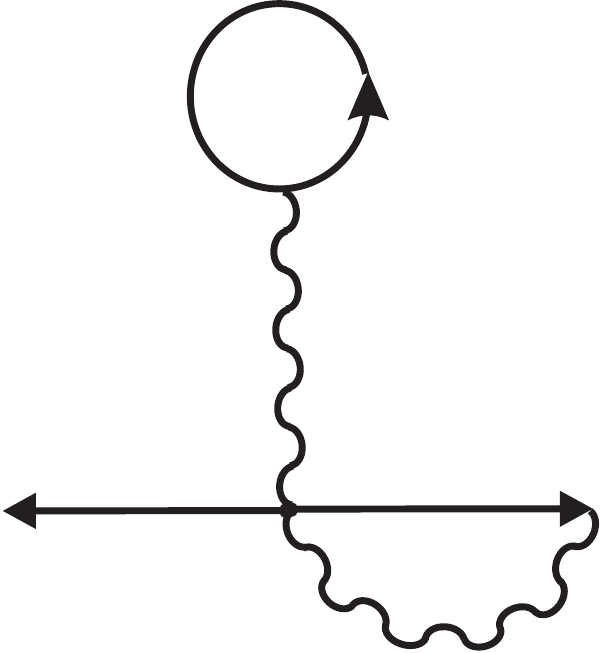}.
\label{feynman-diagram-series1}
\end{equation}
From the form of the first and the second diagrams in expansion of 
(\ref{feynman-diagram-series1}) it follows that the Hartree-Fock approximation 
$\Sigma^{xc}= \Sigma^{x}+V^{self}$ consists of the exchange 
$ \Sigma^{x}$  and the self-consistent Coulomb potential 
$V^{self}$  which is the polarization correction of the first-order on interaction
$V$.


From the another side, the self-energy 
$\hat \Sigma _{self}(\omega,\ k)$ is the interaction energy for the current of charge
carriers with an electromagnetic field 
$\hat A_\mu (\omega,\ k)$:
$$\left\langle 0\right| \hat \Sigma _{self}(\omega,\ k) \left |0 \right\rangle
= \left\langle 0\right| \hat \chi^\dagger e\gamma^\mu \hat\chi
\hat A_\mu (\omega,\ k)  \left |0 \right\rangle \equiv
\left\langle 0\right| \hat j^\mu \hat A_\mu (\omega,\ k)\left |0
\right\rangle.$$
Here 
$\gamma_\mu$ are the Dirac matrices, 
$\left |0 \right\rangle$ is the vacuum state. 
In the Hartree-Fock approximation 
(\ref{feynman-diagram-series1}) 
electron-hole decay is absent:
$\Im m\,\Sigma^{xc} (k) =0$, and respectively
in this approximation we can rewrite imaginary part of the expression 
(\ref{complex-conductivity}) through the photon creation and annihilation operators
$\hat A^\dagger _{\mu}(\omega,\ k)$, $\hat A_{\mu}(\omega,\ k)$ as 
\begin{equation}
\Im m \left.\langle 0\right| \hat j^\mu \hat A^\dagger_\mu
(\omega,\ k)\left |0 \rangle\right. =\Im m
 [(i\hbar)^2 V^2(k) \hat \Pi^{RPA}(\omega,\ k)G_1]=\Im m
 [i\hbar V(k) \hat \Pi^{(1)}(\omega,\ k)G_1],
 \label{complex-conductivity1}
\end{equation}
where 
$\hat \Pi^{(1)}=i\hbar V(k)\hat \Pi^{RPA} $,
$\hat \Pi^{RPA}$ is the polarization operator in the random phases approximation (RPA).
Expressing current component $\hat j_\mu$ through the complex conductivity 
$\sigma^* (\omega, \ k)$ as 
$\hat j_\mu=\sigma^* (\omega, \ k)\hat A
_\mu (\omega,k)$ and using commutation relations, one can transform 
(\ref{complex-conductivity1}) to the following form 
\begin{equation}
\Im m [\sigma^* (\omega, \ k)\left.\langle 0\right| \left[\hat
A^\dagger_\mu (\omega,\ k),\hat A_\mu (\omega,\ k)\right] \left |0
\rangle\right. ] =\Im m [
 \imath \hbar V(k) \hat \Pi^{(1)}(\omega,\ k)G_1 ],
 \label{complex-conductivity2}
\end{equation}
where the vacuum expectation 
$\left.\langle 0\right| \left[\hat
A^\dagger_\mu(\omega,\ k) ,\hat A_\mu (\omega,\ k)\right] \left |0 \rangle\right.$  from the commutator
$\left[\hat A^\dagger_\mu
(\omega,\ k),\hat A_\mu (\omega,\ k)\right] $ is a photon propagator
${1\over \omega^2 -k^2}$. For the case of small spatial dispersion 
$k\to 0$, one has 
$G_1 =1/(\hbar\omega)$
, and the expression 
(\ref{complex-conductivity2}) becomes 
"optical"\ alternating-current (ac) conductivity
(\cite{Kraft-Ropke}
):
\begin{equation}
{\sigma^* (\omega)\over \omega} =\hbar
 (i V)^2(k) \hat \Pi^{RPA}(\omega,\ k) = i(1-\epsilon^* (\omega)),
 \label{complex-conductivity3}
\end{equation}
where 
$\epsilon^* =1- V(k) \hat \Pi^{(1)} $ is a complex dielectric permeability. 

Now, let us consider the case of spatial dispersion of states
$\omega \le k > 0,\ \omega > 0$ at  very small frequencies 
$\omega \ll 1 $. In the model 
$N=2$ of massless pseudo-fermions with spatial dispersion,  the expression 
(\ref{complex-conductivity2}) for the  hole 
conductivity 
$\sigma^*_{min}$ having positive values is transformed to the form 
\begin{equation}
\Im m \ {\sigma^* _{min}(\omega, \ k)} =(\omega - k)\Im m \ [
 i V(k) \hat \Pi^{(1)}(\omega,\ k) ].
 \label{complex-conductivity4}
\end{equation}
Because of 
$\omega - k \le 0$
, in accord with formula 
(\ref{complex-conductivity3}) the left hand side of 
(\ref{complex-conductivity4}) gains physical positive values if the real part of the polarization operator
$\hat \Pi^{(1)}$ in the right hand side equals zero. 
The last means that 
the direct-current (dc) conductivity (minimal dc-conductivity) $\sigma_{min}$
in the model 
$N=2$ is equal to zero. 

In the model 
$N=3$, the action of the operators 
$\left(\Sigma_{rel}^{x}\right)_{AB}$ \eqref{Sigma-AB3} and $\left( \Sigma_{rel}^{x}\right)_{BA}$ \eqref{Sigma-BA3}
provides the squeezing 
$k \to k_{AB}$ of the part of states and respectively the expression for
(\ref{complex-conductivity4})  is transformed to the following: 
\begin{equation}
\Im m \ {\sigma^* _{min}(\omega, \ k_{AB})} = \Im m \
 (i V)^2(k_{AB}) \hat \Pi^{RPA}(\omega,\ k_{AB}) (\omega - k_{AB}).
 \label{complex-conductivity5}
\end{equation}
If  $\omega > v_F k_{AB}$, due to dynamic decrease of spatial dispersion
there could exist part of quasi-particle states 
(\ref{complex-conductivity5}).
It is possible because
$\omega $ can not gain the values
less than vacuum frequency
$\omega_0$. The last means that in the model
$N=3$,  $\sigma_{min}$ can get non-zero value
as well as optical conductivity.
Thus, we prove that contrary to the $N=2$ model, the model
$N=3$ can demonstrate  non-zero
dc-conductivity.

Let us write down the Fourier-Laplace image of the operator
$\vec v_{AB}$ in the coordinate representation of secondary quantization
\begin{eqnarray}
 \int e^{-\imath \omega t}e^{-\imath  \vec p \cdot \vec r}
\chi^{\dagger}_{+\sigma_{_B} } (x) v^i_{x^+x^-} \chi_{+\sigma_{_B}
} (x) dt\, d\vec r . \label{Laplas-Fourier-transform}
\end{eqnarray}
Because of 
the fact that the operator $\vec v_{AB}$ does not depend on time, 
trace "Tr"\ of the Fourier components of 
(\ref{Laplas-Fourier-transform}) in representation,
where the operator $H_0$ is diagonal,
can be rewritten in the matrix form as
\begin{eqnarray}
\mbox{Tr } \int e^{\imath \omega t}e^{-\imath  \vec k \cdot \vec
r} \chi^{\dagger}_{+\sigma_{_B} } (x^+) v^i_{x^+x^-}
\chi_{+\sigma_{_B} } (x^-) dt\, d\vec r 
= {1\over (2\pi)^6} \int e^{-i(\omega^+ -\omega^ - - \omega) t}
e^{\imath  (\vec {p^+} -\vec {p^-} -\vec k) \cdot \vec r}\nonumber \\
\times \mbox{Tr }\ \chi^{\dagger}_{+\sigma_{_B} } (p^+, \omega^+)
v^i_{\vec {p}^+ -\vec {p}^-,\, (\vec {p}^+ +\vec {p}^-)/2}\,
\chi_{+\sigma_{_B} } (p^-, \omega^-) dt\, d\vec r\, d\omega^+ \,
d\omega^- d^2\vec {p}\,^+ d^2\vec {p}\,^- \ \
\nonumber \\
=\mbox{Tr }{1\over (2\pi)^3}  \int d\omega^+  d\omega^- d\vec
{p}\,^+ d\vec {p}\,^- \delta (\omega^+ -\omega^ - - \omega)\
\delta (\vec {p^+} -\vec {p^-} -\vec k)
\chi^{\dagger}_{+\sigma_{_B} } (p^+, \omega^+) v^i_{\vec {p}^+
-\vec {p}^-, p}\,\chi_{+\sigma_{_B} } (p^-, \omega^-)
\nonumber \\
=\mbox{Tr }\ {1\over (2\pi)^3} \int \chi^{\dagger}_{+\sigma_{_B} }
(p^+, \omega^{+-}+\omega') v^i_{\vec {p}^+ -\vec {p}^-,
p}\,\chi_{+\sigma_{_B} } (p^-, \omega ')\,
d^2\vec {p}\
 d\omega'
 \label{Laplas-Fourier-transform1}
\end{eqnarray}
at the following 
conditions:
\begin{eqnarray}
\vec {p\,}^{\pm}=\vec p\pm \vec k/2 ,\quad \omega'= \omega^+
-\omega, \quad \omega^{+-}=\omega^+-\omega^-=\omega
\label{frequency-impuls-condition}
\end{eqnarray}
owing to $\delta$-functions entering the expression (\ref{Laplas-Fourier-transform1}).

Let us perform the Wick turn
(change from time to imaginary time)
in order to find Fourier-Laplace image of the expression 
(\ref{Laplas-Fourier-transform}) at finite temperatures $T$. In this representation, the Fourier image
of one-particle Green function 
$G_1(\vec r_1,\ \vec r\,'_1,\ t_1-t'_1)$ has the form 
\cite{Kraft-Ropke}
\begin{eqnarray}
G_1(\vec r_1,\ \vec r\,'_1,\ z_\nu)= \int_0^{-i\hbar \beta} dt_1
e^{iz_\nu (t_1-t'_1)} G_1(x_1,\ x'_1),
\end{eqnarray}
where 
$\bar{\beta} =1/(cT)$, 
$z_\nu$ are Matsubara frequencies defined by the following expressions 

\begin{eqnarray}
z_\nu ={\pi \nu \over -i \hbar \bar \beta}+{\mu \over \hbar},\ \ \
\nu= \left\{
\begin{array}{c}
\pm 1,\ \pm 3,\ \ldots \mbox{ for fermions}, \\
0, \ \pm 2,\ \pm 4,\ \ldots \mbox{ for bosons}.
\end{array}
\right.
\end{eqnarray}
Analytical continuation of the function 
$G_1(x_1,\ x'_1)$ to interval 
 $(0, \ -i \hbar \bar \beta)$ is an advanced correlation function 
$G^>_1$ (see 
\cite{Kraft-Ropke}
):
\begin{eqnarray}
G_1(\vec r_1,\ \vec r\,'_1,\ z_\nu)= \int_{-\infty} ^{+\infty}
\int_0^{-i\hbar \bar \beta} {d\omega \over 2\pi i \hbar } \ dt_1
e^{i(z_\nu -\omega )(t_1-t'_1)} G^>_1(\vec r_1,\ \vec r\,'_1, \
\omega)
=-\int_{-\infty} ^{+\infty}{d\omega \over 2\pi \hbar } G^>_1(\vec
r_1,\ \vec r\,'_1, \ \omega) {e^{(z_\nu -\omega )\hbar \bar \beta}
- 1\over (z_\nu -\omega )}
\nonumber \\
= -\int_{-\infty} ^{+\infty}{d\omega \over 2\pi \hbar } G^>_1(\vec
r_1,\ \vec r\,'_1, \ \omega) {e^{(\mu -\omega \hbar )\bar \beta}
e^{i\pi \nu}- 1\over (z_\nu -\omega )} . \ \ \
\label{Mazubara-Laplas-Fourier-transform}
\end{eqnarray}
We note that  the right-hand side of the expression
(\ref{Mazubara-Laplas-Fourier-transform}) includes the Fourier image
of the retarded correlation function
$G^>_1(\vec r_1,\ \vec
r\,'_1, \ \omega)$:
\begin{eqnarray}
G_1(\vec r_1,\ \vec r\,'_1,\ z_\nu)= -\int_{-\infty}
^{+\infty}{d\omega \over 2\pi \hbar } {G^<_1(\vec r_1,\ \vec
r\,'_1, \ \omega) e^{i\pi \nu}- G^>_1(\vec r_1,\ \vec r\,'_1, \
\omega)\over (z_\nu -\omega )} \nonumber \\ = \int_{-\infty}
^{+\infty}{d\omega \over 2\pi \hbar } {G^>_1(\vec r_1,\ \vec
r\,'_1, \ \omega) \mp G^<_1(\vec r_1,\ \vec r\,'_1, \ \omega)\over
(z_\nu -\omega )}= \int_{-\infty} ^{+\infty}{d\omega \over 2\pi
\hbar } {\tilde A(\vec r_1,\ \vec r\,'_1, \ \omega)\over (z_\nu
-\omega )}. \label{Mazubara-Laplas-Fourier-transform1}
\end{eqnarray}
In (\ref{Mazubara-Laplas-Fourier-transform1}) the upper sign refers to bosons, the lower sign to fermions; function
$\tilde A(\vec r_1,\ \vec r\,'_1,\ \omega)$ is called the spectral weight 
\begin{eqnarray}
\tilde A(p,\ \omega) = A^{HF}(p,\ \omega)+A^{C}(p,\ \omega)=
G_1(\omega + \imath \epsilon)  -
G_1(\omega - \imath \epsilon) \nonumber \\
={2\over \imath } {\Im m \,\Sigma^C (p, \ \omega)\over [\hbar
\omega - E(p)- \Re e \,\Sigma^C (p, \ \omega)]^2 + [\Im m \,
\Sigma^C (p, \ \omega)]^2 }, \label{Mazubara-spectral-function}
\end{eqnarray}
where $E(p),\ A^{HF}$ is the energy of quasi-particle excitation and the spectral
weight in the Hartree-Fock approximation, 
$A^{C}$ is the many-particle correction to the spectral weight. 
If the life time 
$\tau= 1/(2\, \Im m \, \Sigma^C ) = 1/\Gamma$ of the quasi-particle excitation
is large, then the smallness of the decay rate $\Gamma $ leads to the following relation 
\begin{eqnarray}
\lim_{\Gamma \to 0}\tilde A(p,\ \omega)=
 A^{HF}(p,\ \omega)=
\delta (\hbar \omega - E(p))/\imath
\label{HF-Mazubara-spectral-function} .
\end{eqnarray}

The temperature Green's functions of 2D fermions in the diagonal
matrix representation are obtained by summation of the Fourier-Laplace images on Matsubara frequencies
$z_1$, $z_{12}=z_2+z_1$ with subsequent integration over momenta 
$\vec p$ \cite{Kraft-Ropke}
:
\begin{eqnarray}
G_1(x -x'') ={1\over -\imath \bar{\beta} \hbar }\sum _{\nu_1} \int
{d^2\vec p \over (2\pi )^2} e^{-\imath z_1(t-t'')} e^{-\imath \vec
p\cdot(\vec r - \vec {r''})}G_1(z_1, p),
  \\
G_2(\vec r_1, \vec r_2, \vec {r'_1}, \vec {r'_2}, t-t') ={1\over
-\imath \bar{\beta} \hbar }\sum _{\nu_{12}} \int {d^2\vec p_1
d^2\vec p_2 d^2\vec {p'_1} d^2\vec {p'_2} \over (2\pi )^{8}}
e^{-\imath \left( \vec p_1\cdot\vec r_1 +\vec p_2\cdot\vec r_2
-\vec {p'_2}\cdot\vec {r'_2} -\vec {p'_1}\cdot\vec {r'_1} \right)
}
\nonumber \\ \times
e^{-\imath z_{12}(t-t')} G_2(z_{12}, p_1, p_2, p'_1, p'_2).
\label{two-pointsG_2}
\end{eqnarray}
Here the perturbed Green function 
$G_2(z_{12}, p_1, p_2, p'_1, p'_2)$ is represented in the form of the perturbation theory series
\begin{eqnarray}
G_2(z^{-+}, p^-, p^+, p_1, p_2) ={1\over -\imath \bar{\beta} \hbar
}\sum _{\nu^{+}} \left\{ G_1(z^{-+} -z^+, p^- )G_1(z^+,
p^+)\right.
\nonumber\\
\times \left[(2\pi)^4 \left( \delta ( \vec { p^-} - \vec { p_1})
\delta ( \vec { p^+} - \vec { p_2}) -\delta ( \vec { p^-} - \vec {
p_2}) \delta ( \vec { p^+} - \vec { p_1})
\right)\right. \nonumber \\
\left. \left. +\imath \int V (\vec {\bar{ p_2}} -\vec
p\,^+)(2\pi)^2 \delta \left( \vec {\bar{ p_1}} + \vec {\bar{ p_2}}
-\vec p\,^- -\vec p\,^+ \right) G_2(z^{-+}, \bar{ p_1}, \bar{
p_2}, p_1, p_2) {d\vec {\bar{ p_1}} d\vec {\bar{ p_2}}\over (2\pi
)^{4}} \right] \right\}. \label{perturbation_theory}
\end{eqnarray}
Taking into account (\ref{Laplas-Fourier-transform1},
\ref{frequency-impuls-condition}), the two-point Green function 
$G_1(x ,\ x'')$ entering in (\ref{green-function-representation1})
in the Matsubara representation is written as 
\begin{eqnarray}
G_1(x, x'') ={1\over -\imath \bar{\beta} \hbar }\sum _{\nu^{+-}}
\int {d^2\vec p\,^- \over (2\pi )^2} e^{-\imath z^{+-}(t-t'')}
e^{-\imath \vec p\,^-\cdot(\vec r - \vec {r''})}G_1(p^+, p^-,
z^{+-} ) , \label{two-pointsG_1}
\end{eqnarray}
where $z^{+-}=z^+ - z^-$. Substitution of the Fourier series for 
$v^i_{x^+ x^-}$:
\begin{eqnarray}
v^i_{x^+ x^-} =\int {d^2\vec p \over (2\pi )^2} e^{-\imath \vec
p\cdot(\vec r^+ - \vec r^-)} v^i_{\vec {p}^+ -\vec {p}^-, p} ,
\label{fourier-velosity}
\end{eqnarray}
$G_1(x, x'')$ (\ref{two-pointsG_1}), and 
$G_2(\vec r_1, \vec r_2,
\vec {r'_1}, \vec {r'_2}, t-t')$ (\ref{two-pointsG_2}) into
(\ref{green-function-representation1}) and analytical continuation of
the obtained expression to the whole complex plane allow us to find the contribution
for $j_i^{Ohm}(\omega, \ k)$ which is linear on 
$A_i$:
\begin{eqnarray}
j_i^{Ohm}=\sum_{\mu} e {\chi^\mu}^{\dagger}_{+\sigma_{_B} } (x^+)
v^i_{x^+x^-} \chi^\mu_{+\sigma_{_B} }(x^{-})
={(-1)\imath e^2 A_i\over (\imath)^3c(2\pi)^{14}}
\nonumber \\
\times \mbox{Tr}
\left\{\imath\int d\bar{t}\, d^2 \vec {\bar{ r}}\, dt'\, d^2\vec
{r}\,' dt''\, d^2\vec {r''} \left(1-
  \widetilde {\Sigma_{AB}\Sigma_{BA}}(\vec p_{AB}- e\vec A/c )
   {d^2\widetilde {\Sigma_{AB} \Sigma_{BA}}\over d   p_i^{AB} d p_i^{AB}}
  (0)\right)\right.
 \int d^2\vec p\
e^{-\imath \vec p\cdot(\vec r\,^+  -\vec {\bar r})} {v^\dagger
}^i_{\vec {p}^+  - \vec {\bar p}, \vec p}
\nonumber \\
\times
 {1\over -\imath
\bar{\beta} \hbar }\sum _{\nu_{12}} \int d^2\vec {p^+} d^2\vec
{\bar p} d^2\vec {p'} d^2\vec {p''} e^{-\imath \left( \vec
p^+\cdot\vec r^+ +\vec {\bar p}\cdot\vec {\bar r} -\vec
{p''}\cdot\vec {r''} -\vec {p'}\cdot\vec {r'} \right)
} 
e^{-\imath (z_{12}+ \omega_{12})(t^+ - t^- - t' +\bar t)}
G_2(z_{12}, p^+, \bar p, p', p'')
 \nonumber \\
\times \left. {1\over -\imath \bar{\beta} \hbar }\sum _{\nu^{+-}}
\int d^2\vec p\,^-  e^{-\imath (z^{+-}+\omega^{+-})(t'' - t')}
e^{-\imath \vec p\,^-\cdot(\vec {r''} - \vec {r'})} G_1(\hat p^+,
\hat p^-, z^{+-} )
 \right\} 
 \int d^2\vec {\tilde p}
e^{-\imath \vec {\tilde p} \cdot(\vec r\, ' - \vec r\,^-)}
v^i_{\vec {p}\, ' -\vec {p}^-, \vec{\tilde p}/2}
 \ ,\ \ \
\label{green-function-representation1_1-0}
\end{eqnarray}
where
\begin{eqnarray}
\omega_{12}=2\pi \nu_{12}
,\ \omega^{+-}=2\pi \nu^{+-} 
; \label{mazubara-cyclic-frequency}
\end{eqnarray}
$(p^+ + \bar p)/2=p$, 
$ (p' + p^-)/2 = \tilde p /2 $.
Integrals of the form 
$(2\pi)^{-2}\int \exp(-\imath  \Delta \vec q \cdot  \vec r)d^2 \vec r = \delta
(\Delta \vec q) $, over spatial coordinates lead to the appearance in
(\ref{green-function-representation1_1-0}) the Dirac $\delta$-functions so that
\begin{eqnarray}
j_i^{Ohm}=\sum_{\mu} e {\chi^\mu}^{\dagger}_{+\sigma_{_B} } (x^+)
v^i_{x^+x^-} \chi^\mu_{+\sigma_{_B} }(x^{-})
={(-1)\imath e^2 A_i\over (\imath)^3c(2\pi)^{
10}}
\nonumber \\
\times
\mbox{Tr} \left\{\imath\int
d\bar{t}\, 
dt'\, d^2\vec {r}\,'
dt''\, 
\left(1-
  \widetilde {\Sigma_{AB}\Sigma_{BA}}(\vec p_{AB}- e\vec A/c )
   {d^2\widetilde {\Sigma_{AB} \Sigma_{BA}}\over d   p_i^{AB} d p_i^{AB}}
  (0)\right)\right.
\int d^2\vec p\
e^{-\imath \vec p\cdot 
\vec r^+  
} {v^\dagger }^i_{\vec {p}^+  - \vec {\bar p}, \vec p}
\nonumber \\
\times {1\over
-\imath \bar{\beta} \hbar }\sum _{\nu_{12}} \int \delta (\vec p -
\vec {\bar p}) d^2\vec p\,^+ d^2\vec {\bar p} d^2\vec {p'} d^2\vec
{p''} e^{-\imath \left( \vec {p^+}\cdot \vec r^+
-\vec {p'}\cdot\vec {r'} \right) }
e^{-\imath (z_{12}+\omega_{12})(t^+ - t^- - t' +\bar t)}
G_2(z_{12}, p^+, \bar p, p', p'')
 \nonumber \\
\times \left. {1\over -\imath \bar{\beta} \hbar }\sum _{\nu^{+-}}
\int d^2\vec p\,^-  \delta (\vec {p''} - \vec {p^-}) e^{-\imath
(z^{+-}+\omega^{+-})(t'' - t')}
e^{-\imath \vec {p}\,^-\cdot (
- \vec {r'})} G_1(\hat p^+, \hat p^-, z^{+-} )
 \right\} 
 \int d^2\vec {\tilde p}
e^{-\imath \vec {\tilde p} \cdot(\vec r\, ' - \vec r\,^-)}
v^i_{\vec {p}\, ' -\vec {p}^-, \vec {\tilde p}/2}
\nonumber 
\end{eqnarray}
\begin{eqnarray}
={(-1) \imath e^2 A_i\over (\imath)^3c(2\pi)^{
10}}\mbox{Tr} \left\{\imath\int
d\bar{t}\, 
dt'\, d^2\vec {r}\,'
dt''\, 
\left(1-
  \widetilde {\Sigma_{AB}\Sigma_{BA}}(\vec p_{AB}- e\vec A/c )
   {d^2\widetilde {\Sigma_{AB} \Sigma_{BA}}\over d   p_i^{AB} d p_i^{AB}}
  (0)\right)\right.
\nonumber \\
\times
 \int d^2\vec p\
e^{-\imath \vec p\cdot 
\vec r^+  
}
{v^\dagger }^i_{\vec {p}^+  - \vec {
p}, \vec  p} {1\over -\imath \bar{\beta} \hbar }\sum _{\nu_{12}}
\int 
d^2\vec p\, ^+ 
d^2\vec {p'} 
e^{-\imath \left( \vec p\, ^+ \cdot \vec r^+
-\vec {p'}\cdot\vec {r'} \right) }
e^{-\imath (z_{12}+\omega_{12})(t^+ - t^- - t' +\bar t)}
G_2(z_{12}, p^+, p
, p', p^-)
 \nonumber \\
\times \left. {1\over -\imath \bar{\beta} \hbar }\sum _{\nu^{+-}}
\int d^2\vec p\,^-  
e^{-\imath (z^{+-}+\omega^{+-})(t'' - t')}
e^{-\imath \vec {p}\,^- 
\cdot (
- \vec {r'})} G_1(\hat p^+, \hat p^-, z^{+-} )
 \right\} 
 \int d^2\vec {\tilde p}
e^{-\imath \vec {\tilde p} \cdot(\vec r\, ' - \vec r\,^-)}
v^i_{
\vec {p'} -\vec {p}^-, \vec {\tilde p}/2}
 \ .\ \
\label{green-function-representation1_1-1}
\end{eqnarray}
Integration of 
(\ref{green-function-representation1_1-1}) over the rest spatial variables
$\vec{r'}$ gives 
\begin{eqnarray}
j_i^{Ohm}=\sum_{\mu} e {\chi^\mu}^{\dagger}_{+\sigma_{_B} } (x^+)
v^i_{x^+x^-} \chi^\mu_{+\sigma_{_B} }(x^{-})
={(-1)\imath e^2 A_i\over (\imath)^3c(2\pi)^{
8}}\mbox{Tr} \left\{\imath\int
d\bar{t}\, 
dt'\, 
dt''\, 
\left(1-
  \widetilde {\Sigma_{AB}\Sigma_{BA}}(\vec p_{AB}- e\vec A/c )
\right. \right. \nonumber
\end{eqnarray}
\begin{eqnarray}
\times \left.
   {d^2\widetilde {\Sigma_{AB} \Sigma_{BA}}\over d   p_i^{AB} d p_i^{AB}}
  (0)\right) \int d^2\vec p\
e^{-\imath \vec p\cdot 
\vec r^+  
}
{v^\dagger }^i_{\vec {p}^+  - \vec {
p}, \vec  p} {1\over -\imath \bar{\beta} \hbar }\sum _{\nu_{12}}
\int 
d^2\vec p\, ^+ 
d^2\vec {p'} 
e^{-\imath 
\vec p\, ^+ \cdot \vec r^+
} e^{-\imath (z_{12}+\omega_{12})(t^+ - t^- - t' +\bar t)}
\nonumber \\
\times
G_2(z_{12}, p^+, p
, p', p^-)
\left. {1\over -\imath \bar{\beta} \hbar }\sum _{\nu^{+-}}
\int d^2\vec p\,^-  
e^{-\imath (z^{+-}+\omega^{+-})(t'' - t')}
G_1(\hat p^+, \hat p^-, z^{+-} )
 \right\} %
 \int d^2\vec {\tilde p}\
 \delta (\vec {\tilde p} - \vec {p}\,^- - \vec {p'})
\nonumber \\ \times
e^{-\imath \vec {\tilde p} \cdot(
- \vec r\,^-)}
v^i_{
\vec {p'} -\vec {p}^-, \vec {\tilde p}/2}
={(-1) \imath e^2 A_i\over (\imath)^3c(2\pi)^{
8}}\mbox{Tr} \left\{\imath\int
d\bar{t}\, 
dt'\, 
dt''\, 
\left(1-
  \widetilde {\Sigma_{AB}\Sigma_{BA}}(\vec p_{AB}- e\vec A/c )
   {d^2\widetilde {\Sigma_{AB} \Sigma_{BA}}\over d   p_i^{AB} d p_i^{AB}}
  (0)\right)\right.
\nonumber \\
\times
 \int d^2\vec p\
e^{-\imath \vec p\cdot 
\vec r^+  
}
{v^\dagger }^i_{\vec {p}^+  - \vec {
p}, \vec  p} {1\over -\imath \bar{\beta} \hbar }\sum _{\nu_{12}}
\int 
d^2\vec p\, ^+ 
d^2\vec {p'} 
e^{-\imath 
\vec p\, ^+ \cdot \vec r^+
}
e^{-\imath (z_{12}+\omega_{12})(t^+ - t^- - t' +\bar t)}
G_2(z_{12}, p^+, p
, p', p^-)
 \nonumber \\
\times \left. {1\over -\imath \bar{\beta} \hbar }\sum _{\nu^{+-}}
\int d^2\vec p\,^-  
e^{-\imath (z^{+-}+\omega^{+-})(t'' - t')}
G_1(\hat p^+, \hat p^-, z^{+-} )
 \right\} 
e^{-\imath 
(\vec {p}\,^- + \vec {p'})
\cdot(
- \vec r\,^-)}
v^i_{
\vec {p'} -\vec {p}^-, (\vec {p}\,^- + \vec {p'})/2}
 \ .\ \
\label{green-function-representation1_1-2}
\end{eqnarray}

Taking into account 
(\ref{mazubara-cyclic-frequency}), the integral of the form 
$$
(2\pi)^{-1}\int  \exp(-\imath  \Delta \omega t)\ d t
=(2\pi)^{-2}\int  \exp(-\imath  2\pi \Delta \nu t)\ d (2\pi t) =
(2\pi)^{-1}\delta (\Delta \nu)
$$
over time variables leads again to the Dirac $\delta$-functions
in 
(\ref{green-function-representation1_1-2})  
 $\delta (\Delta \omega) $ and we obtain:
 \begin{eqnarray}
j_i^{Ohm}=\sum_{\mu} e {\chi^\mu}^{\dagger}_{+\sigma_{_B} } (x^+)
v^i_{x^+x^-} \chi^\mu_{+\sigma_{_B} }(x^{-})
={(-1)\imath e^2 A_i\over (\imath)^3c(2\pi)^{
8}}\mbox{Tr} \left\{\imath\int
d\bar{t}\, 
dt'\, 
dt''\, 
\left(1-
  \widetilde {\Sigma_{AB}\Sigma_{BA}}(\vec p_{AB}- e\vec A/c )
   {d^2\widetilde {\Sigma_{AB} \Sigma_{BA}}\over d   p_i^{AB} d p_i^{AB}}
  (0)\right)\right.
\nonumber \\
\times
 \int d^2\vec p\
e^{-\imath \vec p\cdot 
\vec r^+  
}
{v^\dagger }^i_{\vec {p}^+  - \vec {
p}, \vec  p} {1\over -\imath \bar{\beta} \hbar }\sum _{\nu_{12}}
\int 
d^2\vec p\, ^+ 
d^2\vec {p'} 
e^{-\imath 
\vec p\, ^+ \cdot \vec r^+
}
e^{-\imath (z_{12}+2\pi \nu_{1(2)})(t^+ - t^- - t' +\bar t)}
G_2(z_{12}, p^+, p
, p', p^-)
 \nonumber \\
\times \left. {1\over -\imath \bar{\beta} \hbar }\sum _{\nu^{+-}}
\int d^2\vec p\,^-  
e^{-\imath (z^{+-}+2\pi \nu^{+(-)})(t'' - t')}
G_1(\hat p^+, \hat p^-, z^{+-} )
 \right\} 
e^{-\imath 
(\vec {p}\,^- + \vec {p'})
\cdot(
- \vec r\,^-)}
v^i_{
\vec {p'} -\vec {p}^-, (\vec {p}\,^- + \vec {p'})/2}
\nonumber \\
={(-1)\imath e^2 A_i\over (\imath)^3c(2\pi)^{
8}}\mbox{Tr} \left\{\imath {1\over -\imath \hbar \bar \beta}
\int_0^{-\imath \hbar \bar \beta}
d\bar{t}\, 
\left(1-
  \widetilde {\Sigma_{AB}\Sigma_{BA}}(\vec p_{AB}- e\vec A/c )
   {d^2\widetilde {\Sigma_{AB} \Sigma_{BA}}\over d   p_i^{AB} d p_i^{AB}}
  (0)\right)\right.
 \int d^2\vec p\
e^{-\imath \vec p\cdot 
\vec r^+  
}
{v^\dagger }^i_{\vec {p}^+  - \vec {
p}, \vec  p}
\nonumber \\
\times
 {1\over -\imath \bar{\beta} \hbar }\sum _{\nu_{12}}
\int 
d^2\vec p\, ^+ 
d^2\vec {p'} 
e^{-\imath 
\vec p\, ^+ \cdot \vec r^+
}
\delta\left(\hbar \nu_{12}+{\mu_{12}\over 2\pi}+ \hbar
\nu^{+-}+{\mu^{+-}\over 2\pi}\right)
 e^{-\imath (z_{12}+2\pi \nu_{1(2)})(t^+ - t^- 
+\bar t)}
G_2(z_{12}, p^+, p
, p', p^-)
 \nonumber \\
\times \left. {1\over -\imath \bar{\beta} \hbar }\sum _{\nu^{+-}}
\int d^2\vec p\,^-  
\delta\left(\hbar \nu^{+-}+{\mu^{+-}\over 2\pi}\right)
G_1(\hat p^+, \hat p^-, z^{+-} )
 \right\} 
e^{-\imath 
(\vec {p}\,^- + \vec {p'})
\cdot(
- \vec r\,^-)}
v^i_{
\vec {p'} -\vec {p}^-, (\vec {p}\,^- + \vec {p'})/2}
 \ .\ \ \ \ \ \
\label{green-function-representation1_1-3}
\end{eqnarray}
Performing the change 
$\delta\left(\hbar \nu_{12}+{\mu_{12}\over
2\pi}+\hbar \nu^{+-}+ {\mu^{+-}\over
2\pi}\right)\to\delta\left(z_{12}+z^{+-}\right)$, one transforms 
Eq.~(\ref{green-function-representation1_1-3}) as follows
 \begin{eqnarray}
j_i^{Ohm}=\sum_{\mu} e {\chi^\mu}^{\dagger}_{+\sigma_{_B} } (x^+)
v^i_{x^+x^-} \chi^\mu_{+\sigma_{_B} }(x^{-})
\nonumber \\
={(-1)\imath e^2 A_i\over (\imath)^3c(2\pi)^{
8}}\mbox{Tr} \left\{\imath {1\over -\imath \hbar \bar \beta}
\int_0^{-\imath \hbar \bar \beta}
d\bar{t}\, 
\left(1-
  \widetilde {\Sigma_{AB}\Sigma_{BA}}(\vec p_{AB}- e\vec A/c )
   {d^2\widetilde {\Sigma_{AB} \Sigma_{BA}}\over d   p_i^{AB} d p_i^{AB}}
  (0)\right)\right.
\nonumber \\
\times
 \int d^2\vec p\
e^{-\imath \vec p\cdot 
\vec r^+  
}
{v^\dagger }^i_{\vec {p}^+  - \vec {
p}, \vec  p} {1\over -\imath \bar{\beta} \hbar }\sum _{\nu_{12}}
\int 
d^2\vec p\, ^+ 
d^2\vec {p'} 
e^{-\imath 
\vec p\, ^+ \cdot \vec r^+
}
 e^{\imath (z^{+-} + \pi \nu^{+-})(t^+ - t^- 
+\bar t)}
G_2(z^{-+}, p^+, p
, p', p^-)
 \nonumber \\
\times \left. {1\over -\imath \bar{\beta} \hbar }\sum _{\nu^{+-}}
\int d^2\vec p\,^-  
\delta\left(\hbar \nu^{+-}+{\mu^{+-}\over 2\pi}\right)
G_1(\hat p^+, \hat p^-, z^{+-} )
 \right\} 
e^{-\imath 
(\vec {p}\,^- + \vec {p'})
\cdot(
- \vec r\,^-)}
v^i_{
\vec {p'} -\vec {p}^-, (\vec {p}\,^- + \vec {p'})/2}
 \ ,\ \
\label{green-function-representation1_1-4}
\end{eqnarray}
where $z^{-+}=-z^{+-},\ \nu^{-+}=-\nu^{+-}$. Based on 
the following change: 
$$
{1\over -\imath \bar{\beta} \hbar }\sum _{\nu_{12}} \to {1\over
-\imath \bar{\beta} \hbar 2\pi} \int_0^{-\imath \bar{\beta} \hbar}
d(2\pi \nu_{1(2)})= {1\over -\imath \bar{\beta} \hbar 2\pi}
\int_0^{-\imath \bar{\beta} \hbar} d \, \omega_{12} ={1 \over
2\pi},
$$
the subsequent integration on variable 
$\bar t$  in 
(\ref{green-function-representation1_1-4})
$$
\int_0^{-\imath \hbar \bar \beta} d\bar{t}\, \left.
e^{\imath(z^{+-}+\omega^{+-})\bar t}\right|_{z^{+-}\to 0} ={1\over
\imath }\left. {\exp\{[{2\pi\imath \, \nu^{+(-)}\over \hbar \beta}
+\omega^{+-}]\hbar \bar \beta\}-1 \over
(z^{+-}+\omega^{+-})}\right|_{z^{+-}\to 0} ={\hbar \bar \beta\over
\imath }
$$
leads to the expression for the current 
\begin{eqnarray}
j_i^{Ohm}=\sum_{\mu} e {\chi^\mu}^{\dagger}_{+\sigma_{_B} } (x^+)
v^i_{x^+x^-} \chi^\mu_{+\sigma_{_B} }(x^{-})
={(-1)\imath ^2 e^2 A_i\over (\imath)^3c(2\pi)^{9}}\mbox{Tr}
\left\{ {1\over -\imath \bar{\beta} \hbar }\sum _{\nu^{+-}}
e^{\imath (z^{+-} )(t^+ - t^- )} \int d^2\vec p\, ^+ e^{-\imath
\vec p\, ^+ \cdot \vec r^+}  \right. \nonumber
\\ \times
\int d^2\vec p\,^- e^{\imath \vec {p}\,^- \cdot \vec r\,^-} \int
d^2\vec p\ d^2\vec {p'}
 e^{-\imath \vec p\cdot \vec r^+  }e^{\imath \vec {p'} \cdot \vec r\,^-}
\left(1-
  \widetilde {\Sigma_{AB}\Sigma_{BA}}(\vec p_{AB}- e\vec A/c )
   {d^2\widetilde {\Sigma_{AB} \Sigma_{BA}}\over d   p_i^{AB} d p_i^{AB}}
  (0)\right)
\nonumber \\
\times {v^\dagger }^i_{\vec {p}^+  - \vec {p}, \vec  p} \,
 G_2(z^{-+}, p^+, p, p', p^-)
\left.
G_1(\hat p^+, \hat p^-, z^{+-} )
 \right\} 
v^i_{\vec {p'} -\vec {p}^-, (\vec {p}\,^- + \vec {p'})/2}
\delta\left(\hbar \omega^{+-}+ \mu^{+-}\right)
 \ .\ \
\label{green-function-representation1_1-5}
\end{eqnarray}
Substituting 
(\ref{Mazubara-Laplas-Fourier-transform1}, \ref{Mazubara-spectral-function}, \ref{HF-Mazubara-spectral-function}) 
into (\ref{green-function-representation1_1-5}), we find finally 
\begin{eqnarray}
j_i^{Ohm}=\sum_{\mu} e {\chi^\mu}^{\dagger}_{+\sigma_{_B} } (x^+)
v^i_{x^+x^-} \chi^\mu_{+\sigma_{_B} }(x^{-})
={(-1)\imath ^2 e^2 A_i\over (\imath)^3c(2\pi)^{10}\hbar}\mbox{Tr}
\left\{ {1\over -\imath \bar{\beta} \hbar }\sum _{\nu^{+-}}
e^{\imath (z^{+-} )(t^+ - t^- )}  \right. \nonumber
\\ \times
\int d^2\vec p\, ^+ e^{-\imath \vec p\, ^+ \cdot \vec r^+} \int
d^2\vec p\,^- e^{\imath \vec {p}\,^- \cdot \vec r\,^-}
 \int d^2\vec p\ d^2\vec {p'}
 e^{-\imath \vec p\cdot \vec r^+  }e^{\imath \vec {p'} \cdot \vec r\,^-}
\left(1-
  \widetilde {\Sigma_{AB}\Sigma_{BA}}(\vec p_{AB}- e\vec A/c )
   {d^2\widetilde {\Sigma_{AB} \Sigma_{BA}}\over d   p_i^{AB} d p_i^{AB}}
  (0)\right)
\nonumber \\
\times {v^\dagger }^i_{\vec {p}^+  - \vec {p}, \vec  p} \,
 G_2(z^{-+}, p^+, p, p', p^-)
%
\int {d \omega\over \imath (z^{+-} - \omega ) } \left[ \delta
(\hbar \omega - E(\hat p^+, \hat p^-)) \right.
\nonumber \\
\left. \left. + \left( {\Gamma (\hat p^+, \hat p^-, \ \omega)
\over [\hbar \omega - E(\hat p^+, \hat p^-)- \Re e \,\Sigma^C
(\hat p^+, \hat p^-, \ \omega)]^2 + [\Gamma (\hat p^+, \hat p^-, \
\omega)/2]^2} - \delta (\hbar \omega - E(\hat p^+, \hat p^-))
\right) \right]
 \right\} %
v^i_{\vec {p'} -\vec {p}^-, (\vec {p}\,^- + \vec {p'})/2}
\nonumber \\
\times \delta\left(\hbar \omega^{+-}+ \mu^{+-}\right)
\nonumber \\
={(-1)\imath ^2 e^2 A_i\over (\imath)^3c(2\pi)^{10}\hbar}\mbox{Tr}
\left\{ {1\over -\imath \bar{\beta} \hbar }\sum _{\nu^{+-}}
e^{\imath (z^{+-} )(t^+ - t^- )} \int d^2\vec p\, ^+ e^{-\imath
\vec p\, ^+ \cdot \vec r^+} \int d^2\vec p\,^- e^{\imath \vec
{p}\,^- \cdot \vec r\,^-} \right.
 \int d^2\vec p\ d^2\vec {p'}
 e^{-\imath \vec p\cdot \vec r^+  }e^{\imath \vec {p'} \cdot \vec r\,^-}
\nonumber
\\ \times
\left(1-
  \widetilde {\Sigma_{AB}\Sigma_{BA}}(\vec p_{AB}- e\vec A/c )
   {d^2\widetilde {\Sigma_{AB} \Sigma_{BA}}\over d   p_i^{AB} d p_i^{AB}}
  (0)\right)
{v^\dagger }^i_{\vec {p}^+  - \vec {p}, \vec  p} \,
 G_2(z^{-+}, p^+, p, p', p^-)
\left[ {1\over \imath (z^{+-} - \omega (\hat p^+, \hat p^-)) }
\right.\nonumber \\
\left. \left. + 2\pi \left({ \Gamma (\hat p^+, \hat p^-, \ z^{+-})
 \over [\hbar z^{+-}
- E(\hat p^+, \hat p^-)- \Re e \,\Sigma^C (\hat p^+, \hat p^-, \
z^{+-})]^2 + [\Gamma (\hat p^+, \hat p^-, \ z^{+-})/2]^2} - \delta
(\hbar z^{+-} - E(\hat p^+, \hat p^-))\right)\right]
 \right\} %
 \nonumber \\ \times
v^i_{\vec {p'} -\vec {p}^-, (\vec {p}\,^- + \vec {p'})/2}
\delta\left(\hbar \omega^{+-}+ \mu^{+-}\right)
 \ ,\ \
\label{green-function-representation1_1-6}
\end{eqnarray}
where $\omega (\hat p^+, \hat p^-)=E(\hat p^+, \hat p^-)/\hbar $.

The product 
$\delta (\hbar z^{+-} - E(\hat p^+, \hat p^-))$ and $\delta (\hbar \omega^{+-} + \mu^{+-})$ in
(\ref{green-function-representation1_1-6}) gives vanishing contribution 
\begin{eqnarray}
j_i^{Ohm}=\sum_{\mu} e {\chi^\mu}^{\dagger}_{+\sigma_{_B} } (x^+)
v^i_{x^+x^-} \chi^\mu_{+\sigma_{_B} }(x^{-})
\nonumber \\
={(-1)\imath ^2 e^2 A_i\over (\imath)^3c(2\pi)^{10}\hbar}\mbox{Tr}
\left\{ {1\over -\imath \bar{\beta} \hbar }\sum _{\nu^{+-}}
e^{\imath (z^{+-} )(t^+ - t^- )} \int d^2\vec p\, ^+ e^{-\imath
\vec p\, ^+ \cdot \vec r^+} \int d^2\vec p\,^- e^{\imath \vec
{p}\,^- \cdot \vec r\,^-}
 \int d^2\vec p\ d^2\vec {p'}
 e^{-\imath \vec p\cdot \vec r^+  }e^{\imath \vec {p'} \cdot \vec r\,^-}
\right.
\nonumber \\
\times \left(1-
  \widetilde {\Sigma_{AB}\Sigma_{BA}}(\vec p_{AB}- e\vec A/c )
   {d^2\widetilde {\Sigma_{AB} \Sigma_{BA}}\over d   p_i^{AB} d p_i^{AB}}
  (0)\right)
{v^\dagger }^i_{\vec {p}^+  - \vec {p}, \vec  p} \,
 G_2(z^{-+}, p^+, p, p', p^-)
\left[ {1\over \imath (z^{+-} - \omega (\hat p^+, \hat p^-)) }
\right.\nonumber \\
\left. \left. + { 2\pi\Gamma (\hat p^+, \hat p^-, \ z^{+-})
 \over
[\hbar z^{+-} - E(\hat p^+, \hat p^-)- \Re e \,\Sigma^c (\hat p^+,
\hat p^-, \ z^{+-})]^2 + [\Gamma (\hat p^+, \hat p^-, \
z^{+-})/2]^2} \right]
 \right\} %
v^i_{\vec {p'} -\vec {p}^-, (\vec {p}\,^- + \vec {p'})/2}
\delta\left(\hbar \omega^{+-}+ \mu^{+-}\right)
 \ .\ \
\label{green-function-representation1_1-7_0}
\end{eqnarray}
The presence of 
$\delta\left(\hbar \omega^{+-}+ \mu^{+-}\right)$ in
(\ref{green-function-representation1_1-7_0}) allows to state on possibility of
formation of long-time order in considered macroscopic phenomena 
\cite{Callen-et-al-1967}.

The Hamiltonian $H$ satisfies the equations of motion 
\begin{eqnarray}
\imath \hbar {\partial \over \partial t^{\pm}} \chi_p(\vec r,\,
t^{\pm}) = H(p^{\pm})\,
 e^{\imath \omega^{\pm}t^{\pm}}\chi_p(\vec r), \ t^+> 0, \ t^- < 0
\end{eqnarray}
for electrons and holes respectively. 
Therefore the frequencies 
$\hat\omega^{+},\ \hat\omega^{-}$ satisfy the expressions 
$\hat\omega^{+}\equiv  \omega^e = H(p^{+})/\hbar $,
$\hat\omega^{-}\equiv  \omega^h= H^{\dagger}(-p^{-})/\hbar $, that allows to rewrite 
(\ref{green-function-representation1_1-7_0}) as
\begin{eqnarray}
j_i^{Ohm}=\sum_{\mu} e {\chi^\mu}^{\dagger}_{+\sigma_{_B} } (x^+)
v^i_{x^+x^-} \chi^\mu_{+\sigma_{_B} }(x^{-})
\nonumber \\
={(-1)\imath ^2 e^2 A_i\over (\imath)^3c(2\pi)^{10}\hbar}\mbox{Tr}
\left\{ {1\over -\imath \bar{\beta} \hbar }\sum _{\nu^{+-}}
e^{\imath (z^{+-} )(t^+ - t^- )} \int d^2\vec p\, ^+ e^{-\imath
\vec p\, ^+ \cdot \vec r^+} \int d^2\vec p\,^- e^{\imath \vec
{p}\,^- \cdot \vec r\,^-}
 \int d^2\vec p\ d^2\vec {p'}
 e^{-\imath \vec p\cdot \vec r^+  }e^{\imath \vec {p'} \cdot \vec r\,^-}
\right. \nonumber
\\ \times
\left(1-
  \widetilde {\Sigma_{AB}\Sigma_{BA}}(\vec p_{AB}- e\vec A/c )
   {d^2\widetilde {\Sigma_{AB} \Sigma_{BA}}\over d   p_i^{AB} d p_i^{AB}}
  (0)\right)
{v^\dagger }^i_{\vec {p}^+  - \vec {p}, \vec  p} \,
 G_2(z^{-+}, p^+, p, p', p^-)
\left[ {1\over \imath (z^{+-} - \omega^e (\hat p^+) + \omega^h
(-\hat p^-)) }
\right.\nonumber \\
\left. \left. + { 2\pi\Gamma (\hat p^+, \hat p^-, \ z^{+-})
 \over
[\hbar z^{+-} - E(\hat p^+, \hat p^-)- \Re e \,\Sigma^c (\hat p^+,
\hat p^-, \ z^{+-})]^2 + [\Gamma (\hat p^+, \hat p^-, \
z^{+-})/2]^2} \right]
 \right\} %
v^i_{\vec {p'} -\vec {p}^-, (\vec {p}\,^- + \vec {p'})/2}
\delta\left(\hbar \omega^{+-}+ \mu^{+-}\right)
 \ .\ \
\label{green-function-representation1_1-7}
\end{eqnarray}
For strongly correlated systems, which include 
Dirac materials, the energy 
$E(\hat p^+,\ \hat p^-) $ of the pair of a free electron and a hole 
does not exceed the correlational energy 
$\Re e\Sigma^c $
 \cite{MyQuantumMatter2015}
: $E(\hat p^+,\
\hat p^-)< \Re e\Sigma^c $.
The last allows to rewrite the current of electron-hole pairs 
(\ref{green-function-representation1_1-7}) in the form 
\begin{eqnarray}
j_i^{Ohm}=\sum_{\mu} e {\chi^\mu}^{\dagger}_{+\sigma_{_B} } (x^+)
v^i_{x^+x^-} \chi^\mu_{+\sigma_{_B} }(x^{-})
={(-1)\imath ^2 e^2 A_i\over (\imath)^3c(2\pi)^{10}\hbar}\mbox{Tr}
\left\{ {1\over -\imath \bar{\beta} \hbar }\sum _{\nu^{+-}}
e^{\imath (z^{+-} )(t^+ - t^- )} \int d^2\vec p\, ^+ e^{-\imath
\vec p\, ^+ \cdot \vec r^+}  \right. \nonumber
\\ \times
\int d^2\vec p\,^- e^{\imath \vec {p}\,^- \cdot \vec r\,^-} \int
d^2\vec p\ d^2\vec {p'}
 e^{-\imath \vec p\cdot \vec r^+  }e^{\imath \vec {p'} \cdot \vec r\,^-}
\left(1-
  \widetilde {\Sigma_{AB}\Sigma_{BA}}(\vec p_{AB}- e\vec A/c )
   {d^2\widetilde {\Sigma_{AB} \Sigma_{BA}}\over d   p_i^{AB} d p_i^{AB}}
  (0)\right)
\nonumber \\
\times {v^\dagger }^i_{\vec {p}^+  - \vec {p}, \vec  p} \,
 G_2(z^{-+}, p^+, p, p', p^-)
\left[ {\delta\left(\hbar \omega^{-+}+ \mu^{-+}\right)\over
-\imath (z^{-+} +\omega^e (\hat p^+) - \omega^h (-\hat p^-)) }
\right.\nonumber \\
\left. \left. + {2\pi \Gamma (\hat p^+, \hat p^-, \ z^{+-})
\delta\left(\hbar \omega^{+-}+ \mu^{+-}\right)
 \over
[\hbar z^{+-} 
- \Re e \,\Sigma^c (\hat p^+, \hat p^-, \ z^{+-})]^2 + \Gamma^2
(\hat p^+, \hat p^-, \ z^{+-})/4} \right]
 \right\} 
v^i_{\vec {p'} -\vec {p}^-, (\vec {p}\,^- + \vec {p'})/2}
 \ .\ \
\label{green-function-representation1_1-8}
\end{eqnarray}
Since $\Sigma^c (\hat p^+, \hat p^-, \ z^{+-}) =
\int_{-\infty}^{\infty}{d\omega \over 2\pi} {\Gamma (\hat p^+,
\hat p^-, \ \omega)\over z^{+-} - \omega}$, the correction
for the correlation interaction describes the decay 
$\Gamma$ 
and therefore influences only on initial conditions of macroscopic current appearance.
The charged-exciton model $N=3$ is protected by the hexagonal symmetry, therefore in subsequent we calculate the
Fourier component of the current without accounting of the
exciton decay as 
\begin{eqnarray}
j_i^{Ohm}(\vec r^+, \vec r^-, t^+ - t^-) =\sum_{\mu} e
{\chi^\mu}^{\dagger}_{+\sigma_{_B} } (x^+) v^i_{x^+x^-}
\chi^\mu_{+\sigma_{_B} }(x^{-})
\nonumber \\
=
  {1\over -\imath \bar{\beta} \hbar }\sum _{\nu^{+-}}
e^{-\imath z^{-+} (t^+ - t^-)}\int { d^2\vec p\, ^+\over
(2\pi)^{2}} e^{-\imath \vec p\, ^+ \cdot \vec r^+} \int  { d^2\vec
p\,^- \over (2\pi)^{2}} e^{\imath \vec {p}\,^- \cdot \vec r\,^-}
j_i^{Ohm}(z^{-+},\  \vec p\, ^+, \vec p\, ^- ),\ \
\label{green-function-representation1_1-9} \\
j_i^{Ohm}(z^{-+},\  \vec p\, ^+, \vec p\, ^- )
\nonumber \\
={(-1)\imath ^2 e^2 A_i\over (\imath)^3c(2\pi)^{6}}\mbox{Tr}
\left\{  
 \int d^2\vec p\ d^2\vec {p'}
 e^{-\imath \vec p\cdot \vec r^+  }e^{\imath \vec {p'} \cdot \vec r\,^-}
\left(1-
  \widetilde {\Sigma_{AB}\Sigma_{BA}}(\vec p_{AB}- e\vec A/c )
   {d^2\widetilde {\Sigma_{AB} \Sigma_{BA}}\over d   p_i^{AB} d p_i^{AB}}
  (0)\right)\right.
\nonumber \\
\left. \times {v^\dagger }^i_{\vec {p}^+  - \vec {p}, \vec  p} \,
 G_2(z^{-+}, p^+, p, p', p^-)
\left[ {\delta\left(\hbar \omega^{-+}+ \mu^{-+}\right)\over
-\imath \hbar (z^{-+} +\omega^e (\hat p^+) - \omega^h (-\hat p^-))
}
\right]
 \right\} 
v^i_{\vec {p'} -\vec {p}^-, (\vec {p}\,^- + \vec {p'})/2}
 \ .\ \
\label{green-function-representation1_1-10}
\end{eqnarray}

\subsection{Hartree--Fock approximation
}
In the Hartree--Fock approximation, the substitution of Eq.~
(\ref{perturbation_theory}) into
(\ref{green-function-representation1_1-10}) gives 
\begin{eqnarray}
j_i^{Ohm}(z^{-+},\  \vec p\, ^+, \vec p\, ^- )
={\imath^2  e^2 A_i\over c(2\pi)^{6}}\mbox{Tr}
\left\{  
 \int d^2\vec p\ d^2\vec {p'}
 e^{-\imath \vec p\cdot \vec r^+  }e^{\imath \vec {p'} \cdot \vec r\,^-}
\left(1-
  \widetilde {\Sigma_{AB}\Sigma_{BA}}(\vec p_{AB}- e\vec A/c )
   {d^2\widetilde {\Sigma_{AB} \Sigma_{BA}}\over d   p_i^{AB} d p_i^{AB}}
  (0)\right)\right.
\nonumber \\
\times {v^\dagger }^i_{\vec {p}^+  - \vec {p}, \vec  p} \,
 {1\over -\imath \bar{\beta} \hbar }\sum _{\nu^{+}}
 G_1(z^{-+} -z^+, p^- )G_1(z^+, p^+)
(2\pi)^4 \left( \delta (\vec { p'} - \vec { p^+}  ) \delta ( \vec
{ p} -\vec { p^-}  ) -\delta ( \vec { p'} - \vec { p^-}  ) \delta
( \vec { p} - \vec { p^+}  ) \right)
 \nonumber \\ \times
\left.{\delta\left(\hbar \omega^{-+}+ \mu^{-+}\right)\over  \hbar
(z^{-+} +\omega^e (\hat p^+) - \omega^h (-\hat p^-)) }
 \right\} 
v^i_{\vec {p'} -\vec {p}^-, (\vec {p}\,^- + \vec {p'})/2}
 \ .\ \
\label{Laplas-Fourier-transform5}
\end{eqnarray}
The matrix elements of the velocity operator have the following form:
\begin{equation}
\begin{split}
{v^\dagger }^i_{\vec {p}^+  - \vec {p}, \vec  p}= e^{-\imath (\vec
{p}^+  - \vec {p})\cdot \vec r}{v^\dagger }^i(p),\quad v^i_{\vec
{p}^+  - \vec {p}^-, \vec  p}= e^{\imath \vec k\cdot \vec r} v
^i(p)
\end{split}
\label{velocity-matrix-elements}
\end{equation}
where $\vec k= \vec {p}^+  - \vec {p}^-$, $\vec r =\vec {r}^+
-\vec {r}^-$.
For the two-particle Green's function represented by the following Feynman diagram:
\begin{eqnarray}
\begin{array}{c}
\vec {p'}  \longrightarrow \vec {p}^+\\[1mm]
 \ \vec {p} \longleftarrow \vec {p}^-
\end{array} \label{hartry-fock-diagram1}
\end{eqnarray}
the substitution of 
(\ref{velocity-matrix-elements}) into (\ref{Laplas-Fourier-transform5}) and integration over
$\delta(\vec p\,' - \vec p^+)$ first and then with 
$\delta(\vec p - \vec p^-)$ in the expression 
(\ref{green-function-representation1_1-9}, \ref{Laplas-Fourier-transform5}) gives 
\begin{eqnarray}
j_{i,1}^{Ohm}(\vec r^+, \vec r^-, t^+ - t^-) ={\imath ^2  e^2
A_i\over c(2\pi)^{2}}\mbox{Tr} {1\over -\imath \bar{\beta} \hbar
}\sum _{\nu^{+-}} e^{-\imath z^{-+} (t^+ - t^-)}
\int { d^2\vec p\, ^+\over (2\pi)^{2}} e^{-\imath \vec p\, ^+
\cdot \vec r^+} \int  { d^2\vec p\,^- \over 2\pi} e^{\imath \vec
{p}\,^- \cdot \vec r\,^-}
\nonumber \\
\times
\left\{  
 \int d^2\vec p\ \delta ( \vec { p} -\vec { p^-}  )
 e^{-\imath \vec p\cdot \vec r^+  }e^{\imath \vec {p}^+ \cdot \vec r\,^-}
\left(1-
  \widetilde {\Sigma_{AB}\Sigma_{BA}}(\vec p_{AB}- e\vec A/c )
   {d^2\widetilde {\Sigma_{AB} \Sigma_{BA}}\over d   p_i^{AB} d p_i^{AB}}
  (0)\right)\right.
\nonumber \\
\times {v^\dagger }^i_{\vec {p}^+  - \vec {p}, \vec  p} \,
 {1\over -\imath \bar{\beta} \hbar }\sum _{\nu_{+}}
 G_1(z^{-+} -z^+, p - k/2 )G_1(z^+, p^+)
\left.{\delta\left(\hbar \omega^{-+}+ \mu^{-+}\right)\over  \hbar
(z^{-+} +\omega^e (\hat p^+) - \omega^h (-\hat p^-)) }
 \right\} 
v^i_{\vec k, \vec p }
\nonumber\\
= {\imath ^2 e^2 A_i\over c(2\pi)^{2}}\mbox{Tr}
  {1\over -\imath \bar{\beta} \hbar }\sum _{\nu^{+-}}
e^{-\imath z^{-+} (t^+ - t^-)}
\int { d^2\vec p\, ^+\over (2\pi)^{2}} e^{-\imath \vec p\, ^+
\cdot \vec r^+}
e^{\imath \vec {p} \cdot \vec r\,^-}
%
%
\nonumber \\
\times
\left\{  
 \int d^2\vec p\ 
 e^{-\imath \vec p\cdot \vec r^+  }e^{\imath \vec {p}^+ \cdot \vec r\,^-}
 e^{-\imath (\vec {p}^+ -\vec p)\cdot \vec r  }
 e^{\imath  \vec k\cdot \vec r}
\left(1-
  \widetilde {\Sigma_{AB}\Sigma_{BA}}(\vec p_{AB}- e\vec A/c )
   {d^2\widetilde {\Sigma_{AB} \Sigma_{BA}}\over d   p_i^{AB} d p_i^{AB}}
  (0)\right)\right.
\nonumber \\
\times {v^\dagger }^i(p) \,
 {1\over -\imath \bar{\beta} \hbar }\sum _{\nu_{+}}
 G_1(z^{-+} -z^+, p^- )G_1(z^+, p^+)
\times \left.{\delta\left(\hbar \omega^{-+}+ \mu^{-+}\right)\over
\hbar (z^{-+} +\omega^e (\hat p^+) - \omega^h (-\hat p^-)) }
 \right\} 
v^i(p) = {\imath^2  e^2 A_i\over c(2\pi)^{2}}\mbox{Tr}
  {1\over -\imath \bar{\beta} \hbar }
  \nonumber\\ \times
\sum _{\nu^{+-}} e^{-\imath z^{-+} (t^+ - t^-)}
\int { d^2\vec p\, ^+\over (2\pi)^{2}}d^2\vec p\ e^{-\imath \vec
p\, ^+ \cdot (\vec r^+ -\vec r^- +\vec r )}
e^{\imath \vec {p} \cdot (\vec r^+ -\vec r^- -\vec r )}
 e^{\imath \vec k\cdot \vec r  }
%
\left\{  
 \left(1-
  \widetilde {\Sigma_{AB}\Sigma_{BA}}(\vec p_{AB}- e\vec A/c )
   {d^2\widetilde {\Sigma_{AB} \Sigma_{BA}}\over d   p_i^{AB} d p_i^{AB}}
  (0)\right)\right.
\nonumber \\
\times {v^\dagger }^i(p) \,
 {1\over -\imath \bar{\beta} \hbar }\sum _{\nu_{+}}
 G_1(z^{-+} -z^+, p^- )G_1(z^+, p^+)
\left.{\delta\left(\hbar \omega^{-+}+ \mu^{-+}\right)\over  \hbar
(z^{-+} +\omega^e (\hat p^+) - \omega^h (-\hat p^-)) }
 \right\} 
v^i(p). \ \ \ \ \ \
\label{green-function-representation1_1-9-diagram1}
\end{eqnarray}

Similarly, we can find the contribution of the second Feynman diagram for two-particle Green's function:
\begin{eqnarray}
\begin{array}{c}
\ \vec {p'} \ \ \ \ \ \ \vec { p}^+\\[-1mm]
\searrow\\[-4.5mm]
\swarrow\\[-2mm]
\ \ \vec {p}  \ \ \ \ \ \  \vec {p}\,^-
\end{array}  \label{hartry-fock-diagram2}
\end{eqnarray}
into Ohmic current 
\begin{eqnarray}
j_{i,2}^{Ohm}(\vec r^+, \vec r^-, t^+ - t^-) =-{\imath^2  e^2
A_i\over c(2\pi)^{2}}\mbox{Tr}
  {1\over -\imath \bar{\beta} \hbar }\sum _{\nu^{+-}}
e^{-\imath z^{-+} (t^+ - t^-)}
\int { d^2\vec p\, ^+\over (2\pi)^{2}} \delta (\vec { p} - \vec {
p^+}) e^{-\imath \vec p\, ^+ \cdot \vec r^+} \int  { d^2\vec p\,^-
\over 2\pi} e^{\imath \vec {p}\,^- \cdot \vec r\,^-}
%
 \nonumber \\ \times
\left\{  
 \int d^2\vec p 
 e^{-\imath \vec p\cdot \vec r^+  }e^{\imath \vec {p}^- \cdot \vec r\,^-}
\left(1-
  \widetilde {\Sigma_{AB}\Sigma_{BA}}(\vec p_{AB}- e\vec A/c )
   {d^2\widetilde {\Sigma_{AB} \Sigma_{BA}}\over d   p_i^{AB} d p_i^{AB}}
  (0)\right)\right.
\nonumber \\
\times
{v^\dagger }^i_{
\vec {p}+\vec k/2 - \vec {p}, \vec  p} \,
 {1\over -\imath \bar{\beta} \hbar }\sum _{\nu_{+}}
 G_1(z^{-+} -z^+, p^- )G_1(z^+, p+ k/2 
 )
\left.{\delta\left(\hbar \omega^{-+}+ \mu^{-+}\right)\over  \hbar
(z^{-+} +\omega^e (\hat p^+) - \omega^h (-\hat p^-)) }
 \right\}  v^i_{0, \vec {p}\,^- }
= -{\imath ^2 e^2 A_i\over c(2\pi)^{2}}
\nonumber\\
\times \mbox{Tr}
  {1\over -\imath \bar{\beta} \hbar }\sum _{\nu^{+-}}
e^{-\imath z^{-+} (t^+ - t^-)}
\int { d^2\vec p\, ^-\over (2\pi)^{2}} e^{\imath 2 \vec {p}\,^-
\cdot \vec r\,^-}
\left\{
 \int d^2\vec p
 e^{-\imath 2 \vec p\cdot \vec r^+  }
\left(1-
  \widetilde {\Sigma_{AB}\Sigma_{BA}}(\vec p_{AB}- e\vec A/c )
   {d^2\widetilde {\Sigma_{AB} \Sigma_{BA}}\over d   p_i^{AB} d p_i^{AB}}
  (0)\right)\right.
\nonumber \\ \times
{v^\dagger }^i_{
\vec {p}+\vec k/2 - \vec {p}, \vec  p} \,
 {1\over -\imath \bar{\beta} \hbar }\sum _{\nu_{+}}
 G_1(z^{-+} -z^+, p^- )G_1(z^+, p+ k/2 
 )
\left.{\delta\left(\hbar \omega^{-+}+ \mu^{-+}\right)\over  \hbar
(z^{-+} +\omega^e (\hat p^+) - \omega^h (-\hat p^-)) }
 \right\} 
v^i_{0, \vec {p}\,^- }
.\ \ \label{green-function-representation1_1-9-diagram2}
\end{eqnarray}

\subsection{Electrophysical  current }

For the case of spatial dispersion and transitions from the energy level
$E_D=0$ ($\vec p^+\to \vec k$, $ \vec p^- = 0$), the expression 
(\ref{green-function-representation1_1-9-diagram2})
vanishes due to the fact that one of the matrix element of the velocity operator has a constant value
$v^i_{0, \vec {p}\,^- } \to v^i_{0, 0 }$ at $\vec p^+\to \vec k$, $ \vec p^-\to 0$.
Therefore the expression 
(\ref{green-function-representation1_1-9-diagram1}) gives the minimal dc-conductivity in the limit
$p^+ \to k$
:
\begin{eqnarray}
\lim_{  p^+\to  k} \sum_{\mu =1}^2 j_{i,\mu}^{Ohm}(\vec r^+, \vec
r^-, t^+ - t^-) \equiv j_{i, \min }^{Ohm}(\vec r^+, \vec r^-, t^+
- t^-)
\nonumber\\
= {\imath ^2 e^2 A_i\over c(2\pi)^{2}}\mbox{Tr}
  {1\over -\imath \bar{\beta} \hbar }\sum _{\nu^{+-}}
e^{-\imath z^{-+} (t^+ - t^-)}
\int { d^2\vec p\, ^+\over (2\pi)^{2}}d^2\vec p\ e^{-\imath (2\vec
p\, ^+ - \vec k )\cdot \vec r }
%
\left\{  
 \left(1-
  \widetilde {\Sigma_{AB}\Sigma_{BA}}(\vec p_{AB}- e\vec A/c )
   {d^2\widetilde {\Sigma_{AB} \Sigma_{BA}}\over d   p_i^{AB} d p_i^{AB}}
  (0)\right)\right.
\nonumber \\
\times {v^\dagger }^i(p) \,
 {1\over -\imath \bar{\beta} \hbar }\sum _{\nu_{+}}
 G_1(z^{-+} -z^+, p^- )G_1(z^+, p^+)
\left.{\delta\left(\hbar \omega^{-+}+ \mu^{-+}\right)\over  \hbar
(z^{-+} +\omega^e (\hat p^+) - \omega^h (-\hat p^-)) }
 \right\} v^i(p) \nonumber \\
= {\imath ^2 e^2 A_i\over c(2\pi)^{2}}\mbox{Tr}
  {1\over -\imath \bar{\beta} \hbar }\sum _{\nu^{+-}}
e^{-\imath z^{-+} (t^+ - t^-)} \int e^{-\imath \vec k \cdot \vec r
}{ d^2\vec k \over (2\pi)^{2}} \int d^2\vec p\
\left\{  
 \left(1-
  \widetilde {\Sigma_{AB}\Sigma_{BA}}(\vec p_{AB}- e\vec A/c )
   {d^2\widetilde {\Sigma_{AB} \Sigma_{BA}}\over d   p_i^{AB} d p_i^{AB}}
  (0)\right)\right.
\nonumber \\
\times {v^\dagger }^i(p) \,
 {1\over -\imath \bar{\beta} \hbar }\sum _{\nu_{+}}
 G_1(z^{-+} -z^+, p^- )G_1(z^+, p^+)
\left.{\delta\left(\hbar \omega^{-+}+ \mu^{-+}\right)\over  \hbar
(z^{-+} +\omega^e (\hat p^+) - \omega^h (-\hat p^-)) }
 \right\} v^i(p) .
\ \ \label{green-function-representation1_1-9-diagram1_1}
\end{eqnarray}

\subsection{ Optical current }

Transitions from low-lying energy levels $ E \to E_D $ occur at
excitation by electromagnetic radiation of the optical range. Spatial dispersion is small at optical frequencies,
which are high relative to those used in electrophysics.
In the zeroth approximation, the dynamic (optical) conductivity is calculated in the limit
$p^- \to p $ at $ k \to 0$. 
To find  contribution of the first Feynman diagram
(\ref{hartry-fock-diagram1}) into optical current 
 $j_{i}^{opt}$, let us rewrite 
(\ref{green-function-representation1_1-9-diagram1}) in the following form: 
\begin{eqnarray}
\lim_{  k \to 0} j_{i,1}^{Ohm}(\vec r^+, \vec r^-, t^+ - t^-)
\equiv   j_{i, 1}^{opt}(\vec r^+, \vec r^-, t^+ - t^-)
\nonumber\\
= {\imath ^2 e^2 A_i\over c(2\pi)^{2}}\mbox{Tr}
  {1\over -\imath \bar{\beta} \hbar }\sum _{\nu^{+-}}
e^{-\imath z^{-+} (t^+ - t^-)}
\int { [d^2\vec k + d^2\vec p\, ^- ]\over (2\pi)^{2}}d^2\vec p\
e^{-\imath (\vec k + 2\vec p\, ^-   )\cdot \vec r }
%
\left\{
 \left(1-
  \widetilde {\Sigma_{AB}\Sigma_{BA}}(\vec p_{AB}- e\vec A/c )
   {d^2\widetilde {\Sigma_{AB} \Sigma_{BA}}\over d   p_i^{AB} d p_i^{AB}}
  (0)\right)\right.
\nonumber \\
\times {v^\dagger }^i(p) \,
 {1\over -\imath \bar{\beta} \hbar }\sum _{\nu_{+}}
 G_1(z^{-+} -z^+, p^- )G_1(z^+, p^+)
\left.{\delta\left(\hbar \omega^{-+}+ \mu^{-+}\right)\over  \hbar
(z^{-+} +\omega^e (\hat p^+) - \omega^h (-\hat p^-)) }
 \right\} v^i(p)  .
\ \ \ \ \ \
\label{optical-current1}
\end{eqnarray}
Due to equality to zero of 
(\ref{green-function-representation1_1-9-diagram2}) at $p^- = 0 $, the expression 
(\ref{optical-current1}) gives the main contribution to 
$j_{i}^{opt}$ at $p\to 0$:
\begin{eqnarray}
\lim_{  k, p \to  0} \sum_{\mu =1}^2 j_{i,\mu}^{Ohm}(\vec r^+,
\vec r^-, t^+ - t^-) \equiv \lim_{ p,\ p^- \to  0} j_{i,1}^{
opt}(\vec r^+, \vec r^-, t^+ - t^-)
\nonumber \\
= {\imath ^2 e^2 A_i\over c(2\pi)^{2}}\mbox{Tr}
  {1\over -\imath \bar{\beta} \hbar }\sum _{\nu^{+-}}
e^{-\imath z^{-+} (t^+ - t^-)}
\int { [d^2\vec k + d^2(\vec p- \vec k/2 )]\over
(2\pi)^{2}}d^2\vec p\ e^{-\imath \vec k \cdot \vec r }
%
\left\{
 \left(1-
  \widetilde {\Sigma_{AB}\Sigma_{BA}}(\vec p_{AB}- e\vec A/c )
   {d^2\widetilde {\Sigma_{AB} \Sigma_{BA}}\over d   p_i^{AB} d p_i^{AB}}
  (0)\right)\right.
\nonumber \\
\times {v^\dagger }^i(p) \,
 {1\over -\imath \bar{\beta} \hbar }\sum _{\nu_{+}}
 G_1(z^{-+} -z^+, p^- )G_1(z^+, p^+)
\left.{\delta\left(\hbar \omega^{-+}+ \mu^{-+}\right)\over  \hbar
(z^{-+} +\omega^e (\hat p^+) - \omega^h (-\hat p^-)) }
 \right\} v^i(p)
 \nonumber \\
= {\imath ^2 e^2 A_i\over 2 c(2\pi)^{2}}\mbox{Tr}
  {1\over -\imath \bar{\beta} \hbar }\sum _{\nu^{+-}}
e^{-\imath z^{-+} (t^+ - t^-)}
\int {  d^2\vec k \over (2\pi)^{2}}d^2\vec p\ e^{-\imath  \vec k
\cdot \vec r }
%
\left\{  
 \left(1-
  \widetilde {\Sigma_{AB}\Sigma_{BA}}(\vec p_{AB}- e\vec A/c )
   {d^2\widetilde {\Sigma_{AB} \Sigma_{BA}}\over d   p_i^{AB} d p_i^{AB}}
  (0)\right)\right.
\nonumber \\
\times {v^\dagger }^i(p) \,
 {1\over -\imath \bar{\beta} \hbar }\sum _{\nu_{+}}
 G_1(z^{-+} -z^+, p^- )G_1(z^+, p^+)
\left.{\delta\left(\hbar \omega^{-+}+ \mu^{-+}\right)\over  \hbar
(z^{-+} +\omega^e (\hat p^+) - \omega^h (-\hat p^-)) }
 \right\} v^i(p) . \ \ \ \ \ \
\label{optical-current2}
\end{eqnarray}

Contribution (\ref{green-function-representation1_1-9-diagram2}) of the second Feynman diagram
(\ref{hartry-fock-diagram2}) gives 
the following correction to optical current at small values of
$p^- \ll 1$, $k= 0$:
\begin{eqnarray}
j_{i,2}^{ opt}(\vec r^+, \vec r^-, t^+ - t^-) = -{\imath ^2 e^2
A_i\over c(2\pi)^{2}}\mbox{Tr}
  {1\over -\imath \bar{\beta} \hbar }\sum _{\nu^{+-}}
e^{-\imath z^{-+} (t^+ - t^-)}
\int { d^2\vec p\, ^-\over (2\pi)^{2}} e^{-\imath  \vec {p}\,^-
\cdot (\vec r\,^+ -\vec r\,^-)} \nonumber
\end{eqnarray}
\begin{eqnarray}
\times
\left\{  
 \int d^2\vec p
 e^{-\imath  \vec p\cdot (\vec r\,^+ -\vec r\,^-) }
\left(1-
  \widetilde {\Sigma_{AB}\Sigma_{BA}}(\vec p_{AB}- e\vec A/c )
   {d^2\widetilde {\Sigma_{AB} \Sigma_{BA}}\over d   p_i^{AB} d p_i^{AB}}
  (0)\right)\right.
 {1\over -\imath \bar{\beta} \hbar }\sum _{\nu_{+}}
 G_1(z^{-+} -z^+, p^- )G_1(z^+, p^+
 )
 \nonumber
\\
\times
\left.{\delta\left(\hbar \omega^{-+}+ \mu^{-+}\right)\over  \hbar
(z^{-+} +\omega^e (\hat p^+) - \omega^h (-\hat p^-)) }
 \right\} 
v^i_{0, \vec {p}\,^- }
= -{\imath^2  e^2 A_i\over c(2\pi)^{2}}\mbox{Tr}
  {1\over -\imath \bar{\beta} \hbar }\sum _{\nu^{+-}}
e^{-\imath z^{-+} (t^+ - t^-)}
\int { d^2\vec p \over (2\pi)^{2}} e^{-\imath  \vec {p} \cdot \vec
r}
\nonumber \\
\times
\left\{  
 \int d^2\vec {\tilde p}
 e^{-\imath  \vec {\tilde p}\cdot \vec r }
\left(1-
  \widetilde {\Sigma_{AB}\Sigma_{BA}}(\vec p_{AB}- e\vec A/c )
   {d^2\widetilde {\Sigma_{AB} \Sigma_{BA}}\over d   p_i^{AB} d p_i^{AB}}
  (0)\right)\right.
\nonumber \\
\times {v^\dagger }^i({\tilde p}) \,
 {1\over -\imath \bar{\beta} \hbar }\sum _{\nu_{+}}
 G_1(z^{-+} -z^+, {\tilde p}^- )G_1(z^+, {\tilde p}^+
 )
\left.{\delta\left(\hbar \omega^{-+}+ \mu^{-+}\right)\over  \hbar
(z^{-+} +\omega^e (\hat {\tilde p}^+) - \omega^h (-\hat {\tilde
p}^-)) }
 \right\} 
v^i({\tilde p})
, 
\ \ \label{optic-current-correction}
\end{eqnarray}
where $\tilde p = p^-$. Taking into account that 
$\tilde p < p$, and redesignating 
$p$ as 
$k$, 
we obtain the final expression for  $j_{i,2}^{ opt}$
\begin{eqnarray}
j_{i,2}^{ opt}(\vec r^+, \vec r^-, t^+ - t^-) = -{\imath ^2 e^2
A_i\over c(2\pi)^{2}}\mbox{Tr}
  {1\over -\imath \bar{\beta} \hbar }\sum _{\nu^{+-}}
e^{-\imath z^{-+} (t^+ - t^-)}
\int { d^2\vec k \over (2\pi)^{2}} e^{-\imath  \vec {k} \cdot \vec
r}
\nonumber \\
\times
\left\{  
 \int_{-\infty}^{-\vec k/2} d^2\vec {\tilde p}
 e^{-\imath  \vec {\tilde p}\cdot \vec r }
\left(1-
  \widetilde {\Sigma_{AB}\Sigma_{BA}}(\vec p_{AB}- e\vec A/c )
   {d^2\widetilde {\Sigma_{AB} \Sigma_{BA}}\over d   p_i^{AB} d p_i^{AB}}
  (0)\right)\right.
\nonumber \\
\times {v^\dagger }^i({\tilde p}) \,
 {1\over -\imath \bar{\beta} \hbar }\sum _{\nu_{+}}
 G_1(z^{-+} -z^+, {\tilde p}^- )G_1(z^+, {\tilde p}^+
 )
\left.{\delta\left(\hbar \omega^{-+}+ \mu^{-+}\right)\over  \hbar
(z^{-+} +\omega^e (\hat {\tilde p}^+) - \omega^h (-\hat {\tilde
p}^-)) }
 \right\} 
v^i({\tilde p}). \ \ \label{optic-current-correction_end}
\end{eqnarray}
The correction 
(\ref{optic-current-correction_end}) is small due to the exponential multiplier
$e^{-\imath \vec{\tilde p}\cdot \vec r}$. 
This correction takes into account optical transitions from the below
lying negative energy levels
$E< E_D$, and has oscillating character that even more diminishes the value
of optical conductivity.
Oscillating character of optical conductivity of pure graphene
and its decrease at wavelengths less than 500 nm in  visible range has been observed
in \cite{Nair2008Science}.


The comparison of 
(\ref{green-function-representation1_1-9-diagram1_1}) and (\ref{optical-current2}) demonstrates 
that optical current 
for transition from zero-valued Dirac level 
$E_D=0$, in a factor two less than electrophysical current. 

\subsection{
Limit transition from optical to electrophysical current}

In the known approaches 
\cite{Ando2002,Falkovsky,Ziegler},
the value for low-frequency dynamical conductivity
$\sigma_{dyn}(\omega)$ of
pure graphene demonstrates a jump
to $\sigma_{dyn}(0)$ at zero frequency
$\omega =0$ at finite temperature and
$T \to 0$. In the case of vanishing decay rate
$\Gamma =0$, $\sigma_{dyn}(\omega)$ becomes
$e^2/(4\hbar)$ for all
$T$ in the limit
$\omega \to 0$;
conductivity becomes zero:
$\sigma_{dyn}(0)=0$ at $\omega
=0$ and $T \to 0$  contrary to a finite value for minimal dc-conductivity
\cite{Ziegler}. 
Let us show that in the model $ N = 3 $ there is no conductivity jump in the transition from
the case $\omega(0) \to 0$ to
the case $\omega (0) \to  \omega(k) $. To do this,
let us consider the behaviour of the correction
(\ref{green-function-representation1_1-9-diagram2}) in the case of spatial dispersion
$\omega (k)\to 0$, when $\vec p - \vec
p^- \to 0$:
\begin{eqnarray}
\lim_{\omega (k)\to 0} j_{i,2}^{Ohm}(\vec r^+, \vec r^-, t^+ -
t^-) = -{\imath ^2 e^2 A_i\over c(2\pi)^{2}}\mbox{Tr}
  {1\over -\imath \bar{\beta} \hbar }\sum _{\nu^{+-}}
e^{-\imath z^{-+} (t^+ - t^-)}
\int { d^2(\vec p - \vec k/2)\over (2\pi)^{2}} e^{-\imath ( \vec
{p}^- +\vec p ) \cdot \vec r}
\nonumber \\
\times
\left\{  
 \int d^2\vec p
 \left(1-
  \widetilde {\Sigma_{AB}\Sigma_{BA}}(\vec p_{AB}- e\vec A/c )
   {d^2\widetilde {\Sigma_{AB} \Sigma_{BA}}\over d   p_i^{AB} d p_i^{AB}}
  (0)\right)\right.
\nonumber \\
\times
{v^\dagger }^i_{
\vec k/2, \vec  p} \,
 {1\over -\imath \bar{\beta} \hbar }\sum _{\nu_{+}}
 G_1(z^{-+} -z^+, p^- )G_1(z^+, p+ k/2 
 )
\left.{\delta\left(\hbar \omega^{-+}+ \mu^{-+}\right)\over  \hbar
(z^{-+} +\omega^e (\hat p^+) - \omega^h (-\hat p^-)) } v^i_{0,
\vec {p}\,^- } \right\} = -{\imath ^2 e^2 A_i\over c(2\pi)^{2}}
\nonumber\\
\times \mbox{Tr}
  {1\over -\imath \bar{\beta} \hbar }\sum _{\nu^{+-}}
e^{-\imath z^{-+} (t^+ - t^-)}
\int { d^2(\vec p - \vec k/2)\over (2\pi)^{2}} e^{- \imath  (2\vec
p^- +\vec k)  \cdot \vec r}
\left\{  
 \int d^2\vec p
 \left(1-
  \widetilde {\Sigma_{AB}\Sigma_{BA}}(\vec p_{AB}- e\vec A/c )
   {d^2\widetilde {\Sigma_{AB} \Sigma_{BA}}\over d   p_i^{AB} d p_i^{AB}}
  (0)\right)\right.
\nonumber \\
\times {v^\dagger }^i(p) \,
 {1\over -\imath \bar{\beta} \hbar }\sum _{\nu_{+}}
 G_1(z^{-+} -z^+, p^- )G_1(z^+, p^+ 
 )
\left.{\delta\left(\hbar \omega^{-+}+ \mu^{-+}\right)\over  \hbar
(z^{-+} +\omega^e (\hat p^+) - \omega^h (-\hat p^-)) }
v^i(p)\right\} ,\ \omega (k)\to 0.\ \
\label{limit-from-low-frequancy}
\end{eqnarray}
Since the condition
$p,\ p^- \ll k/2$ holds, one can rewrite
(\ref{limit-from-low-frequancy}) as
\begin{eqnarray}
\lim_{\omega (k)\to 0} j_{i,2}^{Ohm}(\vec r^+, \vec r^-, t^+ -
t^-) = {\imath ^2 e^2 A_i\over 2c(2\pi)^{2}}\mbox{Tr}
  {1\over -\imath \bar{\beta} \hbar }\sum _{\nu^{+-}}
e^{-\imath z^{-+} (t^+ - t^-)}
\int { d^2 \vec k\over (2\pi)^{2}} e^{- \imath  \vec k  \cdot \vec
r}
\nonumber \\
\times
\left\{  
 \int d^2\vec p
 \left(1-
  \widetilde {\Sigma_{AB}\Sigma_{BA}}(\vec p_{AB}- e\vec A/c )
   {d^2\widetilde {\Sigma_{AB} \Sigma_{BA}}\over d   p_i^{AB} d p_i^{AB}}
  (0)\right)\right.
\nonumber \\
\times {v^\dagger }^i(p) \,
 {1\over -\imath \bar{\beta} \hbar }\sum _{\nu_{+}}
 G_1(z^{-+} -z^+, p^- )G_1(z^+, p^+ 
 )
\left.{\delta\left(\hbar \omega^{-+}+ \mu^{-+}\right)\over  \hbar
(z^{-+} +\omega^e (\hat p^+) - \omega^h (-\hat p^-)) }
v^i(p)\right\} ,\ \omega (k)\to 0.\ \
\label{limit-from-low-frequancy1}
\end{eqnarray}
Performing summation of
(\ref{optical-current2}) and
(\ref{limit-from-low-frequancy1}), we get the current in dynamical regime in
low-frequency limit
$\omega (0) \to  \omega(k), \
\omega(k) =v_F k/\hbar \to 0$, which  coincides with the expression for
minimal dc-current 
(\ref{green-function-representation1_1-9-diagram1_1}).

\subsection{Ohmic conductivity 
of the model $N=3$}

In (\ref{green-function-representation1_1-9-diagram1_1})
summation of the product of one-particle Green's functions on
Matsubara frequencies can be performed using the residue theorem.
This theorem allows us to replace the integration along the contour
$ C '= \sum_ \nu C' _ \nu $, oriented counterclockwise and
surrounding the Matsubara frequencies $ z_ \nu $, to integration over
another contour $ C $ oriented clockwise and surrounding the poles of functions
$G_1$ \cite{Kadanov}. 
Fermi-Dirac distribution function 
\begin{eqnarray}
f[\beta (z-\mu/\hbar)]= \left\{ \exp[\bar{\beta}\hbar (z-
\mu/\hbar)]+1\right\}^{-1} \label{fermi-dirac-distribution}
\end{eqnarray}
has the poles in the points 
$z_\nu = \pi \nu /(\imath \hbar \bar \beta)+\mu /\hbar$, $\nu= \pm 1, \ \pm 3, \ \pm 5, \ \ldots$
with residues 
$(\hbar \bar \beta)^{-1}$. Therefore the sum entering as a multiplier in
(\ref{green-function-representation1_1-9-diagram1_1}), can be changed on a sum over poles 
$\beta ( z^+-z^- ) -\beta(E(p^-)/\hbar)$ and $\beta(E(p^+)/\hbar)$ of one-particle Green function
$G_1(z^{-+} -z^+, p^-)$ and $G_1(z^+, p^+)$ respectively:
\begin{eqnarray}
{1\over -\imath \bar{\beta} \hbar }\sum _{\nu_{+}} G_1(z^{-+}
-z^+, p^-)G_1(z^+, p^+) = \imath \bar{\beta}\int_C {d [\beta ( z)]
\over 2 \pi \imath}
{f[\beta (z-\mu/\hbar)]\over \beta ( z) - \beta(E(p^+)/\hbar) } {1\over \beta ( z^{-+}
) -  \beta ( z) -\beta(E(p^-)/\hbar)}
\nonumber \\
= -\imath \bar{\beta}\int_C {d [\beta ( z)] \over 2 \pi \imath}
f[\beta (z-\mu/\hbar)]
{1\over \beta(E(p^+)/\hbar) -  \beta ( z)} {1\over \beta ( z^{-+}
) -  \beta ( z) -\beta(E(p^-)/\hbar)}
\nonumber \\
=- \imath  \bar{\beta}  {f[\beta ((E(p^+)-\mu)/\hbar)] - f[\beta (
z^{-+} -E(p^-)-\mu/\hbar)]\over \beta ( z^{-+}) -
\beta(E(p^-)/\hbar) - \beta(E(p^+)/\hbar) }
=- \imath  \bar{\beta}  {f[\beta ((H(p^+)-\mu)/\hbar)] - f[\beta
(H^\dagger (-p^-)-\mu/\hbar)]\over \beta ( z^{-+}) - \beta
(H(p^+)/\hbar) + \beta(H^\dagger (-p^-)/\hbar) }.
\label{counter-integral}
\end{eqnarray}
Here 
$C$ is a counterclockwise contour. 
Taking into account of (\ref{counter-integral}), we find the Fourier image of the Ohmic current
(\ref{green-function-representation1_1-9-diagram1_1}):
\begin{eqnarray}
j_i^{Ohm}(\omega^{-+}, \ k) = {\imath \bar \beta  e^2 A_i\over
c(2\pi)^{2}}\mbox{Tr}
\left\{  
 \int d^2\vec p
 \left(1-
  \widetilde {\Sigma_{AB}\Sigma_{BA}}(\vec p_{AB}- e\vec A/c )
   {d^2\widetilde {\Sigma_{AB} \Sigma_{BA}}\over d   p_i^{AB} d p_i^{AB}}
  (0)\right)\right.
\nonumber \\
\times {v^\dagger }^i(p) \,
 {f[\beta ((H(p^+)-\mu)/\hbar)]
- f[\beta (H^\dagger (-p^-)-\mu/\hbar)]\over \beta ( z^{-+}) -
\beta (H(p^+)/\hbar) + \beta(H^\dagger (-p^-)/\hbar) }
\left.{\delta\left(\hbar \omega^{-+}+ \mu^{-+}\right)\over  \hbar
(z^{-+} +\omega^e (\hat p^+) - \omega^h (-\hat p^-)) }
v^i(p)\right\} ,\ \hbar \omega \ll v_F k \to 0.\ \
\label{Laplas-Fourier-transform6}
\end{eqnarray}
Due to 
(\ref{graphene-quasirel-current}), the coefficient at 
$A_i$ entering  the expression (\ref{Laplas-Fourier-transform6}) after its division on 
$c$, gives Ohmic conductivity of 
SM:
\begin{eqnarray}
\sigma_{ii}^{Ohm}(\omega^{-+}, \ k) = {\imath \bar \beta  e^2
\over (2\pi c)^{2}}\mbox{Tr}
\left\{  
 \int d^2\vec p
 \left(1-
  \widetilde {\Sigma_{AB}\Sigma_{BA}}(\vec p_{AB}- e\vec A/c )
   {d^2\widetilde {\Sigma_{AB} \Sigma_{BA}}\over d   p_i^{AB} d p_i^{AB}}
  (0)\right)\right.
\nonumber \\
\times {v^\dagger }^i(p) \,
 {f[\beta ((H(p^+)-\mu)/\hbar)]
- f[\beta (H^\dagger (-p^-)-\mu/\hbar)]\over \beta ( z^{-+}) -
\beta (H(p^+)/\hbar) + \beta(H^\dagger (-p^-)/\hbar) }
\left.{\delta\left(\hbar \omega^{-+}+ \mu^{-+}\right)\over  \hbar
(z^{-+} +\omega^e (\hat p^+) - \omega^h (-\hat p^-)) }
v^i(p)\right\} ,\ \hbar \omega \ll v_F k \to 0.\ \
\label{conduction2}
\end{eqnarray}

\subsection{ 
Polarization effects }
Let us consider the influence of particle-hole production on the behavior of the
diagonal elements
of complex conductivity $\sigma_{ll}$ 
of SM. This contribution 
in $\sigma_{ll}$ is described 
by the polarization contribution to  current  entering into the current
(\ref{graphene-quasirel-current2}) in main text. The explicit expression
is obtained in a manner similar to previous calculations
and reads 
\begin{eqnarray}
j_l^{Zitterbew} = -{e^2 A_l\over c\,
 \widetilde {\Sigma_{AB}\Sigma_{BA}}(\vec p_{AB}- e\vec A/c )}
 \chi_{+\sigma_{_B} }^\dagger  \chi_{+\sigma_{_B} }
={\imath e^2 A_l\ \mbox{Tr}\over \imath^3 c \widetilde
{\Sigma_{AB}\Sigma_{BA}}(\vec p_{AB}- e\vec A/c )}
\nonumber \\
\times \left\{\chi_{+\sigma_{_B} }^\dagger  \chi_{+\sigma_{_B} } 
-
{(-\imath)
}\int \int \sum_i \left[ {e\over c}  v^i_{x^{+}-x',\, \bar{ x}}
 A_i(\bar{x})
+\widetilde {\Sigma_{AB}\Sigma_{BA}}(\vec p_{AB}- e\vec A/c )
{d\widetilde {\Sigma_{AB} \Sigma_{BA}}\over d   p_i^{AB}} (0) \
{v^i_{x^{+}-x',\, \bar{ x}} }\right. \right. \nonumber
\\
+ \left.\left. {\widetilde {\Sigma_{AB}\Sigma_{BA}}^2(\vec p_{AB}-
e\vec A/c )\over 2} \sum_{j} {d^2\widetilde {\Sigma_{AB}
\Sigma_{BA}}\over d   p_i^{AB} d   p_j^{AB}} (0)\
{v^i_{x^{+}-x',\, \bar{ x}}\, v^j_{x^{+}-x',\ \bar{ x}}
 }
\right]\right.
\nonumber \\ \times \left. G_2(x^{+},\, \bar{x},\, x',\, x'') G_1(
x^{-} -x'') d\bar{t}\, d^2 \vec {\bar{ x}}\, dt'\, d^2\vec {x'}\,
dt''\, d^2\vec {x''} + \ldots \right\} .\ \ \
\label{Zitterbeweg-current-Laplas0-0}
\end{eqnarray}
 $\widetilde {\Sigma_{AB}\Sigma_{BA}}(\vec p_{AB}- e\vec A/c )
 \approx {d\widetilde {\Sigma_{AB} \Sigma_{BA}}\over d   p_i^{AB}} p_i^{AB}$ near 
Majorana zero-energy state, and respectively
the first term in equation (\ref{Zitterbeweg-current-Laplas0-0}) is equal to zero:
\begin{eqnarray}
j_l^{Zitterbew} = -{e^2 A_l\over c \,
 \widetilde {\Sigma_{AB}\Sigma_{BA}}(\vec p_{AB}- e\vec A/c )}
 \chi_{+\sigma_{_B} }^\dagger  \chi_{+\sigma_{_B} }
={\imath e^2 A_l\over \imath^3 c
}
\nonumber \\
\times \mbox{Tr} \left\{
\imath 
\int \int \sum_i \left[ {e\over c}  v^i_{x^{+}-x',\, \bar{ x}}
 \tilde A_i(\bar{x})
+
{d\widetilde {\Sigma_{AB} \Sigma_{BA}}\over d   p_i^{AB}} (0) \
v^i_{x^{+}-x',\, \bar{ x}} \right. \right.
+ \sum_k {d\widetilde {\Sigma_{AB} \Sigma_{BA}}\over d   p_k^{AB}}
(0)
{d\widetilde {\Sigma_{AB} \Sigma_{BA}}\over d   p_i^{AB}} (0) \
v^k_{x^{+}-x',\, \bar{ x}} v^i_{x^{+}-x',\, \bar{ x}} \nonumber
\\
+ \left.\left. {\widetilde {\Sigma_{AB}\Sigma_{BA}} (\vec p_{AB}-
e\vec A/c )\over 2} \sum_{j} {d^2\widetilde {\Sigma_{AB}
\Sigma_{BA}}\over d   p_i^{AB} d   p_j^{AB}} (0)\
{v^i_{x^{+}-x',\, \bar{ x}}\, v^j_{x^{+}-x',\ \bar{ x}}
 }
\right]\right.
\nonumber \\ \times \left. G_2(x^{+},\, \bar{x},\, x',\, x'') G_1(
x^{-} -x'') d\bar{t}\, d^2 \vec {\bar{ x}}\, dt'\, d^2\vec {x'}\,
dt''\, d^2\vec {x''} + \ldots \right\}
\nonumber \\
={  e^2 A_l\over \imath  c} \mbox{Tr} \left\{
\int \int \sum_i \left[ {\widetilde {\Sigma_{AB}\Sigma_{BA}} (\vec
p_{AB}- e\vec A/c )\over 2}   {d^2\widetilde {\Sigma_{AB}
\Sigma_{BA}}\over d   p_i^{AB} d   p_i^{AB}} (0)\
{v^i_{x^{+}-x',\, \bar{ x}}\, v^j_{x^{+}-x',\ \bar{ x}}
 }
\right]\right.
\nonumber \\ \times \left. G_2(x^{+},\, \bar{x},\, x',\, x'') G_1(
x^{-} -x'') d\bar{t}\, d^2 \vec {\bar{ x}}\, dt'\, d^2\vec {x'}\,
dt''\, d^2\vec {x''} + \ldots \right\} .\ \ \
\label{Zitterbeweg-current-Laplas0-2}
\end{eqnarray}
The  current (\ref{Zitterbeweg-current-Laplas0-2})
has the form similar to Ohmic one and can be calculated in the same way
resulting with 
\begin{eqnarray}
j_l^{Zitterbew}(\omega^{-+}, \ k) = {\imath \bar \beta  e^2
A_l\over c(2\pi)^{2}}\mbox{Tr}
\left\{  
 \int d^2\vec p
 \sum_{i=1}^2
{\widetilde {\Sigma_{AB}\Sigma_{BA}}(\vec p_{AB}- e\vec A/c )\over
2} {d^2\widetilde {\Sigma_{AB} \Sigma_{BA}}\over d   p_i^{2}}(0)
\right.
\nonumber \\
\times {v^\dagger }^i(p) \,
 {f[\beta ((H(p^+)-\mu)/\hbar)]
- f[\beta (H^\dagger (-p^-)-\mu/\hbar)]\over \beta ( z^{-+}) -
\beta (H(p^+)/\hbar) + \beta(H^\dagger (-p^-)/\hbar) }
\left.{\delta\left(\hbar \omega^{-+}+ \mu^{-+}\right)\over  \hbar
(z^{-+} +\omega^e (\hat p^+) - \omega^h (-\hat p^-)) } v^i(p)
\right\},\ \hbar \omega \ll v_F k \to 0.\ \
\label{Zitterbeweg-current-Laplas}
\end{eqnarray}
In the expression 
(\ref{Zitterbeweg-current-Laplas}) for 
$j_l^{Zitterbew}$, not only electron moving along an applied electric field $\vec E$
contribute to but also electrons diffusively moving in the direction which is perpendicular to $\vec E$.
Due to 
(\ref{graphene-quasirel-current}), the coefficient at 
$A_i$ entering into the expression (\ref{Zitterbeweg-current-Laplas}) after its division by 
$c$ gives 
the polarization contribution to the  conductivity 
\begin{eqnarray}
\sigma_{ll}^{Zitterbew}(\omega^{-+}, \ k) = {\imath \bar \beta e^2
\over (2\pi\, c)^{2}}\mbox{Tr}
\left\{  
 \int d^2\vec p
 \sum_{i=1}^2
{\widetilde {\Sigma_{AB}\Sigma_{BA}}(\vec p_{AB}- e\vec A/c )\over
2} {d^2\widetilde {\Sigma_{AB} \Sigma_{BA}}\over d   p_i^{2}}(0)
\right.
\nonumber \\
\times {v^\dagger }^i(p) \,
 {f[\beta ((H(p^+)-\mu)/\hbar)]
- f[\beta (H^\dagger (-p^-)-\mu/\hbar)]\over \beta ( z^{-+}) -
\beta (H(p^+)/\hbar) + \beta(H^\dagger (-p^-)/\hbar) }
\left.{\delta\left(\hbar \omega^{-+}+ \mu^{-+}\right)\over  \hbar
(z^{-+} +\omega^e (\hat p^+) - \omega^h (-\hat p^-)) }
v^i(p)\right\}. 
\label{Zitterbewegung_conduction}
\end{eqnarray}

\subsection{ 
Magnetoelectric effects}

The contribution of magnetoelectric effects to the expression (\ref{graphene-quasirel-current2})
for the current
is calculated in the same way, the result reads 
\begin{eqnarray}
j_{2(1)}^{spin-orbit} =
 (-1)^{1(2)} {\imath e \over 2 }
\chi_{+\sigma_{_B} }^\dagger  \sigma_z \,
 v^{1(2)}_{x^+ x^-} \chi_{+\sigma_{_B} }
=(-1)^{1(2)} {\imath  \over 2 }{(-1)\imath e\over (\imath)^3} \
\nonumber 
\end{eqnarray}
\begin{eqnarray}
\times \mbox{Tr}\left\{ -
{(-\imath)
}\int \int \sum_i \left[ {e\over c}  v^i_{x^{+}-x',\, \bar{ x}}
 A_i
+\widetilde {\Sigma_{AB}\Sigma_{BA}}(\vec p_{AB}- e\vec A/c)
{d\widetilde {\Sigma_{AB} \Sigma_{BA}}\over d   p_i^{AB}} (0) \
{v^i_{x^{+}-x',\, \bar{ x}} }\right. \right.
 \nonumber
\\
+{1\over 2}\sum_{j} {d^2\widetilde {\Sigma_{AB} \Sigma_{BA}}\over
d   p_i^{AB} d   p_j^{AB}} (0) \left( \widetilde
{\Sigma_{AB}\Sigma_{BA}}(\vec p_{AB}- e\vec A/c) v^i_{x^{+}-x',\
\bar{ x}} -{e\over c}A_i\right)
\nonumber \\
\left.\left. \times \left( \widetilde
{\Sigma_{AB}\Sigma_{BA}}(\vec p_{AB}- e\vec A/c) v^j_{x^{+}-x',\
\bar{ x}} -{e\over c}A_j\right) \right]\right.
\left. G_2(x^{+},\, \bar{x},\, x',\, x'') G_1( x^{-} -x'')
d\bar{t}\, d^2 \vec {\bar{ x}}\, dt'\, d^2\vec {x'}\, dt''\,
d^2\vec {x''} + \ldots \right\}
\sigma_z v^{1(2)}_{x'x^{-}} \nonumber \\
=(-1)^{1(2)} {\imath  \over 2 }{(-1)\imath e\over (\imath)^3}
\ \mbox{Tr}\left\{
-
{(-\imath)
}\int \int
 \left[
{d^2\widetilde {\Sigma_{AB} \Sigma_{BA}} \over d p_{1(2)}^{AB} d
p_{2(1)}^{AB}} (0) \left( \widetilde {\Sigma_{AB}\Sigma_{BA}}(\vec
p_{AB}- e\vec A/c) v^{1(2)}_{x^{+}-x',\ \bar{ x}}\right) \right.
\right.
\left.\left( -{e\over c}A_{2(1)}\right) \right]  \nonumber
\\
\times \left. G_2(x^{+},\, \bar{x},\, x',\, x'') G_1( x^{-} -x'')
d\bar{t}\, d^2 \vec {\bar{ x}}\, dt'\, d^2\vec {x'}\, dt''\,
d^2\vec {x''} + \ldots \right\} \sigma_z v^{1(2)}_{x'x^{-}}
 \nonumber \\
=(-1)^{1(2)} {\imath  \over 2 }{   e^2\over \imath c} A_{2(1)}
\mbox{Tr}\left\{ \int \int {d^2\widetilde {\Sigma_{AB}
\Sigma_{BA}} \over d p_{1(2)}^{AB} d p_{2(1)}^{AB}} (0) \widetilde
{\Sigma_{AB}\Sigma_{BA}}(\vec p_{AB}- e\vec A/c)
v^{1(2)}_{x^{+}-x',\ \bar{ x}}  \right.
  \nonumber
\\
\times \left. G_2(x^{+},\, \bar{x},\, x',\, x'') G_1( x^{-} -x'')
d\bar{t}\, d^2 \vec {\bar{ x}}\, dt'\, d^2\vec {x'}\, dt''\,
d^2\vec {x''} + \ldots \right\} \sigma_z v^{1(2)}_{x'x^{-}}. \ \ \ \ \ \ \
\label{spin-orbit-current2-1}
\end{eqnarray}
Contrary to the polarization effects 
(\ref{Zitterbeweg-current-Laplas0-2}), the Hall current 
$j_{2(1)}^{spin-orbit}$ (\ref{spin-orbit-current2-1}) is defined by non-diagonal derivatives of the
dynamical mass
$\widetilde
{\Sigma_{AB} \Sigma_{BA}}$. Performing the similar to previous calculation we find the explicit form for
$j_{2(1)}^{spin-orbit}$:
\begin{eqnarray}
j_{2(1)}^{spin-orbit}(\omega^{-+}, \ k) = (-1)^{1(2)} {\imath
\over 2 } {\imath \bar \beta  e^2 A_{2(1)}\over
c(2\pi)^{2}}\mbox{Tr}
\left\{  
 \int d^2\vec p\
 {\widetilde {\Sigma_{AB}\Sigma_{BA}}(\vec p_{AB}- e\vec A/c )
 }
{d^2\widetilde {\Sigma_{AB} \Sigma_{BA}} \over d p_{1(2)}^{AB} d
p_{2(1)}^{AB}} (0) \right.
\nonumber \\
\times {v^\dagger }^{1(2)}(p) \,
 {f[\beta ((H(p^+)-\mu)/\hbar)]
- f[\beta (H^\dagger (-p^-)-\mu/\hbar)]\over \beta ( z^{-+}) -
\beta (H(p^+)/\hbar) + \beta(H^\dagger (-p^-)/\hbar) }
\left.{\delta\left(\hbar \omega^{-+}+ \mu^{-+}\right)\over  \hbar
(z^{-+} +\omega^e (\hat p^+) - \omega^h (-\hat p^-)) }
\sigma_z v^{1(2)} (p) \right\} . 
\label{spin-orbit-current2}
\end{eqnarray}
Again, the coefficient at 
$A_{2(1)}$ entering into the expression (\ref{spin-orbit-current2}) after its division by 
$c$ gives 
the spin-orbit contribution to the  conductivity 
\begin{eqnarray}
\sigma_{12(21)}^{spin-orbit}(\omega^{-+}, \ k)=
  (-1)^{1(2)} {\imath  \over 2 }
{\imath \bar \beta  e^2  \over (2\pi\, c)^{2}}\mbox{Tr}
\left\{  
 \int d^2\vec p\
 {\widetilde {\Sigma_{AB}\Sigma_{BA}}(\vec p_{AB}- e\vec A/c )
 }
{d^2\widetilde {\Sigma_{AB} \Sigma_{BA}} \over d p_{1(2)}^{AB} d
p_{2(1)}^{AB}} (0) \right.
\nonumber \\
\times {v^\dagger }^{1(2)}(p) \,
 {f[\beta ((H(p^+)-\mu)/\hbar)]
- f[\beta (H^\dagger (-p^-)-\mu/\hbar)]\over \beta ( z^{-+}) -
\beta (H(p^+)/\hbar) + \beta(H^\dagger (-p^-)/\hbar) }
\left.{\delta\left(\hbar \omega^{-+}+ \mu^{-+}\right)\over  \hbar
(z^{-+} +\omega^e (\hat p^+) - \omega^h (-\hat p^-)) }
\sigma_z v^{1(2)} (p)
 \right\}. 
 \label{spin-orbit-conduction}
\end{eqnarray}

\subsection{Diagonalized representation of the Hamiltonian 
}
We utilize the following approximation:
\begin{eqnarray}
\vec v_{AB}\approx {\partial H^{(0)}_{AB}\over  \partial \vec p}
\label{velocity-in-diagonal-represenation_0}
\end{eqnarray}
where $H^{(0)}_{AB}$ is the unperturbed Hamiltonian of the problem 
(\ref{rel-from-pseudi-Dirac-whithout-mix3}) in main text without the dynamical mass term:
\begin{eqnarray}
H^{(0)}_{AB}=c \vec \sigma _{2D}^{BA} \cdot \vec p_{AB} .
\label{non-perturbed_Hamiltonian}
\end{eqnarray}

Let indexes 
$a,\ b=1,\ 2$ be for 
valence and conduction bands. 
We designate the eigenvalues of the unperturbed electron Hamiltonian
(\ref{non-perturbed_Hamiltonian}) in a band 
$a(b)$ through 
$E^e_{a ( b)}$. Eigenvalues of hole Hamiltonian 
$H_0^{\dagger}$  in the band 
$a(b)$ are designated by 
$E^h_{a ( b)}$. Due to the validity of the condition 
(\ref{frequency-impuls-condition}),  it is possible to use the following expansion for
$\omega^{\pm}$ :
\begin{eqnarray}
\hbar \omega^{\pm}\equiv \hbar \omega(p^{\pm}) = \left\{
\begin{array}{ll}
E^e_{a(b)}
(p)+ (-1)^i \omega (k) /2 & \mbox{  for electrons}\\
E^h_{a(b)} (-p) -(-1)^i \omega (-k) /2  & \mbox{  for holes},
\end{array}
\right. \label{frequency-condition}
\end{eqnarray}
where indexes 
$i=1,\ 2$ are for energy level of the band 
$a(b)$ for electron 
$"e"$ (
hole $"h"$) with momentum 
$p^-$ before transition and electron (hole) 
$p^+$ after transition respectively. 
The conditions (\ref{frequency-condition}) mean that 
\begin{equation}
\begin{split}
\left| E^{e(h)}_{a(b)} (p^+)- E^{e(h)}_{a(b)}
(p^-)\right| =\omega (k), 
\\
 [E^e_{a(b)} (p^{\pm})- E^h_{a(b)} (p^{\mp})] = E^e_{a(b)} (+p)- E^h_{a(b)}(-p).
\end{split}\label{frequency-condition1}
\end{equation}

Let us perform calculations in the representation where the Hamiltonian of the problem  is diagonal.
The velocity operator $\vec v_{AB(BA)}$ should be transformed by the similarity transformation of the form
$S^{-1}\vec v_{AB} S$ with a matrix $S$ constructed on the eigenvectors $\chi$ of the Hamiltonian.
The corresponding matrix $S$ should be constructed on the eigenvectors of the operator adjoined to the Hamiltonian.
In every $p$-point the particle (hole) Hamiltonian is represented by $2 \times 2$ matrix,
we denote matrix elements of the exchange operator $\left( i\Sigma _{rel}^{x} \right)_{AB(BA)}$ formally
as $\Sigma^{AB(BA)}_{ij}$.
Then the eigenvectors ${\chi}_i,\ i=1,2$ of the Hamiltonian \eqref{non-perturbed_Hamiltonian} being the rows of
the appropriate matrix $S$ can be  expressed in an explicit way:
\begin{equation}
\chi_1=\left\{\frac{i p \sin (\phi ) (\Sigma^{AB}_{11} \Sigma^{AB}_{21}+\Sigma^{AB}_{12} \Sigma^{AB}_{22})+p \cos (\phi )
   (\Sigma^{AB}_{12} \Sigma^{AB}_{22}-\Sigma^{AB}_{11} \Sigma^{AB}_{21})
   -p (\Sigma^{AB}_{12} \Sigma^{AB}_{21}-\Sigma^{AB}_{11}
   \Sigma^{AB}_{22})}{p \left(\left((\Sigma^{AB}_{11})^2-(\Sigma^{AB}_{12})^2\right) \cos (\phi )-i
   \left((\Sigma^{AB}_{11})^2+(\Sigma^{AB}_{12})^2\right) \sin (\phi )\right)},\ 1\right\}
\end{equation}
\begin{equation}
\chi_2=\left\{\frac{i p \sin (\phi ) (\Sigma^{AB}_{11} \Sigma^{AB}_{21}
+\Sigma^{AB}_{12} \Sigma^{AB}_{22})+p \cos (\phi )
   (\Sigma^{AB}_{12} \Sigma^{AB}_{22}-\Sigma^{AB}_{11} \Sigma^{AB}_{21})
   + p (\Sigma^{AB}_{12} \Sigma^{AB}_{21}-\Sigma^{AB}_{11}
   \Sigma^{AB}_{22})}{p \left(\left((\Sigma^{AB}_{11})^2-(\Sigma^{AB}_{12})^2\right) \cos (\phi )-i
   \left((\Sigma^{AB}_{11})^2+(\Sigma^{AB}_{12})^2\right) \sin (\phi )\right)},\ 1\right\}
\end{equation}

In this way we can calculate numerically the velocity operator in every $p$-point
with subsequent its substitution to the conductivity integral.

\subsection{Integral calculations}

\begin{figure}[hbt]
\begin{center}
\includegraphics[width=5.cm,height=5.cm,angle=0]{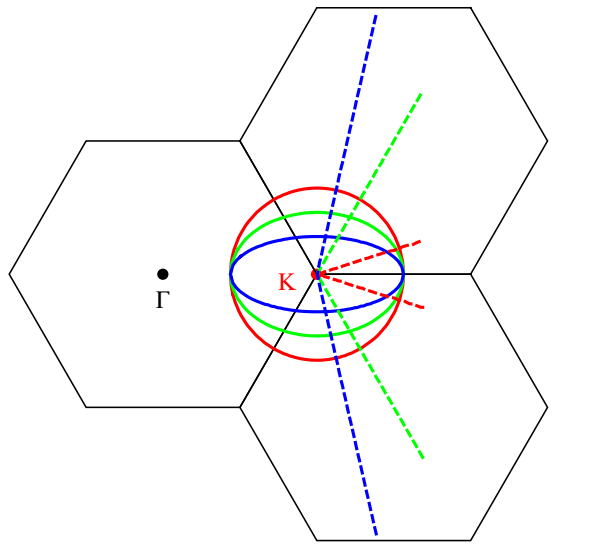}
\end{center}
\caption{Sketch of the poles structure of integrands for intra- (dashed lines)
and inter-(solid lines)  bands contribution to conductivity. Solid lines parameters at
$\omega=1$ (in units of $v_F|\vec K_A|$) are $k/|\vec K_A|=0.1$ (red), $0.7$ (green),
$0.9$ (blue). Dashed lines parameters at
$\omega=0.2$ (in units of $v_F|\vec K_A|$) are $k/|\vec K_A|=0.21$ (red), $0.4$ (green),
$0.9$ (blue).
 }
\label{poles-fig}
\end{figure}

Contributions to conductivity include 2D integrals over the Brillouin zone (BZ). For example, integrals
in the Ohmic contribution given by formulae (\ref{conduction4})  have the
 following form: 
\begin{eqnarray}
\sigma ^{intra}_{ij}(\omega ,\vec{k})=\sum\limits_{a=1,2}\frac{i{{e}^{2}}v_{0}^{2}}{{{\pi
}^{2}}}\int\limits_{\mathrm{BZ}}{{ {d}^{2}}\vec{p}}
\frac{\left(
\left[
{{{{v}}}^{i}_{aa}}(\vec{p}){{{{v}}}^{j}_{aa}} (\vec{p})
\right]
f[{{\varepsilon }_{1}}(\vec{p}-\vec{k}/2)]-
f[{{\varepsilon }_{1}}(\vec{p}+\vec{k}/2)] \right)}{\left(
{{\varepsilon }_{1}}(\vec{p}+\vec{k}/2)-{{\varepsilon
}_{1}}(\vec{p}- \vec{k}/2) \right)\left( \omega - {{\varepsilon
}_{1}}(\vec{p}+\vec{k}/2)+{{\varepsilon }_{1}}(\vec{p}-\vec{k}/2)
\right)}, \label{sigma-in1}
\end{eqnarray}
%
\begin{eqnarray}
\sigma ^{inter}_{ij}(\omega ,\vec{k})=\frac{2i\omega
{{e}^{2}}v_{0}^{2}}{{{\pi
}^{2}}}\int\limits_{\mathrm{BZ}}{{{d}^{2}}\vec{p}}\frac{v_{12}^{i}(\vec{p})
v_{21}^{i}(\vec{p})\left( f[{{\varepsilon
}_{1}}(\vec{p}-\vec{k}/2)]- f[{{\varepsilon
}_{2}}(\vec{p}+\vec{k}/2)] \right)}{\left( {{\varepsilon
}_{2}}(\vec{p}+\vec{k}/2)-{{\varepsilon }_{1}}(\vec{p}- \vec{k}/2)
\right)\left( {{\omega }^{2}}- {{\left( {{\varepsilon
}_{2}}(\vec{p}+\vec{k}/2)-{{\varepsilon }_{1}}(\vec{p}- \vec{k}/2)
\right)}^{2}} \right)} . \label{sigma-out1}
\end{eqnarray}
Here the first integral is for intraband transitions, the second one is for interband ones.
Now, we highlight the pole structure for the integrands for
small but finite $\vec k$, accounting that
$\epsilon_1(\vec p)=-\epsilon_2(\vec p) $.
The first  integral can be rewritten as
\begin{eqnarray}{{\sigma }^{intra}_{ij}}(\omega ,\vec{k})=\sum\limits_{a=1,2}
        \frac{i{{e}^{2}}v_{0}^{2}}{{{\pi }^{2}}}\int\limits_{\mathrm{BZ}}{{{d}^{2}}\vec{p}\left[
{{{{v}}}^{i}_{aa}}(\vec{p }){{{{v}}}^{j}_{aa}}(\vec{p})
\right]}\,\,\left.\frac{df(\varepsilon)}{d \varepsilon}\right|_{\varepsilon=\varepsilon_1(\vec{p})}\frac{1}
{(\omega - \vec{k}\cdot\nabla {{\varepsilon }_{1}}(\vec{p}))}\
, \label{sigma-in2a}
\end{eqnarray}
where we have performed the Taylor series expansion on $|\vec k|$ up to a linear terms.
In the second integral only the second term in denominator can produce  poles, so we expand it
into a power series on $|\vec k|$, make a change to polar coordinates $(p_x,p_y)\to (p,\phi)$ that results
\begin{eqnarray}
\sigma ^{inter}_{ij}(\omega ,\vec{k})=\frac{2i\omega
{{e}^{2}}v_{0}^{2}}{{{\pi
}^{2}}}\int\limits_{\mathrm{BZ}}{pdp\ d\phi}\frac{v_{12}^{i}(\vec{p})
v_{21}^{i}(\vec{p})\left( f[{{\varepsilon
}_{1}}(\vec{p}-\vec{k}/2)]- f[{{\varepsilon
}_{2}}(\vec{p}+\vec{k}/2)] \right)}{\left( {{\varepsilon
}_{2}}(\vec{p}+\vec{k}/2)-{{\varepsilon }_{1}}(\vec{p}- \vec{k}/2)
\right)\left( \omega ^2-4 p^2-k^2 \sin ^2\phi \right)} . \label{sigma-out2}
\end{eqnarray}
Pole structures of (\ref{sigma-in2a}) and (\ref{sigma-out2}) is presented in fig.~\ref{poles-fig}.
In accordance to (\ref{sigma-in2a}) and  fig.~\ref{poles-fig} at $\vec k=0$ this  integral
 is a regular one, whereas at finite $k$
there is a line of poles (dashed lines in figure).
%
At a finite $k$ the pole structure of $\sigma ^{inter}_{ij}(\omega ,\vec{k})$ (\ref{sigma-out2})
is an elliptic one that results
in necessity to account an infinite sum of poles as contribution to conductivity:
$\sigma ^{inter}_{ij}(\omega ,\vec{k})\propto \int\ d\phi \mbox{ Res}({\phi}) $,
where $\mbox{ Res}({\phi}) $ is a residue in the pole located at angle
$\phi$ on
the poles line.
The integral (\ref{sigma-out2}) in the case $k= 0$ holds poles laying at a circumference that can
be effectively reduced to a single one as  $\sigma ^{inter}_{ij}(\omega ,\vec{k})\propto 2\pi \mbox{ Res}_1 $,
where $\mbox{ Res}_1 $ is a residue in arbitrary point
of the circumference.
For very oblate ellipse
at large $k$ the main contribution
to
the integral (\ref{sigma-out2}) gives the only points
touching the circumference
$\sigma ^{inter}_{ij}(\omega ,\vec{k})\sim 2 \mbox{ Res}_1 $. Thus,
the value of
the optical conductivity  decreases with the growth of
$k$.
In experiment the
optical transmission through a graphene monolayer
for normal incidence and respectively the optical conductivity
really decreases in optical wave range in direction of shorter
wave lengths
\cite{Nair2008Science,Falkovsky2011LowTemp.Phys}.

\begin{figure}[hbt]
\begin{center}
\hspace{0cm}(a)\hspace{5cm} (b) \hspace{5cm} (c) \\
\includegraphics[width=5.cm,height=5.cm,angle=0]{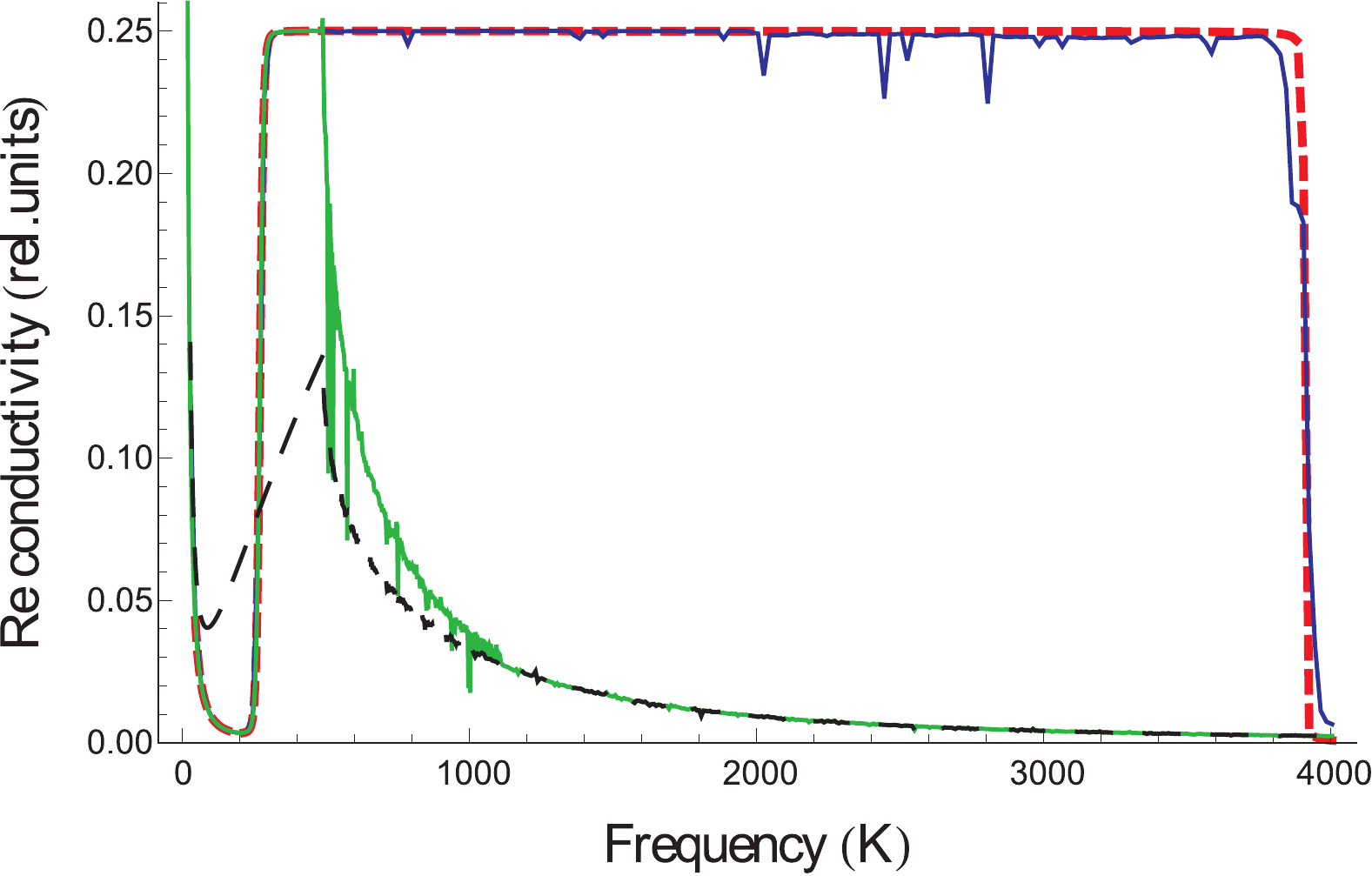}
 \includegraphics[width=5.cm,height=5.cm,angle=0]{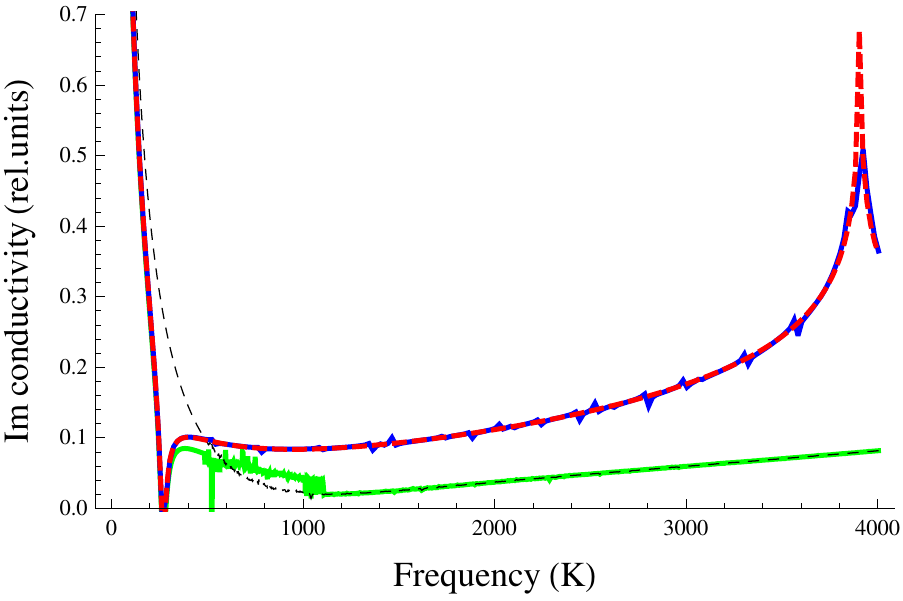}
\hspace{5mm} \includegraphics[width=5.cm,height=5.cm,angle=0]{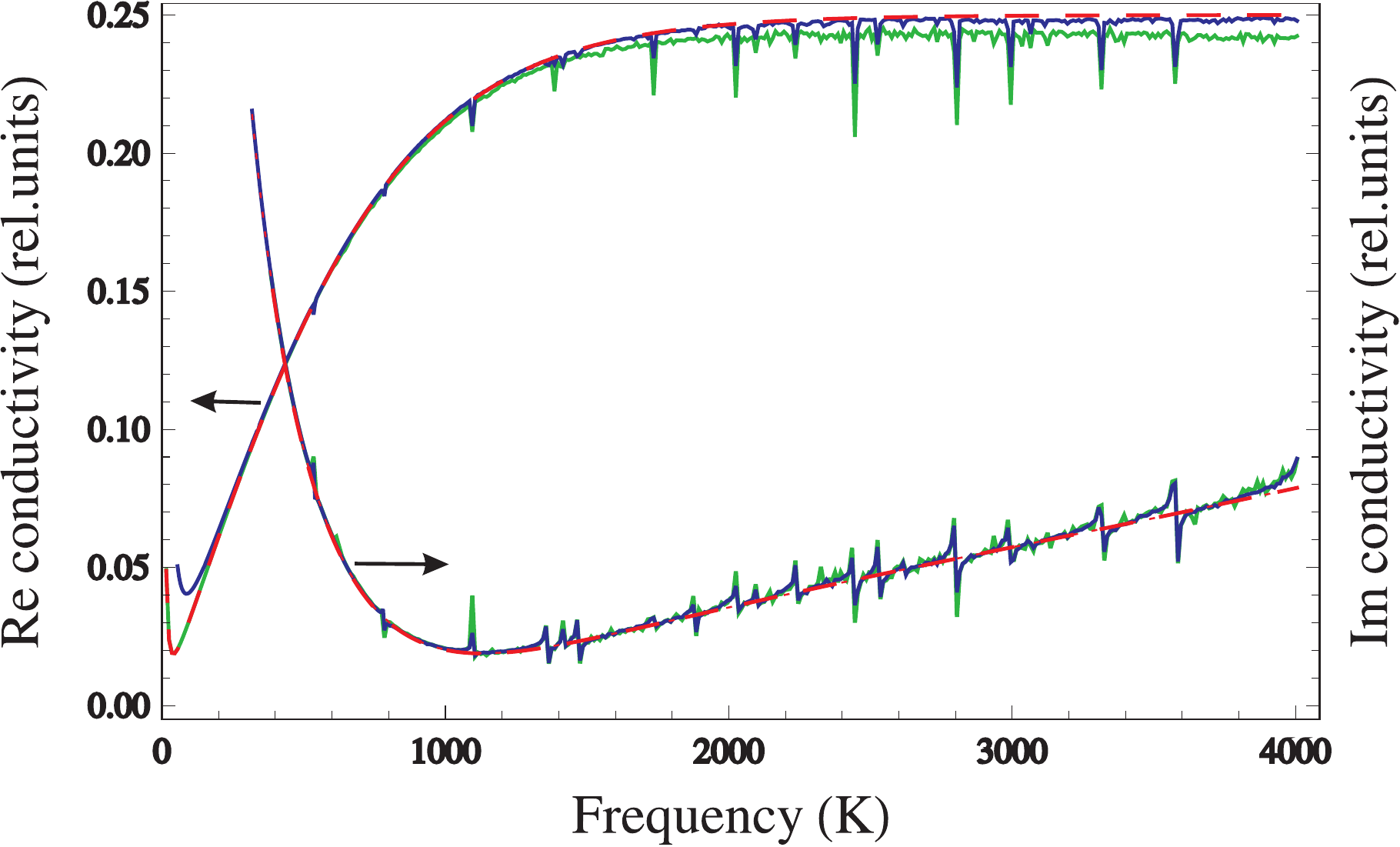}
 \end{center}
\caption{Frequency dependencies of the real  and imaginary parts of the massless Ohmic term of conductivity
at very small wave number $q=10^{-8}|\vec K_A|$ in rel.~units $e^2/\hbar$; a cutting parameter $q_{max}=0.14~K_A$. The model
\cite{Falkovsky,Falkovsky2008PhysUspekhi} is simulated at T=3K, chemical potential $\mu = 135$K
(red dashed  lines in figures (a--b)) and at T=200K, $\mu = 33$K  (red  dashed line in figure (c)).
Numerical results for our  model are  green-solid and   black-dashed lines for  the  first-order
approximation (with zero gauge-phases) \eqref{Majorana-bispinor1-first-approximation}
at T=3K, $\mu = 135$K and T=200K, $\mu = 33$K respectively,
and blue line for the second-order approximation (with non-zero gauge-phases)
\eqref{variational-Majorana-bispinor} at T=3K, $\mu = 135$K in figures (a--b). (c) Dependence of the real  and imaginary
parts of the massless Ohmic term of conductivity on a damping $\gamma$ for
the second-order approximation \eqref{variational-Majorana-bispinor} with
$\gamma=0.1$ (green curve) and $\gamma=1$ (blue curve)  at T=200K, $\mu = 33$K. }
\label{q-max0-14simulation}
\end{figure}

We define the energy limit of applicability
%
1~eV$\approx 10^4$~K of tight-binding approximation
as a 
$\omega _{max}$  of the model with linear dispersion.
Then, the upper integration limit
$q_{max}$ on momentum
is $q_{max}\sim \omega _{max}/v_F$, and respectively
for the model of massless
Dirac fermions the integration should use the range
from 0 to $q<0.14~K_A-0.28~K_A $.
As the simulation results presented
in fig.~\ref{q-max0-14simulation} demonstrate,  the integration within this limits
leads to the conductivity fall in the range
$\omega_{max}\sim 4000-8000$~K.
But, experimentally such a fall starts at much
higher frequencies
($8000$~K)
\cite{Nair2008Science,Falkovsky2011LowTemp.Phys}.
Thus, the range of momenta to predict conductivity
in visible optical range
is outside the limits of applicability
of the massless
Dirac fermion model.

According to the
Table~\ref{majorana-velocity}, the Fermi velocity
$v_M$ for our model of massless
Majorana energy states
decrease in a factor higher than two
for high
wave numbers $q$ in respect to its value in the
Dirac point.
Therefore, the range
$q$, $q  < \omega_{max} /  v_M$, $\omega_{max}<$1~eV increases in a factor two
compared to the
massless Dirac fermions. Since $q<0.28~K_A-0.56~K_A $
for our model, the range of integration over momentum
$|q|<0.44~|K_A|$ is appropriate one and for it
$\omega_{max}> 7000$~K.

In the first-order approximation with zero-phases of the gauge fields, the analytical formulas for
the integrands in separate conductivity contribution terms  have been used. The integrals have been
calculated with adaptive integration steps in both directions ($|k|,\phi$) providing
high calculation accuracy (not less than 0.01\%).

%
In the second order approximation with nonzero-phases of the gauge fields,
we have to calculate numerically by introducing into consideration
a small positive damping constant $\gamma$ for the states as a small imaginary contribution
to the energies. The values
of $\gamma$ define the extend of smoothness of the singular behaviour of integrand
and does not influence
on the general form of the dependency curve in accordance to
fig.~\ref{q-max0-14simulation}c.

All quantities necessary for  calculation of the  complex conductivity have been calculated on a grid
in the space of wave vectors with 200 discretization    in angle $\phi$ for every given
wave number $ q$ and variable step (a denser grid at small wave numbers and larger at large ones) to the
maximum wave number 
$q_{max} = 0.44|K_A|$.
2D interpolation on this grid has been used to integrands evaluation
in the intermediate points that is necessary for conductivity simulations.
An error stipulated by the
interpolation from the grid in wave vectors space has been
roughly estimated by interpolation of the conic spectrum of Dirac
pseudo-fermion model on the same lattice with subsequent
usage of the interpolation data for evaluation of the
conductivity. Its value turns out to be less than $10^{-3}$\% . 

Total estimation of the conductivity error has been performed by variation of the number of points used for
interpolation of the energy band spectrum
(by diminishing this number at factor two and subsequent comparison of the simulation
results in both cases). It turns out to be not exceeding 10\% in the considered frequency  region.
It should be noted that
the error bars for values of the Fermi velocity which  was measured by different techniques including
transport experiments (Shubnikov -- de~Haas oscillations) \cite{NatPhys7-2011Elias},
infrared measurements of the Pauli blocking in graphene \cite{Z.Q.Li_et_al_Nature Phys.4_532(2008)},
magneto-optics \cite{magneto_optics_Fermi_velocity_NatCommun2014},
 were also of the order of 10\% .


\end{document}